\documentclass[11pt]{elsarticle}

\usepackage{fullpage}

\usepackage{wrapfig}
\usepackage{subfigure}
\usepackage{color}
\usepackage{graphicx}
\usepackage{float}
\usepackage{array}
\usepackage{amsmath,amsfonts,amssymb}

\DeclareMathAlphabet{\bi}{OT1}{cmr}{bx}{it}

\newcommand{\B}{{\bi B}}
\newcommand{\D}{{\bi D}}

\newcommand{\wh}[1]{ { \widehat{#1} } }
\newcommand{\wwh}[1]{ {\wh{\wh{#1}}} }
\newcommand{\wwhc}[1]{ {\wwh{\,#1\,}} }

\DeclareMathOperator{\RE}{Re}
\DeclareMathOperator{\IM}{Im}

\newcommand{\Zmax}{ { {\it Z_{\max}}  } }  
\newcommand{\Xmax}{ { {\it X_{\max}}  } } 
\newcommand{\Rmax}{ { {\it \rho_{\max}}  } } 

\newcommand{\nout}{ {n_0^{\text{ext}} } } 
\newcommand{\lri}{\nu}  

\newcommand{\degree}{^{\circ}}

\newcommand{\myeqdef}{\, \stackrel{\rm def}{=} \,}

\newcommand{\oh}[1]{ { \mathcal{O}\left(h^{#1}\right) } }
\newcommand{\CDO}[2]{ { D_{#1}^{ (#2) }  } }

\newcommand{\bvec}[1] {{\bi {#1}}}

\newcommand{\kperpl}{{k_\perp^{(l)}}}
\newcommand{\kparl}{{k_\parallel^{(l)}}}
\newcommand{\EincL}{ { E_{\text{inc}}^{0} } } 
\newcommand{\EincR}{ { E_{\text{inc}}^{Z_{\max}} } } 
\newcommand{\UincLl}{ { u_{\text{inc},\,l}^{0} } } 
\newcommand{\UincRl}{ { u_{\text{inc},\,l}^{Z_{\max}} } } 

\newcommand{\EincLtext}[1]{ { E_{\text{inc}}^{0,\text{#1}} } } 

\newcommand{\F}{\bvec{F}}
\newcommand{\E}{\bvec{E}}
\newcommand{\J}{\bvec{J}}

\DeclareMathOperator{\sech}{sech}
\DeclareMathOperator{\curl}{curl}

\graphicspath{{}}

\begin{document}

\begin{frontmatter}

\title{A High-Order Numerical Method for the Nonlinear Helmholtz    
	Equation in Multidimensional Layered Media}

\makeatletter

\author[Tel-Aviv]{G. Baruch\fnref{ISF}}
\ead{guy.baruch@math.tau.ac.il}
\ead[url]{http://www.tau.ac.il/$\sim$guybar}
\author[Tel-Aviv]{G. Fibich\fnref{ISF}}
\ead{fibich@tau.ac.il}
\ead[url]{http://www.math.tau.ac.il/$\sim$fibich}
\fntext[ISF]{
	The research of these authors was partially supported by the Israel Science
	Fund, Grant \#~123/08
}
\author[Raleigh]{S. Tsynkov\fnref{cor}}
\ead{tsynkov@math.ncsu.edu}
\ead[url]{http://www.math.ncsu.edu/$\sim$stsynkov}
\address[Tel-Aviv]{Department of Applied Mathematics,
School of Mathematical Sciences, Tel Aviv University, Ramat Aviv, Tel Aviv 69978, Israel}
\address[Raleigh]{Department of Mathematics,
North Carolina State University, Box 8205, Raleigh, NC 27695, USA}
\fntext[cor]{Corresponding author. Phone: +1-919-515-1877, Facsimile:
+1-919-513-7336.
The research of this author was supported by the US NSF, Grants \#~DMS-0509695 and 
\#~DMS-0810963,
and by the US Air Force, Grant \#~FA9550-07-1-0170.}

\makeatother

\begin{abstract}
We present a novel computational methodology for solving the scalar nonlinear
Helmholtz equation (NLH) that governs the propagation of laser light in
Kerr dielectrics.
The methodology addresses two well-known challenges in nonlinear optics: 
Singular behavior of solutions when the scattering in the medium is assumed
predominantly forward (paraxial regime),
and the presence of discontinuities in the 
optical properties of the medium.
Specifically, we consider a slab of nonlinear material which may be grated
in the direction of propagation and which is immersed in a linear medium as a whole.
The key components of the methodology are a semi-compact high-order
finite-difference scheme that maintains accuracy across the discontinuities and
enables sub-wavelength resolution on large domains at a tolerable cost, a nonlocal
two-way artificial boundary condition (ABC) that simultaneously facilitates the
reflectionless propagation of the outgoing waves 
and forward propagation of the given incoming waves, and a nonlinear solver
based on Newton's method.

The proposed methodology combines and substantially extends the capabilities
of our previous techniques built
for 1D and for multi-D. 
It facilitates a direct numerical study of nonparaxial
propagation and goes well beyond the approaches in the literature based on the
``augmented'' paraxial models. In particular, it provides the first ever evidence 
that the singularity of the solution  indeed
disappears in the scalar NLH model that includes the nonparaxial effects.
It also enables simulation of the wavelength-width spatial solitons, as well as of
the counter-propagating solitons.
\end{abstract}

\end{frontmatter}

\section{\label{sec:intro}Introduction}

\subsection{Mathematical Models}
\label{sec:intro_model}

The propagation of electromagnetic waves in materials is governed by Maxwell's
equations with appropriately chosen material responses.
The responses characterize the dependence of material properties --- magnetic
permeability, electric permittivity, and conductivity --- on the location and
frequency of the propagating field.
For high intensity radiation, the material quantities may also
depend on the magnitude of the propagating field, which makes the responses 
nonlinear.

In nonlinear optics, one is often interested in studying the propagation of
monochromatic waves (continuous-wave laser beams) through transparent 
dielectrics.
In this case, the generation of higher harmonics and nonlinear coupling between 
different (temporal) frequencies can often be neglected, and accordingly, a
time-harmonic solution can be assumed.
The magnetic field can then be eliminated, and Maxwell's equations transform to
a second-order differential equation with respect to the electric field, known
as the vector Helmholtz equation, see~\cite{Fibich-Ilan-vectorial-PhysicaD}. 
If the material is isotropic and, in addition, the electric field is assumed
linearly polarized, then one arrives at the scalar nonlinear Helmholtz
equation~(NLH):
\begin{equation}
    \Delta E(\bvec{x}) + \frac{\omega_0^2}{c^2}n^2 E = 0,\qquad 
        n^2(\bvec{x},|E|) = n_0^2(\bvec{x}) + 2n_0(\bvec{x})n_2(\bvec{x})|E|^{2\sigma},
    \label{eq:NLH}
\end{equation}
where $\sigma>0$ and $n$ is the refraction index. 
In physical materials one always has $\sigma=1$, so that the dependence of $n^2$
on $|E|$ is quadratic.
In equation~(\ref{eq:NLH}), $\bvec{x} = [x_1,\ldots,x_D]$ are the spatial
coordinates, $E=E(\bvec{x})$ denotes the scalar electric field,
$\omega_0$ is the laser frequency, $c$ is the speed of light in vacuum,
$\Delta=\partial_{x_1}^2+\ldots+\partial_{x_D}^2$ is the $D$-dimensional
Laplacian, $n_0$ is the linear index of refraction, and $n_2$ is the Kerr
coefficient.
Both $n_0$ and $n_2$ are assumed real, so that the medium is
transparent or lossless.
The coordinate $x_D$ will also be denoted by $z$ and
will hereafter be referred to as longitudinal, whereas the remaining
direction(s) $\bvec{x}_\perp=[x_1,\dots,x_{D-1}]$ will be called
transverse.

Our primary physical setup involves a slab of Kerr material
surrounded on both sides by the linear homogeneous medium in which
$n_0\equiv\nout$ and $n_2\equiv 0$, see
Figure~\ref{fig:physical_setup_a}.

\begin{figure}[h]
	\begin{center}
		\subfigure[The three-layer physical setup.]{%
                \scalebox{0.2525}{\input{setup.pstex_t}}%
                 \label{fig:physical_setup_a}%
		}
		\qquad
		\subfigure[The multi-layer physical setup.]{%
                \scalebox{0.25}{\input{setup-grating.pstex_t}}%
                 \label{fig:physical_setup_b}%
		}
	\end{center}
\end{figure}

We introduce the linear wavenumber $k_0=\omega_0\nout/c$ and the normalized
quantities $
	\lri(\bvec{x})=n_0(\bvec{x})/\nout
$ and $
	\epsilon(\bvec{x}) = 2n_0(\bvec{x})n_2(\bvec{x})/(\nout)^2
$, and then recast equation~(\ref{eq:NLH}) as
\begin{equation}
	\label{eq:NLH2}
	\Delta E(\bvec{x})+k_0^2\left(
		\lri^2(\bvec{x})+\epsilon(\bvec{x})|E|^{2\sigma}
	\right) E=0.
\end{equation}

Note that the  Kerr coefficient $n_2(\bvec{x})$ is always discontinuous at 
the interface planes $z=0$ and $z=\Zmax$, see Figure~\ref{fig:physical_setup_a}.
The linear index of refraction $n_0(\bvec{x})$ may also be
discontinuous at the interface planes.
Discontinuities in $n_0(\bvec{x})$ and $n_2(\bvec{x})$ immediately give
rise to those in $\lri(\bvec{x})$ and $\epsilon(\bvec{x})$, see
equation~(\ref{eq:NLH2}).
Thus, for the typical experimental setting that involves a slab of homogeneous
Kerr material,\footnote{
	This setup withstands an easy generalization to the case of multiple
	plane-parallel layers, see Section~\ref{sec:intro_multi}.
} the coefficients of equation~(\ref{eq:NLH2}) are piecewise-constant:
\begin{equation}
\label{eq:homogeneous-material} 
        \lri(z,\bvec{x}_\perp)=\begin{cases}
1, & z<0,\\
\lri^\text{int},  &  0\leq z \leq Z_{\max},\\
1, & z>Z_{\max},  
\end{cases}
\qquad\text{and}\qquad
        \epsilon(z,\bvec{x}_\perp)=\begin{cases}
0, & z<0,\\
\epsilon^\text{int},  &  0\leq z \leq Z_{\max},\\
0, & z>Z_{\max}.  
\end{cases}
\end{equation}
Discontinuities in the coefficients~(\ref{eq:homogeneous-material}) imply
that additional conditions will be required for the NLH~(\ref{eq:NLH2}) 
at the interfaces $z=0$ and $z=\Zmax$.
These conditions can be obtained by analyzing the corresponding Maxwell's
equations.
They reduce to the continuity of the field~$E(z)$ and its
first normal derivative~$\frac{\partial E}{\partial z}$,
see Appendix~\ref{app:interface}.
When building a numerical approximation, the presence of material
discontinuities requires special attention~(Sections~\ref{sec:1D}
and~\ref{sec:23D}).

\begin{figure}[H]
        \begin{center}
        \scalebox{0.3}{\input{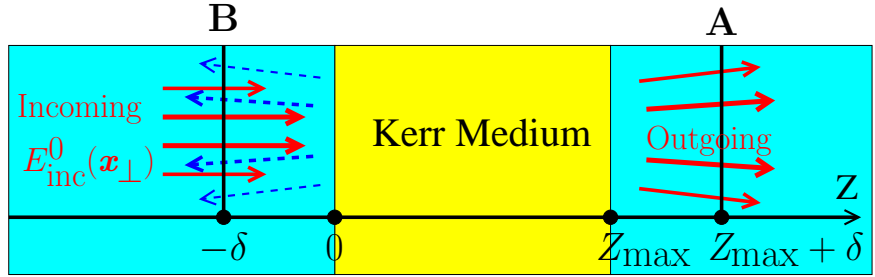}}
                \caption{\label{fig:BC_schematic}
                        Schematic of the boundary conditions in the longitudinal direction:
                        A) One-way radiation boundary condition at $z=\Zmax+\delta$;
                        B) Two-way radiation boundary condition at $z=-\delta$.
        }
        \end{center}
\end{figure}

The problem is driven by a laser beam that impinges on the Kerr material from
the outside and causes a local increase in the overall index of refraction as it
propagates through, see Figure~\ref{fig:physical_setup_a}.
Since light rays bend toward the areas with higher refraction index, the
impinging beam self-focuses inside the Kerr medium.
The material discontinuities at $z=0$ and $z=\Zmax$ 
reflect a portion of the forward propagating wave, resulting in
backward propagating waves.
Moreover, the nonlinearly induced nonuniformities of the refraction index may
also scatter the radiation backwards.
The presence of waves propagating in opposite directions implies that the
boundary conditions for the NLH~(\ref{eq:NLH2}) must ensure the reflectionless
propagation of all the outgoing waves (regardless of their direction of travel
and the angle of incidence at the outer boundary) and at the same time correctly
prescribe the given incoming beam at the boundary, see
Figure~\ref{fig:BC_schematic}.
Such boundary conditions are called {\em two-way boundary conditions}
\cite{FT:01}, see
Section~\ref{sec:1D} for 1D and Section~\ref{sec:23D} for multi-D.

One can also consider a simplified model that would account only for the
forward propagating component of the field.
Let $z\equiv x_D$ be the direction of the impinging laser beam, and let us also
consider the simplest case of $\lri\equiv1$ and $
	\epsilon\equiv \epsilon^{\text{int}}
$ inside the Kerr medium.
Then, introducing the ansatz $E=e^{ik_0z}\phi$, where $\phi=\phi(\bvec{x})$ is
assumed to vary slowly compared with the fast carrier oscillation $e^{ik_0z}$, one
can neglect the small $\phi_{zz}$ term (paraxial approximation), and reduce the
NLH~(\ref{eq:NLH2}) to the nonlinear Schr\"odinger equation (NLS):
\begin{equation}
\label{eq:NLS}
	2ik_0\phi_z(z,\bvec{x}_\perp) + \Delta_\perp\phi
		+k_0^2 \epsilon|\phi|^{2\sigma}\phi=0,
		\qquad 0\leq z \leq \Zmax,
\end{equation}
which governs the envelope $\phi$.
In equation~(\ref{eq:NLS}),
$\Delta_\perp=\partial_{x_1}^2+\ldots+\partial_{x_{D-1}}^2$ denotes the transverse
Laplacian. 
The NLS~(\ref{eq:NLS}) supports only forward propagation because the assumption
of slow variation of $\phi$ does not leave room for any $\sim e^{-ik_0z}$
components in the solution.
Equation~(\ref{eq:NLS}) is first order in $z$ and, unlike the NLH, requires a Cauchy
problem to be formulated and solved with the ``initial'' data provided by the
impinging wave and specified, say, at $z=0$ (see, e.g.,~\cite{Sulem,Boyd} for
detail).

It is well known that solutions of the NLS~(\ref{eq:NLS}) exist globally  when
$\left. \sigma (D-1) < 2\right.$, the subcritical NLS, but can become singular,
i.e., collapse at finite propagation distances, when either 
$\left.\sigma (D-1) > 2\right.$, the supercritical NLS, or 
$\left.\sigma (D-1) = 2\right.$, the critical NLS~\cite{Sulem}.
As shown by Weinstein~\cite{Weinstein-83}, a necessary condition for singularity
formation in the critical NLS is that the input power  
exceeds the critical power $P_{\rm c}$. 
The value of $P_{\rm c}$ is equal to the power of the ground-state solitary wave
solution of the NLS; it can be calculated analytically for $D=2$ and numerically
for $D >2$.

A question that has been open in the literature for over forty years is whether
the more comprehensive NLH model for nonlinear self-focusing eliminates the
singular behavior that characterizes collapsing solutions of the critical and
supercritical NLS.
Unfortunately, the fundamental issue of solvability of the NLH and regularity of
its solutions still remains unaddressed for many important settings. 
Only the one-dimensional case, when equation~(\ref{eq:NLH2}) becomes an ODE, has
been studied extensively, and exact solutions have been obtained using a
combination of analytical and numerical means 
\cite{Marburger-Felber:75,Wilhelm,chen-mills-PRB:87,chen-mills-PRB-ML:87,%
baghdasaryan-99,midrio-01,kwan-lu-04,petracek-06}. 
In multi-D, there have been indications that solutions of the NLH may exist even
when the corresponding NLS solutions become singular, based on both numerical
study of ``modified'' NLS equations
\cite{Akhmediev-93,Akhmediev-93b,Feit-Fleck}, and on asymptotic
analysis~\cite{Fibich-PRL:96}, but these studies did not account for backscattering.
Recently, Sever employed a Palais-Smale type argument and has shown that 
the multi-D NLH is solvable in the sense of $H^1$ and that the solution 
is not unique~\cite{sever:06}.
His argument, however, only applies to self-adjoint operators, whereas the 
physical setups considered in this study require radiation boundary conditions.

\subsection{Numerical Method}
\label{sec:intro_method}

The new computational methodology for the NLH that we present 
builds up on our previous work~\cite{BFT:07,FT:01,FT:04,BFT:05} 
and extends it substantially. We introduce a new 
semi-compact discretization and a new Newton's solver, and the ensuing capabilities
include an explicit demonstration of the removal of singularity that ``plagues'' the NLS,
and the computation of narrow nonparaxial solitons.

Specifically,
we solve the NLH~(\ref{eq:NLH2}) for two different cases.
The first one corresponds to the critical NLS ($\sigma(D-1)=2$).
We consider both the two-dimensional quintic nonlinearity $D=2$ and 
$\sigma=2$ (planar waveguides), and the three-dimensional cubic nonlinearity $D=3$ and
$\sigma=1$ (bulk Kerr medium, for which we additionally assume cylindrical symmetry).
As $\sigma(D-1)=2$ for either setting, one can expect that the role of
nonparaxiality and backscattering will be similar.
This study goes beyond the investigation of the
``modified'' NLS's
\cite{Akhmediev-93,Akhmediev-93b,Feit-Fleck,Fibich-PRL:96}, and the results 
reported in Section~\ref{ssec:arrest} 
{\em provide the first ever numerical evidence that the collapse of focusing
nonlinear waves is indeed arrested in the NLH model, which incorporates the
nonparaxiality and backscattering.}

The second case we analyze is that of a planar waveguide with cubic
nonlinearity ($D=2$  and $\sigma = 1$).
In this subcritical case, solutions to the NLS do not collapse.
Instead, the laser beam can propagate in the Kerr medium over very long
distances without changing its profile\footnote{
	In this case, self-focusing balances diffraction exactly.
}
--- the type of behavior often referred to as  spatial soliton. 
Solitons have been studied extensively as solutions to the NLS.
For beams that are much wider than the optical wavelength, it is
generally expected that the ``subcritical'' NLH will have similarly looking
solutions.
However, it was not until our paper~\cite{FT:04} that it has become actually
possible to study the effect of nonparaxiality and backscattering on solitons.
The methodology proposed in this paper allows us to go further and demonstrate
numerically {\em the existence and sustainability over long distances of very
narrow spatial solitons for the NLH,} basically as narrow as one carrier
wavelength $\lambda=2\pi/k_0$, see Section~\ref{ssec:solitons}.
Furthermore, the NLH appears particularly well suited for modeling interactions
between counter-propagating solitons, as a boundary value problem can
naturally be formulated.
In the NLS framework, on the other hand, the two
counter-propagating solitons will imply two opposite directions of
marching.\footnote{
	Counter-propagating beams have been simulated using two coupled 
	NLSs~\cite{Cohen-counterprop:02}, but this approach involves some approximations 
	which are not needed in the NLH, and whose validity is unclear.
}

The discrete approximation of the NLH 
must be high order so as to minimize the number of points
per wavelength required for solving equation~(\ref{eq:NLH2}) with
sub-wavelength resolution on a large domain, and for resolving the
small-scale phenomenon of backscattering against a background of the
forward-propagating wave.
It must also maintain its accuracy across the material
discontinuities.
As the geometry is simple, and the discontinuities are only in 
the longitudinal direction, we can 
approximate the NLH by finite differences on a rectangular grid.
In the case $D=2$, it will be a Cartesian grid of coordinates~$(x,z)$.
In the case $D=3$, we still want to have only two independent spatial 
variables and hence employ cylindrical symmetry. 
The NLH~(\ref{eq:NLH2}) is then approximated on the rectangular grid of
cylindrical coordinates $(\rho,z)$, where $\rho=(x^2+y^2)^{1/2}$.
In doing so, the discontinuities that are confined to transverse planes
will always be aligned with the grid.

\begin{figure}[ht!]
\begin{center}
\includegraphics[width=0.6\textwidth,clip=true]{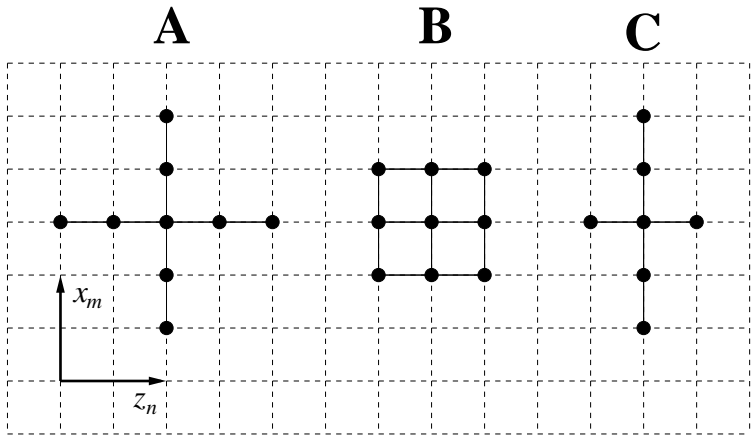}
        \caption{\label{fig:stencils}
                Stencils in $2D$:
                A) Standard central difference fourth-order stencil, as in our
                previous work~\cite{FT:01,FT:04,BFT:05};
                B) Compact $3\times3$ fourth-order stencil for linear operators,
                   as, e.g., in~\cite{harari-turkel-95,singer-turkel-98};
                C) Semi-compact stencil used in this work.
        }
\end{center}
\end{figure}

In our work~\cite{FT:01,FT:04}, we used the standard fourth-order
central differences (five node stencil in each coordinate direction) to
approximate the NLH~(\ref{eq:NLH2}) on a rectangular grid, see
Figure~\ref{fig:stencils}A.
While this approach works well in the regions of smoothness, it deteriorates to
second-order accuracy in regions of material discontinuities.
In the recent paper~\cite{BFT:07}, we discretized the one-dimensional NLH with
fourth-order accuracy using compact finite volumes and a three node stencil.
This discretization handled the material discontinuities with no deterioration
of accuracy and was also extended to higher orders in the linear
case~\cite{BFTT:07}.
However, the extension of the scheme of~\cite{BFT:07} to multi-D is not
straightforward.
Therefore, in the current paper we adopt a hybrid approach. 
We use the standard fourth-order central differences in the transverse
direction, and a compact fourth-order finite difference 
discretization on three nodes in the
longitudinal direction, see Figure~\ref{fig:stencils}C.

The five node transverse stencil does not impair the
accuracy because there are no discontinuities in that direction.
The three node longitudinal part of the scheme is supplemented by
one-sided differences that implement the required interface conditions
at the points of discontinuity. In doing so,
the compact stencil eliminates the need to use
those special differences anywhere except at the 
discontinuities themselves.
Another advantage of having a three node compact stencil in the longitudinal 
direction is that it leads to matrices with a narrower bandwidth.

The interior discretization is supplemented by nonlocal two-way artificial
boundary conditions (ABCs) set at $z=-\delta$ and $z=Z_{\max}+\delta$, see
Figure~\ref{fig:BC_schematic}, and by local radiation boundary conditions at the
transverse far-field boundaries.
The discrete ABCs are similar to those of \cite{FT:04}, but having a three node
compact stencil greatly simplifies their
construction because, unlike in the case of a five node stencil, there are
no additional evanescent modes in the discretization, 
see Section~\ref{ssec:discrete-LBCs}.

The solver employed in~\cite{FT:01,FT:04,BFT:05} was of a fixed-point type. 
On the outer iteration loop, the nonlinearity in equation~(\ref{eq:NLH2}) was
frozen, and a linear Helmholtz equation with variable coefficients was obtained.
This linear equation was then solved iteratively on the inner loop, essentially by
building a sequence of Born approximations~\cite{born-wolf}.
This double-loop iterative method was shown to converge for (subcritical) solitons
and for ``mild'' critical cases, but has never been able to produce convergent
solutions for incoming beams that become singular in the NLS model.

In~\cite{BFT:07}, we have demonstrated that the iterations' convergence 
in~\cite{FT:01,FT:04,BFT:05} breaks down far below the power
threshold for non-uniqueness of the one-dimensional problem.
This suggested that the convergence difficulties 
in~\cite{FT:01,FT:04,BFT:05} were not related to the loss of uniqueness by the
solution \cite{sever:06,chen-mills-PRB:87}, but rather to
the deficiencies of the iteration scheme itself.
The latter may be (partially) accounted for by 
the known convergence
limitations of the Born approximations,
because they can be
interpreted as a Neumann series~\cite{tricomi:85} for the corresponding integral
operator \cite[Section 13.1.4]{born-wolf}.

An alternative iteration proposed in~\cite{BFT:07} is based on Newton's
method.
As, however,  the Kerr nonlinearity is Frech\'et nondifferentiable 
for complex-valued $E$,  for Newton's method to apply
the NLH has to be recast as a system of two equations with real
unknowns. 
The one-dimensional numerical experiments of~\cite{BFT:07}
demonstrate robust convergence of Newton's iterations for a wide range of
input powers.
Therefore, in this paper we implement Newton's method for solving the
multi-dimensional NLH~(\ref{eq:NLH2}), see Section~\ref{sec:Newton}.
As shown in Section~\ref{ssec:arrest}, the method converges for initial
conditions that lead to singularity formation in the critical NLS model,
for both $D=2$ and $D=3$.

\subsection{Extension to the Multi-Layer Case}
\label{sec:intro_multi}

Instead of having a homogeneous Kerr material in the nonlinear region as
shown in Figure~\ref{fig:physical_setup_a}, we can analyze the case of a layered
(grated) material as shown in Figure~\ref{fig:physical_setup_b}.
In doing so, the linear material outside of the Kerr slab still remain homogeneous.

The corresponding extension of the mathematical model is straightforward. It
amounts to introducing a fixed partition of the interval $[0,Z_{\max}]$:
\begin{subequations}
        \label{eqs:grated-material}
        \begin{equation} \label{eq:grating}
        0 = \tilde{z}_0 < \dots < \tilde{z}_l < \dots < \tilde{z}_L = Z_{\max},
        \end{equation}
so that the material characteristics are constant within each sub-interval:
        \begin{equation}\label{eq:grating_b}
                \lri(z,\bvec{x}_\perp)\equiv\tilde{\lri}_l,\quad
        \epsilon(z,\bvec{x}_\perp)\equiv\tilde{\epsilon}_l, \quad \text{for} \quad
        z\in\left(\tilde{z}_l, \tilde{z}_{l+1}\right),
        \end{equation}
\end{subequations}
whereas at the interfaces~(\ref{eq:grating}) they may undergo jumps.
Altogether, this leaves the coefficients of equation~(\ref{eq:NLH2}) 
piecewise constant in $z$. 
The additional interface conditions required by equation~(\ref{eq:NLH2})
are the same as before --- continuity of $E$ and 
$\frac{\partial E}{\partial z}$ between the layers, see Appendix~\ref{app:interface}.

\subsection{Structure of the Paper}
\label{sec:intro_struc}

In Section~\ref{sec:1D}, we illustrate the main concepts of the continuous
formulation and the discretization for the one-dimensional NLH. 
In Section~\ref{sec:23D}, we describe the continuous formulation of the problem
and the discretization for the two-dimensional Cartesian NLH and for the three
dimensional NLH with cylindrical symmetry.
In Section~\ref{sec:Newton}, we introduce Newton's solver for the resulting
system of nonlinear equations on the grid.
Section~\ref{sec:method-summary} provides a summary on the numerical method, 
Section~\ref{sec:interface} relates the input beams for the NLH and the corresponding
NLS models, and Section~\ref{sec:results} contains the results of simulations.
Finally, Section~\ref{sec:discussion} presents our conclusions and outlines 
directions for future work.
Note also that some of the results shown hereafter were previously reported 
in~\cite{BFT-opex:08}.
That paper, however, did not contain any description of the numerical method.

\section{\label{sec:1D}The NLH in One Space Dimension}

In this section we consider the one-dimensional NLH with constant material
coefficients $\lri^2$ and $\epsilon$ for $0<z<\Zmax$, which means that
there are two discontinuities at  $z=0$ and $z=\Zmax$, but no
discontinuities in the interior of the Kerr slab, see Figure~\ref{fig:physical_setup_a}.
In Section~\ref{ssec:continuous-1D}, we present the continuous formulation of
the problem, and in Section~\ref{ssec:discrete-1D} we introduce a compact
discrete approximation.
In Section~\ref{sssec:grated-1D}, we briefly discuss the extension to the 
multi-layer case outlined is Section~\ref{sec:intro_multi}.
This simple one-dimensional case illustrates the key ideas and notations that will
be used later in the more complex multi-dimensional cases.

\subsection{\label{ssec:continuous-1D}Continuous Formulation}

Consider a homogeneous slab of the Kerr material immersed in an infinite linear medium.
\begin{subequations}
	\label{eqs:1DNLH-system}
	The propagation of the electric field is governed by the 1D~NLH 
	equation inside the Kerr material:
	\begin{equation}
	    \frac{d^2E(z)}{dz^2}
	    + k_0^2\left(\lri^2 + \epsilon\left|E\right|^{2\sigma}\right) E = 0,
		 \qquad 0 < z < \Zmax,
		\label{eq:1DNLH} 
	\end{equation}
	and by the linear Helmholtz equation outside the Kerr material:
	\begin{equation}
	    \frac{d^2E(z)}{dz^2} + k_0^2 E = 0,
		\qquad z< 0 \ \ \text{or} \ \ z > \Zmax. 
		\label{eq:1DLH}
	\end{equation}
	At the material interfaces $z=0$ and $z=\Zmax$, the field and its first
	derivative must be continuous~\cite{BFT:07}:
	\begin{equation}
		\begin{gathered}
			\label{eq:1DNLH-continuity}
		   	E(0+)=E(0-), \qquad 
		   	\frac{dE}{dz}(0+)=\frac{dE}{dz}(0-),  \\[1mm]
		   	E(\Zmax+)=E(\Zmax-), \qquad 
		   	\frac{dE}{dz}(\Zmax+)=\frac{dE}{dz}(\Zmax-). 
		\end{gathered}
	\end{equation}
\end{subequations}

We consider the case of two incoming waves with known characteristics that
travel toward the Kerr region $[0,\Zmax]$ from $z=-\infty$ to the right
and from $z=+\infty$ to the left.\footnote{
	Hereafter, we slightly generalize the schematic depicted in
	Figure~\ref{fig:BC_schematic}, in that we allow for incoming waves to impinge
	on {\em both} interfaces, at $z=0$ and $z=\Zmax$.
}
The overall field may also have scattered components, which are outgoing with
respect to the domain $[0,\Zmax]$ and which are not known ahead of time.
The general solution to equation~(\ref{eq:1DLH}) outside $[0,\Zmax]$ is a
superposition of the right-propagating wave $e^{ik_0z}$ and the left-propagating
wave $e^{-ik_0z}$.
Consequently, the field outside $[0,\Zmax]$ shall be sought for in the form:
\begin{equation}
	\label{eq:continuous-E-outside}
	E(z) = \begin{cases}
		\EincL e^{ik_0z} + C_1 e^{-ik_0z}, & -\infty<z \leq 0, \\
		C_2 e^{ik_0(z-\Zmax)} + \EincR e^{-ik_0(z-\Zmax)}, & 
			\Zmax\leq z < \infty,
	\end{cases}
\end{equation}
where $\EincL$ is a given amplitude of the incoming wave that travels to the
right from $z=-\infty$ and impinges on the Kerr medium at $z=0$,
whereas $C_1$ is the amplitude of the outgoing wave traveling to the left
toward $z=-\infty$, which is not known ahead of time\footnote{
	Physically, the left-traveling outgoing wave $C_1e^{-ik_0z}$ may have two
	sources: A portion of the right-traveling wave $\EincL e^{ik_0z}$ may get
	{\em scattered} to the left by the Kerr material slab, and a portion of the
	left-traveling wave $\EincR e^{-ik_0z}$ may be {\em transmitted} through by the
	Kerr material slab. 
	In the nonlinear problem, these phenomena are coupled and cannot be easily
	distinguished from one another.
}.
Likewise, $\EincR$ is a given amplitude of the incoming wave that travels to
the left from $z=+\infty$ and impinges on the Kerr medium at $z=\Zmax$,
whereas $C_2$ is the amplitude of the outgoing right-traveling wave, which is
not known ahead of time.

Representation~(\ref{eq:continuous-E-outside}) is to be enforced by the 
ABCs that should prescribe the given values of $\EincL$ and $\EincR$ and at the same
time allow for the arbitrary values of $C_1$ and $C_2$.
In~\cite{BFT:07}, we have set such ABCs precisely at the material interfaces, and
have shown that they were given by the inhomogeneous Sommerfeld type
relations:
\[
	\left.
		\left( \frac{d}{dz}+ik_0 \right) E 
	\right|_{z=0} 
		= 2ik_0 \EincL,
	\qquad
	\left.
		\left( \frac{d}{dz}-ik_0 \right) E 
		\right|_{z=\Zmax} 
		= -2ik_0 \EincR.
\]
In this paper, we set equivalent ABCs at a certain distance $\delta>0$ away from
the interfaces, see Figure~\ref{fig:BC_schematic}, and inside the linear regions:
\begin{equation}
    \label{eq:1DTWBCs}
	\begin{aligned}
	\left.
		\left( \frac{d}{dz}+ik_0 \right) E 
	\right|_{z=-\delta} 
		= &\> 2ik_0 e^{-ik_0\delta} \EincL, \\
	\left.
		\left( \frac{d}{dz}-ik_0 \right) E 
	\right|_{z=\Zmax+\delta} 
		= &\> -2ik_0 e^{-ik_0\delta} \EincR.
	\end{aligned}
\end{equation} 
As we shall see, the separation between the material interfaces $z=0$ and
$z=\Zmax$ and artificial boundaries $z=-\delta$ and $z=\Zmax+\delta$ simplifies
the discretization of the problem, because the continuity
conditions~(\ref{eq:1DNLH-continuity}) and the boundary
conditions~(\ref{eq:1DTWBCs}) can be discretized independently of each other,
see Sections~\ref{sssec:discrete-1D-interface} and
\ref{sssec:discrete-1DTWBCs}, respectively.

\subsection{\label{ssec:discrete-1D}Discrete Approximation}

The one-dimensional problem~(\ref{eqs:1DNLH-system}),~(\ref{eq:1DTWBCs}) will be
approximated using compact fourth-order finite differences.
We first discuss the discrete approximation of equations~(\ref{eq:1DNLH})
and~(\ref{eq:1DLH}), then the approximation of the interface
condition~(\ref{eq:1DNLH-continuity}), and finally the approximation of the
two-way ABCs~(\ref{eq:1DTWBCs}).
In what follows, we introduce some notations that will be particularly helpful
in multi-D.

We begin with setting up a uniform grid of $N+7$ nodes on $[-\delta,\Zmax+\delta]$: 
\begin{equation}
	\label{eq:grid-1D}
	z_n = n\cdot h, \qquad
	h = \frac{\Zmax}{N}, \qquad
	n= -3,-2,\dots,N+2,N+3,
\end{equation}
so that \[
	z_0 = 0,\qquad z_N = \Zmax, \qquad \delta = 3h.
\]
We also  denote by $E_n$ and $P_n = |E_n|^{2\sigma}E_n$ the values of $E$ and of
$|E|^{2\sigma}E$ at the grid nodes $z_n$.
Finally, we introduce a special notation $D$ for central difference operators, with
the order of accuracy in the superscript and the differentiation variables in
the subscript.
For example, 
\begin{equation*}
	\CDO{zz}{2} E \myeqdef 
		\frac{E_{n+1}-2E_n+E_{n-1}}{h^2} 
		= \left. \frac{d^2E}{dz^2} \right|_{z=z_n} + \oh{2}.
\end{equation*}

\subsubsection{\label{sssec:discrete-1D-equation}Approximation of the Equation}

Inside the Kerr medium, i.e., for $n=1,\dots,N-1$, the material coefficients
$\lri^2$ and $\epsilon$ are constant, and hence the field $E(z)$ is smooth.
Using Taylor's expansion of the field, we obtain from the 
standard second-order central difference
approximation:
\begin{equation}
	\label{eq:A-1D}
	\CDO{zz}{2} E = 
		\frac{E_{n+1}-2E_n+E_{n-1}}{h^2}
		= \partial_{zz}E_n + \frac{h^2}{12}\partial_{zzzz}E_n + \oh{4}.
\end{equation}
Then, recasting the one-dimensional NLH~(\ref{eq:1DNLH}) as 
\[
	\partial_{zz}E_n = -k_0^2\left(
		\lri^2+\epsilon|E_n|^{2\sigma}
	\right) E_n
	= -k_0^2(\lri^2 E_n+\epsilon P_n),
\]
we can approximate the term $\partial_{zzzz}E_n$ on the right-hand side
of~(\ref{eq:A-1D}) with second-order accuracy as
\[
	\partial_{zzzz}E_n = \CDO{zz}{2} \partial_{zz} E_n + \oh{2}
		= -k_0^2 \CDO{zz}{2} \left(
			\lri^2 E_n+ \epsilon P_n 
		\right)+ \oh{2}.
\]
This yields a compact fourth-order approximation for the second derivative:
\[
	\partial_{zz}E_n = \CDO{zz}{2} E_n
		+ \frac{h^2k_0^2}{12} \CDO{zz}{2} 
		\left( \lri^2 E_n + \epsilon P_n \right)
	+\oh{4}.
\]
Then, the resulting  scheme for the one-dimensional NLH~(\ref{eq:1DNLH}) at the
interior nodes reads:
\begin{equation}\begin{gathered}
	\label{eq:discrete-1DNLH}
	\CDO{zz}{2} E_n 
	+ k_0^2\left(1+\frac{h^2}{12} \CDO{zz}{2}\right) 
		\left( \lri^2 E_n + \epsilon P_n \right) = 0, \\
	n = 1,\dots,N-1.
\end{gathered}\end{equation}
This approach is sometimes called 
{\em an equation-based approximation}~\cite{singer-turkel-98}.

Outside the Kerr medium, i.e., for $n<0$ and $n>N$, the foregoing derivation is
repeated with $\lri^2\equiv 1$ and $\epsilon\equiv 0$, which yields a compact
fourth-order approximation of the linear Helmholtz equation~(\ref{eq:1DLH}):
\begin{equation}
	\label{eq:discrete-1DLH}
        \begin{gathered}
	\left(1+\frac{k_0^2h^2}{12}\right) \CDO{zz}{2} E_n 
	+ k_0^2 E_n = 0, \\
	n = -3,-2,-1,\quad\text{and}\quad n=N+1,N+2,N+3.
        \end{gathered}
\end{equation}
Note that equation~(\ref{eq:discrete-1DLH}) for the outermost grid nodes $n=-3$
and $n=N+3$ will involve the ghost values $E_{-4}$ and $E_{N+4}$, respectively.
These ghost values will be determined from the discrete two-way ABCs, see
Section~\ref{sssec:discrete-1DTWBCs}.

\subsubsection{\label{sssec:discrete-1D-interface}Approximation at the
Interfaces}

At the material interfaces $z=0$ and $z=\Zmax$ (i.e., grid nodes $n=0$ and
$n=N$) the discretized field is given by $E_0$ and $E_N$,
respectively.
Hence, the continuity of $E$ at the interface is automatically guaranteed, and only
the continuity of $E_z$, see formula~(\ref{eq:1DNLH-continuity}), requires
special attention.
The latter is enforced by approximating the derivatives at the interfaces with
fourth-order one-sided finite differences.
We again use the differential equation to eliminate one grid point from the 
one-sided stencil and reduce it from the conventional five nodes to four. 
While reducing the size of the stencil at the interface is not as important as
in the interior and exterior of the Kerr material, numerical observations show
that in some cases it may bring down the truncation error at the interface by a
factor of two.

Using Taylor's expansion and the one-dimensional NLH~(\ref{eq:1DNLH}), we can write:
\begin{align*}
	\left. \frac{dE}{dz} \right|_{z=0+}
	=& \>
	\frac{
		-85E_0 + 108E_1 - 27E_2 + 4E_3
	}{ 
		66 h
	}
	- \frac{3h}{11} \left. 
		\frac{d^2E}{dz^2} 
	\right|_{z=0+} 
	+\oh{4}\\
	=&\>
	\frac{
		-85 E_0 + 108E_1 - 27E_2 + 4E_3
	}{ 
		66 h
	}
	+\frac{3k_0^2 h}{11} 
		\left( \lri^2_{0+} E_0 + \epsilon_{0+}P_{0+} \right)  +\oh{4}.
\end{align*}
Repeating the calculation for $E_z(0-)$ and equating the resulting
approximations for $E_z(0-)$ and $E_z(0+)$, we have:
\begin{gather}
		\frac{
			4E_{-3} - 27E_{-2} + 108E_{-1} -170E_0 + 108E_1 - 27E_2 +4E_3 
		 } {66 h} \nonumber \\
	\label{eq:discrete-1D-interface-general}
		 +  \frac{6hk_0^2 }{11}
			\left(
				\frac{\lri^2_{0-}+\lri^2_{0+}}{2}E_0 
				+ \frac{\epsilon_{0-}+\epsilon_{0+}}{2} P_0
			\right) = 0.
\end{gather}
\begin{subequations}
	\label{eqs:discrete-1D-interface}
	Then, substituting $
		\lri_{0-}=1,\; 
		\epsilon_{0-}=0,\;
		\lri_{0+}=\lri
	$ and $	 
		\epsilon_{0+}=\epsilon,
	$ we obtain:
	\begin{gather}
			\frac{
				4E_{-3} - 27E_{-2} + 108E_{-1} -170E_0 + 108E_1 - 27E_2 +4E_3 
			 } {66 h} \nonumber \\
			 +  \frac{6hk_0^2 }{11}
				\left(
					\frac{1+\lri^2}{2}E_0 + \frac\epsilon 2 P_0
				\right) = 0.
	\end{gather}
	A similar equation is obtained for the interface at $n=N$:
	\begin{gather}
			\frac{
				4E_{N-3} - 27E_{N-2} + 108E_{N-1} -170E_N
					+ 108E_{N+1} - 27E_{N+2} +4E_{N+3} 
			 } {66 h} \nonumber \\
			 +  \frac{6hk_0^2 }{11}
				\left(
					\frac{1+\lri^2}{2}E_N + \frac \epsilon 2 P_N
				\right) = 0.
	\end{gather}
\end{subequations}

\subsubsection{\label{sssec:discrete-1DTWBCs}Two-Way Boundary Conditions}

At the exterior nodes $n<0$ and $n>N$, the one-dimensional Helmholtz
equation~(\ref{eq:1DLH}) is approximated with fourth-order accuracy by the 
constant coefficient homogeneous difference equation~(\ref{eq:discrete-1DLH}).
This equation can be recast as 
\begin{equation}
\label{eq:1Dscheme}
	\frac{
		E_{n+1}-2 E_n+E_{n-1}
	}{ h ^2 } 
	+ k^2 = 0, \quad\text{where}\quad  k^2 = \frac{1}{1+k_0^2h^2/12}k_0^2.
\end{equation}
The general solution of equation~(\ref{eq:1Dscheme}) is of the form 
$E_n=C_+ q^n + C_-q^{-n}$, where \[
	q = r + i\sqrt{1-r^2}\quad\text{and}\quad
	q^{-1} = r - i\sqrt{1-r^2}
\]
are roots of the corresponding characteristic equation $q^{-1} -2/r + q = 0$,
and $r =(1 - k^2h^2/2)^{-1}$.
These roots are complex conjugate and have unit magnitudes. 
Moreover, they satisfy $q = e^{ik_0h}\left(1+\oh{5}\right)$ and 
$q^{-1} = e^{-ik_0h}\left(1+\oh{5}\right)$.
Hence, the discrete solution $q^n$ approximates the right-going wave
$e^{ik_0nh}=e^{ik_0z}$, and the discrete solution $q^{-n}$ approximates 
the left-going wave $e^{-ik_0nh}=e^{-ik_0z}$, with fourth-order accuracy.
Consequently, the discrete counterpart of
equation~(\ref{eq:continuous-E-outside}) is 
\begin{equation}
	\label{eq:discrete-E-outside}
	E_n = \begin{cases}
		\EincL q^n + C_1 q^{-n}, &	
			-\infty< n \leq 0, \\
		C_2 q^{n-N} + \EincR e^{-(n-N)}, &  
			N \leq n < \infty.
	\end{cases}
\end{equation}
	Applying equation~(\ref{eq:discrete-E-outside}) at $n=-3$ and $n=-4$,
we can eliminate the unknown constant $C_1$ and 
	express the value of the field at the ghost node $n=-4$ as 
\begin{subequations}
	\label{eqs:discrete-1DTWBCs}
	\begin{equation}
	\label{eqs:discrete-1DTWBCs_a}
		E_{-4} = (q^{-1}-q)q^{-3}\EincL + qE_{-3}.
	\end{equation}
Likewise, applying equation~(\ref{eq:discrete-E-outside}) at
 $n=N+3$ and $n=N+4$, we obtain: 
	\begin{equation}
	\label{eqs:discrete-1DTWBCs_b}
		E_{N+4} = (q^{-1}-q)q^{-3}\EincR + qE_{N+3}.
	\end{equation}
\end{subequations}
Relations~(\ref{eqs:discrete-1DTWBCs}) provide a fourth-order accurate
approximation to the boundary conditions~(\ref{eq:1DTWBCs}) for $\delta=3h$. 
Relation~(\ref{eqs:discrete-1DTWBCs_a}) is substituted into
equation~(\ref{eq:discrete-1DLH}) for $n=-3$ and
relation~(\ref{eqs:discrete-1DTWBCs_b}) is substituted into
equation~(\ref{eq:discrete-1DLH}) for $n=N+3$.
This eliminates the ghost values from scheme~(\ref{eq:discrete-1DLH})
and closes the system of difference equations on the grid~(\ref{eq:grid-1D}).

\subsection{\label{sssec:grated-1D}Extension to the Multi-Layer Material}

In the case of a grated Kerr material  described in Section~\ref{sec:intro_multi},
there are additional discontinuity
points  defined by formula~(\ref{eq:grating}).
The interface conditions at each discontinuity
 point $\tilde z$ are the same as at $z=0$ and
$z=\Zmax$:
\[
  	E(\tilde z+)=E(\tilde z-), \qquad 
   	\frac{dE}{dz}(\tilde z+)=\frac{dE}{dz}(\tilde z-).
\]
Hence, 
in the simple case when $\tilde z$ happens to be at one of the grid
nodes, the discrete continuity condition at $\tilde z$ is given
by the same expression as~(\ref{eq:discrete-1D-interface-general}).
If the discontinuity point does not coincide with any grid node, one
can construct a separate uniform grid for each sub-interval, 
and the extension to the multi-layer case will then be
straightforward.

\section{\label{sec:23D}The NLH in Two and Three Space Dimensions}

\subsection{\label{ssec:continuous-23D}Continuous Formulation}

Here, we build a continuous formulation for the case of a
homogeneous slab of the Kerr material which occupies the region
$0\leq z \leq \Zmax$, see Figure~\ref{fig:physical_setup_a}.
As in the one-dimensional setting, we will later generalize the method to the
multi-layer case, see Section~\ref{ssec:grated-23D}.

We first consider the two-dimensional Cartesian geometry case 
$\bvec x \equiv (z,x)$.
This case models the physical case of propagation in planar waveguides, where
the dynamics in $y$ can be neglected.
In this case, the computational domain is truncated in the transverse direction
to $x\in [-\Xmax,\Xmax]$.
In the longitudinal direction, we truncate the computational domain at a
certain distance $\delta$ from the interfaces,
to $z\in \left[-\delta, \Zmax+\delta\right]$.
\begin{subequations}
	As before, the electric field is governed by the scalar NLH equation inside
	the Kerr medium [cf.\ equation~(\ref{eq:NLH2})]:
	\label{eqs:2DNLH-system}
	\begin{equation}\begin{gathered}
	    E_{zz}(z,x) + E_{xx}
	    + k_0^2\left( 
			\lri^2 + \epsilon \left|E\right|^{2\sigma} 
		\right) E = 0,\\
		(z,x) \in (0,\Zmax)\times [-\Xmax,\Xmax],
		\label{eq:2DNLH}
	\end{gathered}\end{equation}
	and by the linear Helmholtz equation outside the Kerr medium (where
	$\lri\equiv 1$ and $\epsilon\equiv 0$):
	\begin{equation}\begin{gathered}
	    E_{zz}(z,x) + E_{xx}  + k_0^2 E = 0,\\
		(z,x) \in 
		 	\left\{
				[-\delta,0)\cup(\Zmax,\Zmax+\delta]
			\right\}\times[-\Xmax,\Xmax].
		\label{eq:2DLH}
	\end{gathered}\end{equation}
	At the material interfaces $z=0$ and $z=\Zmax$, the field $E$ and its 
	normal derivative $E_z$ are continuous for all $x\in[-\Xmax,\Xmax]$:
	\begin{equation}
		\label{eq:2DNLH-continuity}
           \begin{gathered}
	   			E(0+,x)=E(0-,x), \quad 
	   			E_z(0+,x)=E_z(0-,x),\\
	   			E(\Zmax+,x)=E(\Zmax-,x), \quad 
			   	E_z(\Zmax+,x)=E_z(\Zmax-,x). 
           \end{gathered}
	\end{equation}
\end{subequations}

We also consider the case of three spatial dimensions, which models the 
propagation in bulk medium.
In order to reduce the computational costs, we assume that the
field is cylindrically symmetric $E(\bvec x)\equiv E(z,\rho)$, where 
$\rho = |\bvec x_\perp|=\sqrt{x^2+y^2}$.
This enables us to solve the problem with only two independent spatial 
variables.
In this case, the computational domain in the transverse direction is
$\rho \in [0,\Rmax]$,
	and the scalar NLH equation inside the Kerr medium is
\begin{subequations}
	\label{eqs:3DNLH-system}
	\begin{equation}\begin{gathered}
	    E_{zz}(z,\rho) + E_{\rho\rho} + \frac 1 \rho E_\rho
	    + k_0^2\left( 
			\lri^2 + \epsilon \left|E\right|^{2\sigma}
		\right) E = 0,\\
		(z,\rho) \in (0,\Zmax)\times[0,\Rmax].
		\label{eq:3DNLH}
	\end{gathered}\end{equation}
	The linear Helmholtz outside the Kerr medium is
	\begin{equation}\begin{gathered}
	    E_{zz}(z,\rho) + E_{\rho\rho} +\frac 1 \rho E_\rho + k_0^2 E = 0,\\
		(z,\rho) \in 
		 	\left\{
				[-\delta,0)\cup(\Zmax,\Zmax+\delta]
			\right\}\times[0,\Rmax],
		\label{eq:3DLH}
	\end{gathered}\end{equation}
	and the continuity conditions at the planar interfaces are
	\begin{equation}
		\label{eq:3DNLH-continuity}
           \begin{gathered}
				E(0+,\rho)=E(0-,\rho), \quad 
				E_z(0+,\rho)=E_z(0-,\rho),\\
				E(\Zmax+,\rho)=E(\Zmax-,\rho), \quad 
				E_z(\Zmax+,\rho)=E_z(\Zmax-,\rho). 
           \end{gathered}
	\end{equation}
\end{subequations}

We shall sometimes find it convenient to adopt a general notation for both
cases, by denoting the scalar transverse coordinate as $
	x_\perp = |\bvec x_\perp|
$ and its domain by $\Omega_\perp$.
In the Cartesian case we have $x_\perp \equiv x$ and 
$\Omega_\perp = [-\Xmax,\Xmax]$, while in the cylindrically symmetric case we
have $x_\perp \equiv \rho$ and $\Omega_\perp = [0,\Rmax]$.
We shall also find it convenient to decompose the Laplacian as 
$\Delta = \partial_{zz} + \Delta_\perp$, where $\Delta_\perp = \partial_{xx}$ 
in the Cartesian case and
$\Delta_\perp = \frac{1}{\rho}\partial_\rho(\rho\partial_\rho)\equiv
\partial_\rho^2 + \frac{1}{\rho}\partial_\rho$ in the cylindrically symmetric
case.
Physically, the transverse Laplacian term $\Delta_\perp E$ leads to diffraction.

Using this notation, the Cartesian system~(\ref{eqs:2DNLH-system}) and the
cylindrically symmetric case system~(\ref{eqs:3DNLH-system}) are universally 
represented as
\begin{subequations}
	\label{eqs:23DNLH-system}

	\begin{equation}\begin{gathered}
	    E_{zz}(z,x_\perp) + \Delta_\perp E
	    + k_0^2\left(
			\lri^2 + \epsilon\left|E\right|^{2\sigma}
		\right) E = 0,\\
		(z,x_\perp) \in (0,\Zmax)\times\Omega_\perp,
		\label{eq:23DNLH}
	\end{gathered}\end{equation}
	
	\begin{equation}\begin{gathered}
	    E_{zz}(z,x_\perp) + \Delta_\perp E  + k_0^2 E = 0,\\
		(z,x_\perp) \in 
		 	\left\{
				[-\delta,0)\cup(\Zmax,\Zmax+\delta]
			\right\}\times\Omega_\perp,
		\label{eq:23DLH}
	\end{gathered}\end{equation}
	
	\begin{equation}\begin{gathered}
	   	E(0+,x_\perp)=E(0-,x_\perp), \quad 
	   	E_z(0+,x_\perp)=E_z(0-,x_\perp),\\
	   	E(\Zmax+,x_\perp)=E(\Zmax-,x_\perp), \quad 
	   	E_z(\Zmax+,x_\perp)=E_z(\Zmax-,x_\perp). 
		\label{eq:23DNLH-continuity}
	\end{gathered} \end{equation}
\end{subequations}

\subsubsection{\label{sssec:continuous-23D-trans-BCs}Local Transverse Boundary
Conditions}

\begin{subequations}
	\label{eqs:continuous-trans-BCs}
	Following the approach first used in~\cite{FT:04}, we set locally
	one-dimensional radiation boundary conditions of the Sommerfeld type in the
	transverse 	direction $x_\perp$.
	To do so, we assume that the beam is localized around $x_\perp=0$, so that 
	far from the beam center the nonlinearity becomes negligible, i.e.,  \[
		\epsilon |E|^{2\sigma} \ll \lri^2
		,\qquad\quad |x| \geq \Xmax 
		\quad \text{or} \quad  
		\rho \geq \Rmax.
	\]
	Therefore, the field (approximately) satisfies the constant coefficient
	equation: \[
		\Delta E + \lri_0^2 k_0^2 E = 0
		,\qquad |x| \geq \Xmax
		 \quad \text{or} \quad 
		\rho \geq \Rmax.
	\]
	We further assume that for $|x|\gtrsim \Xmax$ ($\rho\gtrsim\Rmax$) the
	field is composed predominantly of the outgoing plane 
	(cylindrical) waves with nearly normal incidence on the boundary 
$|x|= \Xmax$ ($\rho=\Rmax$).
	This leads to the following radiation boundary conditions in
	the $2D$ Cartesian case~\cite{FT:04}:
	\begin{equation}
		\label{eq:continuous-trans-SBCs-2D}
			E_x - ik_0 \lri_0 E
		\big|_{x = \Xmax} = 0 ,\qquad
			E_z + ik_0 \lri_0 E
		\big|_{x = -\Xmax} = 0.
	\end{equation} 
	In the $3D$ cylindrically symmetric case the local radiation boundary
	condition at $\rho = \Rmax$ reads~\cite{BFT:05}:
	\begin{equation}
		\label{eq:continuous-trans-SBC-3D}
		\left.
		E_\rho - \alpha E
		\right|_{\rho = \Rmax} = 0,\qquad
		\alpha =  \frac{
			\frac{d}{d\rho}
			H^{(1)}_0(\lri_0 k_0 \Rmax)
		}{
			H^{(1)}_0(\lri_0 k_0 \Rmax)
		},
	\end{equation}
	where $H^{(1)}_0$ is the Hankel function of the first kind.
	The symmetry condition at the axis $\rho=0$ is
	\begin{equation}
		\label{eq:continuous-trans-BC-3D-axis}
	\frac{\partial}{\partial\rho}E(z,0) = 0.
	\end{equation} 
\end{subequations}
We emphasize that these transverse boundary conditions are valid as long as the
beam is localized around the axis and remains ``far" from the transverse boundary
at $x=\Xmax$ or $\rho=\Rmax$.

\subsubsection{\label{sssec:continuous-23D-TWBCs}Nonlocal Longitudinal Boundary 
Conditions}

Similarly to the one-dimensional case (see Section~\ref{ssec:continuous-1D}),
the boundary conditions in the longitudinal direction $z$ will be set in the
linear regions at $z=-\delta$ and $z=\Zmax+\delta$.
They should render the boundaries transparent for all the outgoing waves, i.e.,
eliminate any non-physical reflections, and at the same time correctly prescribe
the given incoming wave(s), see Figure~\ref{fig:BC_schematic}.
Unlike in the one-dimensional case, however, a two-way Sommerfeld boundary
condition of type~(\ref{eq:1DTWBCs}), 
which is local in the configuration space, 
cannot be transparent for all the outgoing waves, because these waves travel
with different longitudinal velocities that depend on their angle of incidence.

Therefore, to accommodate all angles of incidence,
we first separate the variables in the linear Helmholtz
equation~(\ref{eq:23DLH}) by expanding its solution with respect to the
eigenfunctions of the transverse Laplacian.
These eigenfunctions solve the ordinary differential equation:
\begin{equation}
\label{eq:eigenf}
	\Delta_\perp \psi^{(l)}(x_\perp) = -(\kperpl)^2 \psi^{(l)},
\end{equation}
subject to the transverse boundary conditions~(\ref{eqs:continuous-trans-BCs}).
The resulting eigenvalue problem is not of the classical Sturm-Liouville type,
since its operator is not self-adjoint (because of the radiation boundary
 conditions).
As a result, the eigenfunctions are not orthogonal. 
Nevertheless, these eigenfunctions are bi-orthogonal \cite[Volume I]{morse} or,
alternatively, real orthogonal, and still form a
complete system.
A comprehensive discussion on completeness of eigensystems arising in 
the diffraction theory, and on convergence of the corresponding series, can
be found in~\cite{Agranovich}.

Since the system of eigenfunctions $\{\psi^{(l)}\}$ is complete, we can expand 
the field $E$ and the incoming beams $\EincL$ and $\EincR$ as
\begin{equation}
	\label{eq:continuous-trans-decomp}
	\begin{gathered}
	E(z,x_\perp) = \sum_{l=0}^\infty u_l(z) \psi^{(l)}(x_\perp)\\
	\EincL(x_\perp) = \sum_{l=0}^\infty \UincLl \psi^{(l)}(x_\perp),
	\qquad 
	\EincR(x_\perp) = \sum_{l=0}^\infty \UincRl \psi^{(l)}(x_\perp).
	\end{gathered}
\end{equation}
In the transformed space, the linear Helmholtz equation~(\ref{eq:23DLH})
reduces to a system of uncoupled one-dimensional linear Helmholtz equations
(ODEs):
\begin{equation}
	\label{eq:continuous-uncoupled-system}
		\left(
			\frac{d^{2}}{dz^{2}}+(\kparl)^{2}
		\right)u_l(z)= 0, \quad 
	(\kparl)^{2}=k_0^{2}-(\kperpl)^2,\qquad 
	l=0,1,\ldots,\infty.
\end{equation}
Each of the uncoupled equations~(\ref{eq:continuous-uncoupled-system}) 
formally coincides with equation~(\ref{eq:1DLH}) and has the same general
solution composed of two waves one of which can be interpreted as propagation in
the positive $z$ direction and the other one --- in the negative $z$ direction.
Unlike in equation~(\ref{eq:1DLH}), however, the quantity $(\kparl)^{2}$ in
equation~(\ref{eq:continuous-uncoupled-system}) may have a negative real part,
in which case the waves become evanescent. It may also have a non-trivial
imaginary part, which is due to the non-self-adjoint transverse (radiation) boundary
conditions (see~\cite{FT:04} for more detail).
Regardless of the particular shape that the waves may assume, the longitudinal
boundary conditions have to ensure that the field in the region $z\leq-\delta$
be of the form [cf.\ formula~(\ref{eq:continuous-E-outside})]:
\begin{equation*}
	u(z) = \UincLl e^{i \kparl z} + C_1 e^{-i \kparl z}.	
\end{equation*}
Therefore, the two-way ABC at $z=-\delta$ can be written as
\begin{subequations}
	\label{eqs:2DTWBCs}
	\begin{equation}
	\label{eqs:2DTWBCs_a}
		\left.
			\left( \frac{d}{dz}+i \kparl \right) u_l
		\right|_{z=-\delta} 
			= 2i \kparl e^{-i \kparl \delta} \UincLl.
	\end{equation}
Similarly, at the opposite boundary, $z=\Zmax+\delta$, we obtain:
	\begin{equation}
	\label{eqs:2DTWBCs_b}
		\left.
			\left( \frac{d}{dz} - i \kparl \right) u_l
		\right|_{z=\Zmax+\delta} 
			= -2i \kparl e^{-i \kparl \delta} \UincRl.
	\end{equation} 
\end{subequations}
Boundary conditions~(\ref{eqs:2DTWBCs}) are local in the transformed space
$\{u_l(z)\}_{l=0}^\infty$.
In this space, the two-way one-dimensional Sommerfeld conditions are applied
independently for each individual mode
defined by~(\ref{eq:continuous-uncoupled-system}).
The equivalent of relations~(\ref{eqs:2DTWBCs}) after the inverse transformation
of~(\ref{eq:continuous-trans-decomp}) will result in a nonlocal
pseudodifferential operator in the original space $\{E(z,x)\}$,
see~\cite{FT:04} or~\cite{Tsynkov:ANM-review} for more details.
{\em Therefore, the  resulting boundary conditions are nonlocal two-way artificial BCs.}

\subsection{\label{ssec:discrete-23D}Discrete Approximation}

We build a semi-compact scheme for the Cartesian
problem~(\ref{eqs:2DNLH-system}),~(\ref{eq:continuous-trans-SBCs-2D}),~(\ref{eqs:2DTWBCs})
in Section~\ref{sssec:discrete-2D},
and for the cylindrically symmetric problem
(\ref{eqs:3DNLH-system}),~(\ref{eq:continuous-trans-SBC-3D}),~(\ref{eq:continuous-trans-BC-3D-axis}),~(\ref{eqs:2DTWBCs}) in Section~\ref{sssec:discrete-3D}.
As in the one-dimensional case, we discretize the governing equations inside and
outside the Kerr material, and then obtain a discretization at the material
interfaces.
The discrete transverse boundary conditions and the discrete two-way ABCs for both problems
are
described in Section~\ref{ssec:discrete-TBCs} and
Section~\ref{ssec:discrete-LBCs}, respectively.

\subsubsection{\label{sssec:discrete-2D} $2D$ Cartesian Case}
On the rectangle $[-3h_z,\,\Zmax+3h_z]\times[-\Xmax,\Xmax]$,
we introduce a uniform Cartesian grid of $(N+7)\times M$ nodes as
\begin{equation}
	\label{eq:grid-2D}
	\begin{gathered}
		z_n = n\cdot h_z, \qquad
		h_z = \frac{\Zmax}{N}, \qquad
		n= -3,-2,\dots,N+2,N+3, \\
		x_m = -\Xmax + (m+1/2) h_x, \qquad
		h_x = \frac{2\Xmax}{M}, \qquad
		m= 0,1,\dots,M-1, \\
	\end{gathered}
\end{equation}
so that \[
	z_0 = 0,\qquad z_N = \Zmax, \qquad 
	x_{-1/2}=-\Xmax, \qquad x_{M-1/2} = \Xmax.
\]
In this paper, we keep $h_z \sim h_x$ so that 
all $\mathcal{O}\left(h_z^jh_x^{k-j}\right)$ terms can be treated as 
terms of the same order $k$ and denoted by $\oh{k}$.
For convenience, we also introduce the following notations for the field and the
Kerr nonlinearity at the grid nodes: 
\[
	E_{n,m} \myeqdef E(z_n,x_m), \qquad
	P_{n,m} \myeqdef |E_{n,m}|^{2\sigma}E_{n,m}.
\]
Finally, we use the previous notation $D$ for central difference operators, with the
order of accuracy in the superscript and the differentiation variables in the
subscript.
For example,
\begin{equation*}
	\CDO{xx}{2} E \myeqdef 
		\frac{E_{n,m+1}-2E_{n,m}+E_{n,m-1}}{h_x^2} 
		= \partial_{xx} E_{n,m} + \oh{2}.
\end{equation*}
Other notations for central differences are listed in Appendix \ref{app:CDOs}.

To build a semi-compact approximation of the NLH~(\ref{eq:2DNLH}) at
the interior points $n=1,\dots,N-1$, we first introduce the following 
mixed order discrete Laplacian:
\begin{align}
	\nonumber
	\CDO{zz}{2}  E_{n,m} + &\> \CDO{xx}{4} E_{n,m}
	= \frac{E_{n-1,m}-2E_{n,m}+E_{n+1,m}}{h_z^2} \\
	\label{eq:mixed-app-2D}
	+ &\> \frac{
		-E_{n,m-2} + 16 E_{n,m-1} -30 E_{n,m} +16 E_{n,m+1} -E_{n,m+2} 
	}{12 h_x^2}\\
	\nonumber
	 = &\> \Delta E_{n,m} + \frac{h_z^2}{12}\partial_{zzzz} E_{n,m} +\oh{4}.
\end{align}
In order to remove the $\oh{2}$ term on the right-hand side
of~(\ref{eq:mixed-app-2D}), we consider the following expression that contains
fourth-order derivatives with respect to both $z$ and $x$, and approximate it to
second-order accuracy using central differences:
\begin{equation}
\label{eq:who_knows}
(\partial_{zzzz}-\partial_{xxxx}) E_{n,m} = (\partial_{zz}-\partial_{xx})\Delta E_{n,m}
= \left(\CDO{zz}{2}-\CDO{xx}{2}\right) \Delta E_{n,m} + \oh{2}.
\end{equation}
Then, we employ the NLH~(\ref{eq:2DNLH}) itself and substitute the expression
\begin{equation*}
	\Delta E_{n,m} 
		= - k_0^2\left( \lri^2 E_{n,m} + \epsilon P_{n,m} \right)
\end{equation*}
into formula~(\ref{eq:who_knows}).
Next, we approximate the derivative $\partial_{xxxx}E$ in
formula~(\ref{eq:who_knows}) to second-order accuracy using central differences,
and altogether obtain:
\begin{equation}
\label{eq:who_knows2}
	\partial_{zzzz} E_{n,m} = 
		-k_0^2 \left(\CDO{zz}{2}-\CDO{xx}{2}\right) 
			\left( \lri^2 E_{n,m} + \epsilon P_{n,m}\right) 
			+\CDO{xxxx}{2} E_{n,m} + \oh{2}.
\end{equation}
Substitution of~(\ref{eq:who_knows2}) into~(\ref{eq:mixed-app-2D})
yields
a semi-compact fourth-order discretization of the Laplacian,
which leads to the following fourth-order scheme for the NLH~(\ref{eq:2DNLH}):
\begin{equation}
	\label{eq:discrete-2DNLH}
	\begin{gathered}
		\left(
			\CDO{zz}{2}
			+ \CDO{xx}{4} E_{n,m}
			- \frac{h_z^2}{12} \CDO{xxxx}{2}
		\right)E_{n,m} \\
		+  k_0^2\left(
			1
			+\frac{h_z^2}{12} \CDO{zz}{2}
			-\frac{h_z^2}{12} \CDO{xx}{2}
		\right)
		\left( \lri^2 E_{n,m} + \epsilon P_{n,m}\right) 
		= 0, \\
	 n = 1,\dots,N-1,\qquad m=0,\dots,M-1.
	\end{gathered}
\end{equation}
To obtain a similar fourth-order scheme for the linear Helmholtz equation~(\ref{eq:2DLH}),
we repeat the previous derivation with $\epsilon P_{n,m}=0$ and $\lri=1$,
which yields:
\begin{equation}
	\label{eq:discrete-2DLH}
	\begin{gathered}
	\left[
		\left( 1+ \frac{k_0^2h_z^2}{12} \right)
		\CDO{zz}{2} 
		+ \left(
			\CDO{xx}{4} 
			- \frac{k_0^2h_z^2}{12} \CDO{xx}{2} 
			- \frac{h_z^2}{12} \CDO{xxxx}{2} 
			\right)  + k_0^2 
		\right] E_{n,m} = 0, \\
		  n = -3,\dots,-1,N+1,\dots,3,\qquad m=0,\dots,M-1.
	\end{gathered}
\end{equation}
Next, we consider material interfaces at the nodes $n=0$ and $n=N$.
Using Taylor's expansion, we can write: \[
	\frac{
		-85 E_{0,m} + 108 E_{1,m} -27 E_{2,m} + 4E_{3,m}
	}{ 66 h_z } = 
		\partial_zE_{0+,m} 
		+ \frac{3h_z}{11} \partial_{zz}E_{0+,m} + \oh{4}.
\]
Then, approximating the derivative $\partial_{zz}E_{0+,m}$ with fourth-order 
accuracy: \[ 
	\partial_{zz}E_{0+,m} = \Delta E_{0+,m} - \partial_{xx}E_{0+,m} 
		= -k_0^2(\lri^2_{0+} E_{0,m}+\epsilon_{0+} P_{0,m}) - 
                 \CDO{xx}{4} E_{0,m} +\oh{4},
\]
we obtain: 
\begin{gather*}
	\partial_zE_{0+,m} = 
		\frac{
			-85 E_{0,m} + 108 E_{1,m} -27 E_{2,m} + 4E_{3,m}
		}{ 66 h_z } \\
		+ \frac{3h_zk_0^2}{11} 
               \left( \lri^2_{0+} E_{0,m}+\epsilon_{0+} P_{0,m} \right) 
		+ \frac{3h_z}{11} \CDO{xx}{4} E_{0,m} +\oh{4}.
\end{gather*}
Deriving a similar formula for $\partial_zE_{0-,m}$ and equating the resulting
expressions for $\partial_zE_{0+,m}$ and $\partial_zE_{0-,m}$, we get a
fourth-order accurate approximation of the continuity condition
$E_z(0-)=E_z(0+)$:
\begin{gather}
		\frac{
			4E_{-3,m} - 27E_{-2,m} + 108E_{-1,m}	-170 E_{0,m}
			+ 108E_{1,m} - 27E_{2,m} +4E_{3,m}
		}{66 h_z} \nonumber
		\\
		\label{eq:discrete-2D-interface-general}
		+  \frac{6h_zk_0^2 }{11} \left(
			\frac{ 	\lri^2_{0-} + \lri^2_{0+} 	}{2}E_{0,m} 
			+\frac{	\epsilon_{0-} + \epsilon_{0+}	}{2} P_{0,m} 
		\right)
		+ \frac{6h_z}{11} \CDO{xx}{4} E_{0,m} 
		= 0. 
\end{gather}
\begin{subequations}
		\label{eqs:discrete-2D-interface}
Finally, substituting $
	\lri_{0-,m}=1, 
	\epsilon_{0-,m}=0,
	\lri_{0+,m}=\lri $ and $
	\epsilon_{0+,m}=\epsilon
$, we have: 
\begin{gather}
		\frac{
			4E_{-3,m} - 27E_{-2,m} + 108E_{-1,m}	-170 E_{0,m}
			+ 108E_{1,m} - 27E_{2,m} +4E_{3,m}
		}{66 h_z} \nonumber
		\\
		+  \frac{6h_zk_0^2 }{11} \left(
			\frac{1+\lri^2}{2}E_{0,m} 
			+\frac{ \epsilon}{2} P_{0,m} 
		\right)
		+ \frac{6h_z}{11} \CDO{xx}{4} E_{0,m} 
		= 0. 
\end{gather}
A similar equation is obtained for the interface at $n=N$:
\begin{gather}
		\frac{
			4E_{N-3,m} - 27E_{N-2,m} + 108E_{N-1,m}	-170 E_{N,m}
			+ 108E_{N+1,m} - 27E_{N+2,m} +4E_{N+3,m}
		}{66 h_z} \nonumber
		\\
		+  \frac{6h_zk_0^2 }{11} \left(
			\frac{1+\lri^2}{2}E_{N,m} 
			+\frac{ \epsilon}{2} P_{N,m} 
		\right)
		+ \frac{6h_z}{11} \CDO{xx}{4} E_{N,m} 
		= 0. 
\end{gather}
\end{subequations}

\subsubsection{\label{sssec:discrete-3D}Cylindrically Symmetric Case}
We use the same grid~(\ref{eq:grid-2D}), except that in the transverse direction
we now have:
\begin{equation} \label{eq:grid-3D}
	\rho_m = (m+1/2) h_\rho, \qquad h_\rho = \frac{\Rmax}{M},
	\qquad m = 0,\dots,M-1,
\end{equation}
so that \[
	\rho_{-1/2} = 0\quad\text{and}\quad \rho_{M-1/2} = \Rmax.
\]
We also keep $h_z \sim h_\rho$ so that all
$\mathcal{O}\left(h_z^jh_\rho^{k-j}\right)$ terms appear of the same order
$\oh{k}$.

To approximate the NLH~(\ref{eq:3DNLH}) at
the interior points $n=1,\dots,N-1$, we begin by introducing a
mixed order discretization of the cylindrical Laplacian
$\Delta = \partial_{zz} + \Delta_\rho\equiv \partial_{zz} + \partial_\rho^2 + 
\frac{1}{\rho}\partial_\rho$ :
\begin{equation}
	\label{eq:mixed-app-3D}
	\left( 
		\CDO{zz}{2} 
		+ \CDO{\rho\rho}{4}  
		+ \frac{1}{\rho_m} \CDO{\rho}{4}
	\right) E_{n,m} 
	=  \Delta E_{n,m} + \frac{h_z^2}{12}\partial_{zzzz} E_{n,m} +\oh{4}.
\end{equation}
To remove the $\oh{2}$ term on the right-hand side of~(\ref{eq:mixed-app-3D}),
we start with the second-order central difference approximation of the
expression $\left.
	(\partial_{zzzz}-\Delta^2_{\rho})E_{n,m} 
	=(\partial_{zz}-\Delta_{\rho}) \Delta E_{n,m}
\right.$, where $\left. \Delta^2_{\rho} =
                \rho^{-3} \partial_\rho
                -\rho^{-2} \partial_{\rho\rho}
                +2\rho^{-1} \partial_{\rho\rho\rho}
                +\partial_{\rho\rho\rho\rho}
        \right.$, and using the NLH~(\ref{eq:3DNLH}) itself, obtain: 
\begin{equation}
\label{eq:who_knows4}
 \begin{split}
	\partial_{zzzz} E_{n,m} = &
		-k_0^2 \left(
			\CDO{zz}{2} 
			-\CDO{\rho\rho}{2} 
			-\frac{1}{\rho_m}\CDO{\rho}{2} 
		\right) \left( 
			\lri^2 E_{n,m} + \epsilon P_{n,m}
		\right) \\
		& + \left(
			\rho_m^{-3} \CDO{\rho}{2} 			
			-\rho_m^{-2} \CDO{\rho\rho}{2}		
			+2\rho_m^{-1} \CDO{\rho\rho\rho}{2} 	
			+\CDO{\rho\rho\rho\rho}{2}			
		\right)E_{n,m} + \oh{2}.
\end{split} 
\end{equation}
Substitution of (\ref{eq:who_knows4}) into~(\ref{eq:mixed-app-3D})
yields a semi-compact fourth-order discretization of the cylindrical Laplacian,
which leads to the following fourth-order scheme for the NLH~(\ref{eq:3DNLH}):
\begin{equation}
	\label{eq:discrete-3DNLH}
\begin{gathered}
		\left( 
			\CDO{zz}{2} 
			+ \CDO{\rho\rho}{4}  
			+ \frac{1}{\rho_m} \CDO{\rho}{4}
		\right) E_{n,m} 
		 -  \frac{h_z^2}{12} 
		\left(
			\rho_m^{-3} \CDO{\rho}{2} 			
			-\rho_m^{-2} \CDO{\rho\rho}{2}		
			+2\rho_m^{-1} \CDO{\rho\rho\rho}{2} 	
			+\CDO{\rho\rho\rho\rho}{2}			
		\right)E_{n,m} \\ 
		 +  k_0^2 \left[ 
			1+ \frac{h_z^2}{12} \left(
				\CDO{zz}{2} 
				-\CDO{\rho\rho}{2} 
				-\frac{1}{\rho_m}\CDO{\rho}{2} 
			\right) 
		\right]
		\left( 	\lri^2 E_{n,m} + \epsilon P_{n,m} \right) 
		= 0, \\
	 n =  1,\dots,N-1,\qquad m=0,\dots,M-1.
\end{gathered}
\end{equation}
To obtain a similar fourth-order scheme for the linear Helmholtz
equation~(\ref{eq:3DLH}), we repeat the previous derivation with
$\epsilon\equiv0$ and $\lri\equiv 1$, which yields:
{\allowdisplaybreaks
\begin{equation}
	\label{eq:discrete-3DLH}
\begin{gathered}
		\left( 
			\CDO{zz}{2} 
			+ \CDO{\rho\rho}{4}  
			+ \frac{1}{\rho_m} \CDO{\rho}{4}
		\right) E_{n,m} 
		 -  \frac{h_z^2}{12} 
		\left(
			\rho_m^{-3} \CDO{\rho}{2} 			
			-\rho_m^{-2} \CDO{\rho\rho}{2}		
			+2\rho_m^{-1} \CDO{\rho\rho\rho}{2} 	
			+\CDO{\rho\rho\rho\rho}{2}			
		\right)E_{n,m} \\ 
		 +  k_0^2 \left[ 
			1+ \frac{h_z^2}{12} \left(
				\CDO{zz}{2} 
				-\CDO{\rho\rho}{2} 
				-\frac{1}{\rho_m}\CDO{\rho}{2} 
			\right) 
		\right]
		E_{n,m}= 0, \\
		  n =  -3,\dots,-1,N+1,\dots,3,\quad m=0,\dots,M-1.
\end{gathered}
\end{equation}
}%
The analysis of material interfaces at $n=0$ and $n=N$ is very similar to that
of Section~\ref{sssec:discrete-2D}, and we arrive at the following
fourth-order accurate approximation of the continuity condition
$E_z(0-)=E_z(0+)$:
\begin{gather}
		\frac{
			4E_{-3,m} - 27E_{-2,m} + 108E_{-1,m}	-170 E_{0,m}
			+ 108E_{1,m} - 27E_{2,m} +4E_{3,m}
		}{66 h_z} \nonumber
		\\
		\label{eq:discrete-3D-interface-general}
		+  \frac{6h_zk_0^2 }{11} \left(
			\frac{\lri^2_{0-}+\lri^2_{0+}}{2} E_{0,m} 
			+ \frac{\epsilon_{0-}+\epsilon_{0}}{2} P_{0,m} 
		\right)
		+  \frac{6h_z}{11} \left( 
			\CDO{\rho\rho}{4} 
			+\frac{1}{\rho_m} \CDO{\rho}{4} 
		\right) E_{0,m} = 0. 
\end{gather}
\begin{subequations}
		\label{eqs:discrete-3D-interface}
Substituting $
	\lri_{0-,m}=1, 
	\epsilon_{0-,m}=0,
	\lri_{0+,m}=\lri $ and $
	\epsilon_{0+,m}=\epsilon
$ into (\ref{eq:discrete-3D-interface-general}), we have:
\begin{gather}
		\frac{
			4E_{-3,m} - 27E_{-2,m} + 108E_{-1,m}	-170 E_{0,m}
			+ 108E_{1,m} - 27E_{2,m} +4E_{3,m}
		}{66 h_z} \nonumber
		\\
		+  \frac{6h_zk_0^2 }{11} \left(
			\frac{1+\lri^2}{2} E_{0,m} 
			+ \frac{\epsilon}{2} P_{0,m} 
		\right)
		+  \frac{6h_z}{11} \left( 
			\CDO{\rho\rho}{4} 
			+\frac{1}{\rho_m} \CDO{\rho}{4} 
		\right) E_{0,m} = 0. 
\end{gather}
A similar equation is obtained for the interface at $n=N$:
\begin{gather}
		\frac{
			4E_{N-3,m} - 27E_{N-2,m} + 108E_{N-1,m}	-170 E_{N,m}
			+ 108E_{N+1,m} - 27E_{N+2,m} +4E_{N+3,m}
		}{66 h_z} \nonumber
		\\
		+  \frac{6h_zk_0^2 }{11} \left(
			\frac{1+\lri^2}{2} E_{N,m} 
			+ \frac{\epsilon}{2} P_{N,m} 
		\right)
		+  \frac{6h_z}{11} \left( 
			\CDO{\rho\rho}{4} 
			+\frac{1}{\rho_m} \CDO{\rho}{4} 
		\right) E_{N,m} = 0. 
\end{gather}
\end{subequations}

\subsection{\label{ssec:discrete-TBCs}Local Transverse Boundary Conditions}

In this section, we briefly describe a discrete approximation of the transverse
boundary conditions~(\ref{eqs:continuous-trans-BCs}).
In doing so, we follow the approach of~\cite{BFT:05}, where additional details
can be found.
Let us first consider the radiation boundary
conditions~(\ref{eq:continuous-trans-SBCs-2D})
and~(\ref{eq:continuous-trans-SBC-3D}) at the ``upper'' boundary $m = M-1/2$.
We will use their discrete counterparts to express the values of  the field at
the	ghost nodes $E_{n,M}$ and $E_{n,M+1}$ via the values at the inner nodes
$E_{n,M-3}$, $E_{n,M-2}$, and $E_{n,M-1}$, and thus eliminate the ghost nodes.
A fourth-order approximation of either Cartesian or cylindrical radiation
boundary condition centered around $m=M-1/2$ (which corresponds to 
$x = \Xmax$ or $\rho=\rho_{\max}$) is given by
\begin{gather*}
                \frac{
                        E_{n,M-2} -27E_{n,M-1} +27 E_{n,M} - E_{n,M+1}
                }{
                        24 h_\perp
                }\\
                -\alpha 
                \frac{
                        -E_{n,M-2} + 9E_{n,M-1} + 9E_{n,M} - E_{n,M+1}
                }{
                        16
                } = 0,
\end{gather*}
where in the Cartesian case $\alpha=i\lri_0 k_0$, see
formula~(\ref{eq:continuous-trans-SBCs-2D}), and in the cylindrical case
$\alpha$ is defined in~(\ref{eq:continuous-trans-SBC-3D}), see~\cite{BFT:05}.
Equivalently, we can write:
	  \begin{gather*}
		\begin{bmatrix}
			c_{-2}, & \dots, & c_{1} 
		\end{bmatrix}
		\cdot
		\begin{bmatrix}
			E_{n,M-2} \\
			\vdots \\
			E_{n,M+1} 
		\end{bmatrix}
		=0, \\
\intertext{where}
		\begin{bmatrix}
			c_{-2}, & \dots, & c_{1} 
		\end{bmatrix} = 
		\begin{bmatrix}
			1, & -27, & 27, & -1 
		\end{bmatrix}
		- \frac{2\alpha h_\perp}{3}
		\begin{bmatrix}
			-1, & 9, & 9, & -1 
		\end{bmatrix}
		.
	\end{gather*}
However, specifying this boundary condition alone is not 
sufficient, because the fourth-order finite difference equation that we
use in the $x_\perp$ direction requires an additional boundary condition. 
The choice of the latter allows for more flexibility	as long as the resulting
method is fourth-order accurate and stable.\footnote{%
	Stability of these approximations can be studied by the methodology
	of~\cite{Ryaben'kii-64}.
}
Hereafter, we choose this second condition as the fourth-order accurate
extrapolation of the ghost value $E_{n,M+1}$ via $\left\{
	E_{n,M-3},\dots,E_{n,M}
\right\}$, which can be conveniently written as
\[
	E_{n,M+1}  =  \sum_{j=-3}^{0}(-1)^{j}\binom{4}{1-j} E_{n,M+j}.
\]
Combining the two discrete boundary conditions as \[
	\begin{bmatrix}
		0, & c_{-2}, & c_{-1}, & c_{0}, & c_{1} \\
		-1 & 4 & -6 & 4 & -1
	\end{bmatrix}
	\begin{bmatrix}
		E_{n,M-3}\\
		\vdots\\
		E_{n,M+1}
	\end{bmatrix} = 
	\begin{bmatrix} 0 \\ 0 \end{bmatrix},
\]
we can express the ghost values $E_{n,M}$ and $E_{n,M+1}$ in terms of the
interior values:
\begin{subequations}
\label{eq:rbc}
	\begin{equation}
\label{eq:rbc_a}
		\begin{bmatrix} 
			E_{n,M} \\ 
			E_{n,M+1} 
		\end{bmatrix} 
		=
		-\,\frac{1}{ c_0+4c_1 }
		\begin{bmatrix} 
			-c_1 & \,c_{-2}+4c_1\, & c_{-1}-6c_1 \\
			c_0 & \,4c_{-2}-4c_0\, & 4c_{-1}+6c_0 
		\end{bmatrix}
		\cdot
		\begin{bmatrix}
			E_{n,M-3}\\
			E_{n,M-2}\\
			E_{n,M-1}
		\end{bmatrix}.
	\end{equation}
	In the Cartesian case, the derivation is repeated to obtain the discrete 
	discrete radiation boundary condition at $x=-\Xmax$, i.e., at $m=0$:
	\begin{equation}
\label{eq:rbc_b}
		\begin{bmatrix} 
			E_{n,-2} \\
			E_{n,-1}  
		\end{bmatrix} 
		=
		-\,\frac{1}{ c_0+4c_1 }
		\begin{bmatrix} 
			4c_{-1}+6c_0 & \,4c_{-2}-4c_0\, & c_0 \\
			c_{-1}-6c_1 & \,c_{-2}+4c_1\, & -c_1 
		\end{bmatrix}
		\cdot
		\begin{bmatrix}
			E_{n,0}\\
			E_{n,1}\\
			E_{n,2}
		\end{bmatrix}.
	\end{equation}
\end{subequations}
	In the cylindrical case, the symmetry
	(\ref{eq:continuous-trans-BC-3D-axis}) is enforced as follows:
	\begin{equation}
		\label{eq:sbc}
		E_{n,-1} = E_{n,0}, \qquad E_{n,-2} = E_{n,1}.
	\end{equation}

Note also that in the Cartesian case there is an alternative way of
building the discrete transverse boundary conditions. It does not require
a finite difference approximation of the continuous boundary conditions
(\ref{eq:continuous-trans-SBCs-2D}), and is rather based on analyzing the
roots of the fourth-order characteristic equation that corresponds to the
five node discretization in the $x$ direction. The idea is similar to
that behind boundary conditions~(\ref{eqs:discrete-1DTWBCs}), and the reader is
referred to \cite{FT:04} for more detail.

\subsection{\label{ssec:discrete-LBCs}Nonlocal Longitudinal Boundary Conditions}

In this section, we construct a discrete counterpart for the two-way
ABCs~(\ref{eqs:2DTWBCs}).
In the continuous case of
Section~\ref{ssec:continuous-23D}, we separated the variables in the
linear Helmholtz equation~(\ref{eq:23DLH}) outside the Kerr region, and then
obtained the ABCs in the transformed space.
In the discrete case, we also begin by separating the variables in the
Cartesian~(\ref{eq:discrete-2DLH}) and cylindrical~(\ref{eq:discrete-3DLH})
difference Helmholtz equation at the exterior grid nodes:
\[
	m=0,\dots,M-1,\qquad n=0,-1,-2\quad n= N,N+1,N+2.
\]
Subsequently, we derive the ABCs in the transformed space.
This derivation is identical for the Cartesian geometry of
Section~\ref{sssec:discrete-2D} and the cylindrical geometry of
Section~\ref{sssec:discrete-3D}.

We first identify the transverse components in the finite difference
operators of~(\ref{eq:discrete-2DLH}) and~(\ref{eq:discrete-3DLH}).
The transverse part of the discrete Laplacian for the Cartesian case is
\begin{subequations}
\label{eq:trans}
\begin{equation}
\label{eq:trans_a}
	L^\perp = 
		\CDO{xx}{4} 
		- \frac{k_0^2h_z^2}{12} \CDO{xx}{2} 
		- \frac{h_z^2}{12} \CDO{xxxx}{2}, 
\end{equation}
whereas for the cylindrically symmetric case it is given by 
\begin{equation}
\label{eq:trans_b}
\begin{gathered}
	L^\perp = 
		\CDO{\rho\rho}{4} + \frac{1}{\rho_m} \CDO{\rho}{4} 
		- \frac{k_0^2 h_z^2}{12} \left(
			\CDO{\rho\rho}{2} 
			+\frac{1}{\rho_m}\CDO{\rho}{2} 
		\right) \\
		 - 
		\frac{k_0^2h_z^2}{12} 
		\left(
			\rho_m^{-3} \CDO{\rho}{2} 			
			-\rho_m^{-2} \CDO{\rho\rho}{2}		
			+2\rho_m^{-1} \CDO{\rho\rho\rho}{2} 	
			+\CDO{\rho\rho\rho\rho}{2}			
		\right). 
\end{gathered}
\end{equation}
\end{subequations}
The separation of variables in equations~(\ref{eq:discrete-2DLH})
and~(\ref{eq:discrete-3DLH}) will be rendered by expanding the solution with
respect to the transverse eigenvectors 
$
	\psi^{(l)}=
		\left[
			\psi^{(l)}_0, \psi^{(l)}_1, \ldots, \psi^{(l)}_{M-1}
		\right]^T.
$
Each eigenvector $\psi^{(l)}$ satisfies the following difference equation on the
grid [cf.\ equation~(\ref{eq:eigenf})]:
\begin{equation}
\label{eq:eigenv}
L^\perp\psi^{(l)}_m=-(\kperpl)^2\psi^{(l)}_m,\qquad m=0,1,\ldots,M-1.
\end{equation}
In the Cartesian case, the operator $L^\perp$ in~(\ref{eq:eigenv}) is defined by
formula~(\ref{eq:trans_a}), and the solution $\psi^{(l)}$ is subject to boundary
conditions~(\ref{eq:rbc_a}),~(\ref{eq:rbc_b}).
In the cylindrically symmetric case, the operator $L^\perp$ in~(\ref{eq:eigenv})
is defined by formula~(\ref{eq:trans_b}), and the solution $\psi^{(l)}$ is
subject to boundary conditions~(\ref{eq:rbc_a}),~(\ref{eq:sbc}).
The argument behind linear independence of $\{\psi^{(l)}\}$ in the Cartesian 
case is based on bi-orthogonality (real orthogonality) 
of the eigenvectors and can be found 
in~\cite{FT:04}.
For the cylindrically symmetric case, the continuous eigenfunctions are
also real orthogonal, but the discrete eigenvectors are not,
see~\cite{BFT:05}.
Yet we observe numerically that they are linearly independent.

The $M$ linearly independent eigenvectors 
are convenient to arrange as a column matrix:
\begin{equation*}
\Psi\stackrel{\rm def}{=}\left[\psi^{(0)},\psi^{(1)},\ldots,\psi^{(M-1)}\right]
=\begin{bmatrix} \psi^{(0)}_0 & \cdots & \psi^{(M-1)}_0\\
\vdots & \ddots & \vdots \\
\psi^{(0)}_{M-1} & \cdots & \psi^{(M-1)}_{M-1} \end{bmatrix}
\end{equation*}
that will diagonalize the discrete transverse Laplacian, i.e., 
$L^\perp\Psi = \Psi \Lambda$, where 
\begin{equation*}
\Lambda={\rm diag}\left\{-(k^{(0)}_\perp)^2,-(k^{(1)}_\perp)^2,\ldots,
-(k^{(M-1)}_\perp)^2\right\},
\end{equation*}
and the eigenvalues $-\left(k^{(l)}_\perp\right)^2$ are defined
in~(\ref{eq:eigenv}).

It will also be convenient to consider the following $M$-dimensional vectors:
\begin{equation*}
{\mathcal E}_n \stackrel{\rm def}{=} \left[E_{n,0},E_{n,1},\ldots,E_{n,M-1}\right]^T
\end{equation*}
that contain the values of the field arranged in the transverse direction. 
With this notation, we can recast both scheme~(\ref{eq:discrete-2DLH}) and
scheme~(\ref{eq:discrete-3DLH}) in the vector form:
\begin{equation}
\label{eq:scheme_vec}
\left(1+\frac{k_0^2h_z^2}{12}\right)\frac{{\mathcal E}_{n+1}-2{\mathcal E}_n+
{\mathcal E}_{n-1}}{h_z^2}
+L^\perp{\mathcal E}_n + k_0^2{\mathcal E}_n=0.
\end{equation}

\begin{subequations} \label{eqs:discrete-trans-decomp}
	For each $n$, let us introduce the vector variable
	\begin{equation}
		U_n=\Psi^{-1}{\mathcal E}_n.
	\end{equation}
	Equality ${\mathcal E}_n=\Psi U_n$ is
	the expansion of ${\mathcal E}_n$ with respect to the eigenvectors
	$\psi^{(l)}$, where the coefficients are given by the components of $U_n$.
	Similarly, we can expand the incoming beam profiles:
	\begin{align}
		U_{\text{inc}}^{0} = &\>
			\Psi^{-1}{\mathcal E}^{0}_{\text{inc}}
		,  & 
		U_{\text{inc}}^{0}
		\stackrel{\rm def}{=} &\>
		\begin{bmatrix}
			u_{\text{inc},0}^{0} \\
			\vdots \\
			u_{\text{inc},M-1}^{0}
		\end{bmatrix}  
		, &
		{\mathcal E}^{0}_{\text{inc}}
		\stackrel{\rm def}{=} &\>
		\begin{bmatrix}
			E^{0}_{{\text{inc}},0} \\
			\vdots \\
			E^{0}_{{\text{inc}},M-1}
		\end{bmatrix}, \\
		 U_{\text{inc}}^{\text{Zmax}} = &\>
			\Psi^{-1}{\mathcal E}^{\text{Zmax}}_{\text{inc}}
		, &
		U_{\text{inc}}^{\text{Zmax}}
		\stackrel{\rm def}{=} &\>
		\begin{bmatrix}
			u_{\text{inc},0}^{\text{Zmax}} \\
			\vdots \\
			u_{\text{inc},M-1}^{\text{Zmax}}
		\end{bmatrix}  
		, &
		{\mathcal E}^{\text{Zmax}}_{\text{inc}}
		\stackrel{\rm def}{=} &\>
		\begin{bmatrix}
			E^{\text{Zmax}}_{{\text{inc}},0} \\
			\vdots \\
			E^{\text{Zmax}}_{{\text{inc}},M-1}
		\end{bmatrix}. 
	\end{align}
\end{subequations}
Formulae~(\ref{eqs:discrete-trans-decomp}) are discrete counterparts
of~(\ref{eq:continuous-trans-decomp}).
Substituting expansions~(\ref{eqs:discrete-trans-decomp}) into 
equation~(\ref{eq:scheme_vec}) and diagonalizing  $L^\perp$: \
$L^\perp {\mathcal E}_n = L^\perp \Psi U_n =
	\Psi \Lambda U_n$,
we have: 
\begin{equation}
\label{eq:scheme_vec2}
\Psi\left(1+\frac{k_0^2h_z^2}{12}\right)\frac{U_{n+1}-2U_n+
U_{n-1}}{h_z^2}
+\Psi\Lambda U_n + \Psi k_0^2U_n=0.
\end{equation}
Finally, multiplying equation~(\ref{eq:scheme_vec2}) by the inverse matrix
$\Psi^{-1}$ from the left we separate the variables.
Recasting the result via individual components of 
$U_n=\left[u_{n,0},u_{n,1},\ldots,u_{n,M-1}\right]^T$, 
we obtain:
\begin{equation}
\label{eq:scheme_trans}
\begin{gathered}
\left(1+\frac{k_0^2h_z^2}{12}\right)\frac{u_{n+1,l}-2u_{n,l}+
u_{n-1,l}}{h_z^2}
-(\kperpl)^2u_{n,l} +  k_0^2u_{n,l}=0,\\
l=0,1,\ldots,M-1.
\end{gathered}
\end{equation}
Formula~(\ref{eq:scheme_trans}) is a system of $M$ uncoupled ordinary difference
equations, which is a discrete counterpart of the continuous uncoupled
system~(\ref{eq:continuous-uncoupled-system}).

Each of the uncoupled difference equations~(\ref{eq:scheme_trans})
is identical to the one-dimensional
difference equation~(\ref{eq:1Dscheme}) if we redefine $k^2$ of~(\ref{eq:1Dscheme})
as 
\begin{equation*}
k^2=\frac{k_0^2-(\kperpl)^2}{1+\frac{k_0^2h_z^2}{12}}\equiv
\frac{(\kparl)^2}{1+\frac{k_0^2h_z^2}{12}}.
\end{equation*}
Therefore, similarly to~(\ref{eqs:discrete-1DTWBCs_a}) we can write for the 
ghost node $n=-4$:
\[
	u_{-4,l} = 
		 \left(q_l^{-1}-q_l\right) q_l^{-3} \UincLl
		+ q_l u_{-3,l},
\] 
where $q_l$ and $q_l^{-1}$ denote roots of the characteristic equation for a
given $l$, and the incoming components $\UincLl$ are defined
in~(\ref{eqs:discrete-trans-decomp}).
Recasting the previous equality in the matrix form and 
transforming back into the configuration space, 
${\mathcal E}=\Psi U$, we obtain the two-way discrete ABCs:
\begin{subequations}
\label{eq:2Dabcs}
\begin{equation}
\label{eq:2Dabcs_a}
	{\mathcal E}_{-4} = 
		 \Psi
		\begin{bmatrix}
			{\displaystyle	\frac{q_0^{-1}-q_0}{q_0^3}} && \\
			&\ddots& \\
			&& {\displaystyle \frac{q_{M-1}^{-1}-q_{M-1}}{q_{M-1}^3} }
		\end{bmatrix} \Psi^{-1}{\mathcal E}^0_{{\text{inc}}}
		+ \Psi 
		\begin{bmatrix}
			q_0 && \\
			&\ddots& \\
			&& q_{M-1} 
		\end{bmatrix} \Psi^{-1}{\mathcal E}_{-3}.
\end{equation}
Likewise,
for the ghost node $n=N+4$ we write similarly to~(\ref{eqs:discrete-1DTWBCs_b}):
\[
	u_{N+4,l} = 
		 \left(q_l^{-1}-q_l\right) q_l^{-3} \UincRl
		+ q_l u_{N+3,l},
\] 
and arrive at the following two-way discrete ABCs:
\begin{equation}
\label{eq:2Dabcs_b}
	{\mathcal E}_{N+4} = 
		 \Psi
		\begin{bmatrix}
			{\displaystyle	\frac{q_0^{-1}-q_0}{q_0^3}} && \\
			&\ddots& \\
			&& {\displaystyle \frac{q_{M-1}^{-1}-q_{M-1}}{q_{M-1}^3} }
		\end{bmatrix} \Psi^{-1}{\mathcal E}^{\text{Zmax}}_{{\text{inc}}}
		+ \Psi 
		\begin{bmatrix}
			q_0 && \\
			&\ddots& \\
			&& q_{M-1} 
		\end{bmatrix} \Psi^{-1}{\mathcal E}_{N+3}.
\end{equation}
\end{subequations}
Relations~(\ref{eq:2Dabcs}) provide a fourth-order accurate approximation to the
boundary conditions~(\ref{eqs:2DTWBCs}) for $\delta=3h_z$. 
In the Cartesian or cylindrical case, relation~(\ref{eq:2Dabcs_a}) is substituted into
equation~(\ref{eq:discrete-2DLH}) or (\ref{eq:discrete-3DLH}), respectively,
for $n=-3$, and relation~(\ref{eq:2Dabcs_b}) is
substituted into equation~(\ref{eq:discrete-2DLH}) or (\ref{eq:discrete-3DLH}), respectively,
 for $n=N+3$.
This eliminates the ghost values from the schemes
(\ref{eq:discrete-2DLH}) and (\ref{eq:discrete-3DLH})
and thus
closes the system of finite-difference equations on the grids~(\ref{eq:grid-2D}) 
and (\ref{eq:grid-3D}).

\subsection{\label{ssec:grated-23D}Extension to the Multi-Layer Material}

In the case of a grated Kerr material  described in Section~\ref{sec:intro_multi},
there are additional discontinuity
points  defined by formula~(\ref{eq:grating}).
The interface conditions at each discontinuity 
 point $\tilde z$ are the same as at $z=0$ and
$z=\Zmax$:
\[
  	E(\tilde z+,x_\perp)=E(\tilde z-,x_\perp), \qquad 
   	\frac{\partial E}{\partial z}(\tilde z+,x_\perp)=
		\frac{\partial E}{\partial z}(\tilde z-,x_\perp).
\]
In the simple case when $\tilde z$ coincides with one of the grid
nodes, the discrete approximation of the continuity conditions
is given by the same formula as 
(\ref{eq:discrete-2D-interface-general}) for the Cartesian case
and by the same formula as~(\ref{eq:discrete-3D-interface-general}) for the
cylindrical case.
Hence, to solve  the one-dimensional NLH for the multi-layer
case one needs to apply the corresponding discrete interface condition
of type~(\ref{eq:discrete-2D-interface-general}) 
or~(\ref{eq:discrete-3D-interface-general}) at each plane~(\ref{eq:grating}).
The extension to the case when a discontinuity plane does not coincide
with any of the uniform grid surfaces~(\ref{eq:grid-2D}) or~(\ref{eq:grid-3D}) 
can be obtained by building separate grids for different layers.

\section{\label{sec:Newton}Newton's Solver}

Here we briefly outline our approach to building a Newton type solver
for the NLH. The reader is referred to
\cite[Section 3]{BFT:07} for a detailed description.
The schemes for the NLH that we constructed in Sections~\ref{sec:1D} 
and~\ref{sec:23D} lead to systems of nonlinear difference equations that we
symbolically write as
\[
	\bvec{F}(\bvec{E}) = 0,
\]
where the quantities $\bvec{F}$ and $\bvec{E}$ are complex.
In the one-dimensional case, they are vectors of dimension $N+7$, which is
the dimension  of grid~(\ref{eq:grid-1D}),
$\bvec{F},\bvec{E}\in \mathbb C^{N+7}$.
In the two-dimensional case, $\bvec{F}$ and $\bvec{E}$ can be interpreted as
matrices of dimension $(N+7)\times M$, which is the dimension of
grid~(\ref{eq:grid-2D}), and for convenience we reshape them as
$(N+7)M$-dimensional vectors: $
	\bvec{F},\bvec{E}\in \mathbb C^{(N+7)M}.
$

To linearize the transformation $\bvec{F}(\bvec{E})$, we first notice that the
Kerr nonlinearity $P=|E|^{2\sigma}E$ is Frech\'et nondifferentiable 
as long as $E$ is complex.
To overcome this, we separate the real and imaginary parts and recast the field
$\bvec{E}$ and mapping $\bvec{F}$ as real vectors of twice the dimension: 
\begin{gather*}
	\bvec{E} \in \mathbb C^{(N+7)M}
	\quad \longrightarrow \quad 
	\bvec{\wh{E}} \in \mathbb R^{2(N+7)M}
	,\\
	\bvec{F}(\bvec{E}): \mathbb C^{(N+7)M} \mapsto \mathbb C^{(N+7)M}
	\quad \longrightarrow \quad 
	\bvec{\wh{F}}(\bvec{\wh{E}}) : \mathbb R^{2(N+7)M} \mapsto \mathbb R^{2(N+7)M}.
\end{gather*}
The new transformation $\bvec{\wh{F}}(\bvec{\wh{E}})$ is differentiable in the
conventional real sense.
Differentiation results in Newton's linearization of $\bvec{\wh{F}}$ that
involves the Jacobian $\wwhc{\J}$, and leads to Newton's iterations:
\begin{equation}
\label{eq:Newton}
	\wh{\E}^{(j+1)} - \wh{\E}^{(j)}  \stackrel{\rm def}{=}
		\delta \E^{(j+1)} =
		 - \left[ \wwhc{\J}( E^{(j)} ) \right]^{-1}
		\wh{\F} ( \E^{(j)} ).
\end{equation}

The convergence of Newton's method is known to be very sensitive to the
choice of the initial guess. In our experiments, 
we take the simplest initial guess $\E^{(0)}\equiv0$.
We have also observed numerically that the algorithm was 
more likely to converge if, during
the first stage of the iteration process, when the iterations $\bvec E^{(j)}$
are ``far" from the solution, we introduce the relaxation mechanism:
\begin{equation}
	\label{eq:Newton-iterates}
	\wh{\E}^{(j+1)} - \wh{\E}^{(j)}  =
		\frac \omega {\max \left\{
			1,\Vert \delta \E^{(j+1)} \Vert_\infty
		\right\}}  
	\delta \E^{(j+1)},  
\end{equation}
where $\omega\in (0,1]$; typically
$\omega=0.5$.
While this mechanism enables the algorithm to converge for a wider range of
 cases, it also
slows down the convergence (from quadratic to linear rate).
Therefore, once the iterates $\bvec E^{(j)}$ are ``sufficiently close" to the
solution so that $\Vert \delta \E^{(j)} \Vert_\infty<0.01$, we change back to
$\omega=1$, thereby reverting to the original Newton's method~(\ref{eq:Newton}).
The criterion for convergence that we employ is the inter-iteration distance
threshold $|\delta \E^{(j)}|<10^{-12}$.

\section{\label{sec:method-summary}Summary of the Numerical Method}

The NLH~(\ref{eqs:23DNLH-system}) subject to local 
transverse boundary conditions~(\ref{eqs:continuous-trans-BCs}) and 
nonlocal two-way longitudinal boundary conditions~(\ref{eqs:2DTWBCs}) 
is discretized on the grid~(\ref{eq:grid-2D}) or~(\ref{eq:grid-3D}).
In the Cartesian case, we obtain semi-compact
schemes~(\ref{eq:discrete-2DNLH}) and~(\ref{eq:discrete-2DLH}) in the
interior and exterior of the Kerr material, respectively, and
discretization~(\ref{eqs:discrete-2D-interface}) for the continuity conditions at
the interface.
In the cylindrically symmetric case, we arrive at the semi-compact 
schemes~(\ref{eq:discrete-3DNLH}) and~(\ref{eq:discrete-3DLH}) 
at the interior and exterior nodes, respectively, and 
discretization~(\ref{eqs:discrete-3D-interface}) for the continuity conditions
at the interface.
For both geometries, we also employ discretization~(\ref{eq:rbc}) for the
local transverse radiation boundary condition, and  
discretization~(\ref{eq:2Dabcs}) for the non-local two-way boundary conditions
at $z=-3h_z$ and $z=\Zmax+3h_z$. 
In addition, discretization~(\ref{eq:sbc}) is used at the axis of the
cylindrical system.

The resulting system of nonlinear difference equations with respect
to complex unknowns $E_{n,m}$ is recast in the real form at the expense of
doubling its dimension. Then, Newton's linearization is applied, see
formula~(\ref{eq:Newton}),
which yields a $2(N+7)M\times 2(N+7)M$ sparse Jacobian matrix, with the
bandwidth of $2M$ for the interior and exterior grid points, where $n\neq 0,N$.
For the points at the interfaces, where $n=0$ or $n=N$, the bandwidth is $6M$.

At each Newton's iteration~(\ref{eq:Newton}) or~(\ref{eq:Newton-iterates}), 
this Jacobian needs to be inverted.
Currently, we are using a sparse direct solver to invert the Jacobians.
This entails an
${\cal O}\left(N\cdot M^2\right)$ memory cost and hence imposes a fairly
strict limit on the grid dimension.
For example, a typical grid dimension of $N\times M = 1000 \times 320$
results in the memory requirement of about 6Gb.

\section{\label{sec:interface}Finding an NLS-Compatible Incoming Beam}

As indicated in Section~\ref{sec:intro_model}, the NLH is the simplest
nonparaxial model that generalizes the NLS.
Accordingly, one of our key goals is to investigate how the addition of 
nonparaxiality affects the solution.
In order to do so, we shall use the NLS solutions as  ``benchmarks," and compare
them with NLH solutions computed for ``similar" input parameters.

We note that for an incoming beam $\EincLtext{NLH}$ which impinges on the material
interface at $z=0-$, only a part of it that we denote by 
$E_{\text{refracted}}$ passes through whereas the rest gets reflected.
In contradistinction to that, in the NLS framework 
all of the incoming beam $\EincLtext{NLS}$
propagates forward. 
Therefore, in order to have comparable incoming beams for these two models,
we should choose the NLH incoming beam $\EincLtext{NLH}$ so that the 
{\em refracted} part of it at $z=0+$ be close to the NLS initial
data $\EincLtext{NLS}$, i.e., 
\begin{equation} \label{eq:refract-NLH-NLS}
	E_{\text{refracted}}^{\text{NLH}}(0+,\bvec{x}_\perp) 
		\approx \EincLtext{NLS}(\bvec x_\perp).
\end{equation}
\noindent
A comprehensive solution to this problem is nontrivial, because the reflection
at the nonlinear interface $z=0$ depends on the NLH solution itself for $z>0$.
Therefore, in this paper we use an approximate treatment which experimentally
proves sufficient.

\begin{subequations}
	\label{eqs:single-interface-1D}
	In order to present this approximate treatment, let us first consider the
	one-dimensional linear problem:
	\begin{equation}
		\frac{d^2E(z)}{dz^2} + \lri^2(z) E = 0,
		\qquad \lri(z) = \begin{cases}
			1,	&	z<0, \\
			\lri,	&	z>0,
		\end{cases} 
	\end{equation}
	with the wave $\EincLtext{NLH}e^{iz}$ impinging on
	the interface from the left.
	The overall field has the form: 
	\begin{equation}
	E(z) = 
		\begin{cases}
			\EincLtext{NLH} e^{iz} + R e^{-iz}, & -\infty<z \leq 0, \\
			T e^{i\lri z} , & 
				0 \leq z < \infty,
		\end{cases}
	\end{equation}
	where $R$ and $T$ are the reflection and transmission (refraction) 
	coefficients.
	The values of $R$ and $T$ are obtained from the continuity condition at the
	interface
	\begin{equation}
		E(0-)=E(0+),\qquad 
		\frac{dE}{dz}(0-) = \frac{dE}{dz}(0+),
	\end{equation}
\end{subequations}
which yields: 
\begin{equation}
	\label{eq:single-interface-1D-refraction}
	|E_{\text{refracted}}^{\text{NLH}}(0+)| = 
	|T| = \frac{
		2
	}{
		1+\lri
	} |\EincLtext{NLH}|.
\end{equation}
Formula~(\ref{eq:single-interface-1D-refraction}) is a standard
result for the transmission of plane waves with normal incidence
at a single linear interface,
see, e.g.,~\cite[Section~7.3, eq.~(7.42)]{Jackson-98}.

We shall use this simple refraction formula to approximate the refracted beam
of our weakly nonlinear multi-dimensional problem:
\[
	E_{\text{refracted}}^{\text{NLH}}(0+,\bvec{x}_\perp) \approx
	\frac{
		2
	}{
		1+\sqrt{ 
			\lri^2 + \epsilon\left|
				E(0+,\bvec{x}_\perp)
   		 	\right|^{2\sigma}
		}
   	}
	\EincLtext{NLH}(\bvec{x}_\perp).
\]
Next, we assume that the NLH solution is close to the
refracted incoming beam,
$
	E(0+,\bvec{x}_\perp) \approx E_{\text{refracted}}(0+,\bvec{x}_\perp),
$ and obtain: 
\begin{equation*}
	E_{\text{refracted}}^{\text{NLH}}(0+,\bvec{x}_\perp) \approx
	\frac{
		2
	}{
		1+\sqrt{ 
			\lri^2 + \epsilon\left|
				E_{\text{refracted}}(0+,\bvec{x}_\perp)
   		 	\right|^{2\sigma}
		}
   	}
	\EincLtext{NLH}(\bvec{x}_\perp).
\end{equation*}
Finally, requirement~(\ref{eq:refract-NLH-NLS}) implies:
\begin{equation} \label{eq:NLH-IC-approx}
	\EincLtext{NLH}(\bvec{x}_\perp)
	= \frac{
		1+\sqrt{
			\lri^2+ \epsilon\left|
				\EincLtext{NLS}(\bvec{x}_\perp)
			\right|^{2\sigma}
		}
	}{2}
	\EincLtext{NLS}(\bvec{x}_\perp).
\end{equation}
Equation~(\ref{eq:NLH-IC-approx}) will be used throughout Section~\ref{sec:results}
for all collimated incoming beams.

\section{\label{sec:results}Numerical Experiments}

\subsection{\label{ssec:solitons} $2D$ Cubic NLH (Solitons)}

\subsubsection{\label{sssec:single-soliton}A Single Collimated Beam
	(Nonparaxial Soliton)}
The Cartesian configuration ($D=2$) models propagation in planar waveguides.
In the case of a cubic nonlinearity ($\sigma=1$) and $\lri=1$, the
one-dimensional NLS~(\ref{eq:NLS}) has solitary wave solutions: 
\begin{equation}
	\label{eq:NLS-soliton-general}
	E(z,x) = \left(
		\frac{2 f^2}{\epsilon}
	\right)^{1/2}
	\frac{
		\exp\left(
			ik_0z (1+\frac{ f^2}{2}) 
		\right)
	}{
		\cosh\left( f k_0 x \right)
	}
	= \frac{\sqrt 2 }{k_0 r_0 \sqrt\epsilon}
	\frac{
		\exp\left(
			ik_0z (1+\frac{1}{2} (k_0r_0)^{-2}) 
		\right)
	}{
		\cosh\left( x/r_0 \right)
	}
\end{equation}
which are called solitons. In formula~(\ref{eq:NLS-soliton-general}),
$r_0$ is the soliton width and $f=(k_0r_0)^{-1}$ is the 
nonparaxiality parameter, which can also be interpreted
as the reciprocal beam width measured in 
linear wavelengths: $2\pi f = \lambda_0/r_0$.

We solve the Cartesian NLH~(\ref{eqs:2DNLH-system}) on an elongated domain:
$\Zmax=240$, $\Xmax=12$, and for $k_0=2\pi/\lambda_{0}=4$, 
$\lri=1$, and 
$\epsilon = k_0^{-2}$.
The problem is driven by the incoming beam
\[
	\EincL(x)=\frac{ 1+\sqrt{ 
		1+\epsilon \sech^2\left( x/\sqrt 2 \right) 
	} }{2}
	\sech\left( x/\sqrt 2 \right),
\] for which the refracted beam is (approximately)
an NLS soliton profile of the width $r_0=\sqrt 2$: \[
	E^0_\text{refracted} \approx 
		\sech\left( x/\sqrt 2 \right),
\] see formula~(\ref{eq:NLH-IC-approx}). 
The corresponding nonparaxiality parameter is $f=1/\sqrt{32}\approx 0.177$, which means
$r_0 = 0.90\cdot \lambda_0$ and  which is considered a very narrow beam.

In this simulation, the field was assumed symmetric with respect to the
$x$-axis, $E(x)=E(-x)$. 
This allows us to increase the resolution in the $x$ direction by a factor of
two.
A non-symmetric simulation at half the resolution provides very similar results.
The grid dimension that we took was $N\times M=4480\times112$, 
which translates into the resolution of $\lambda_0/h_z=30$ and
$\lambda_0/h_x=15$, i.e., $30$ grid points per linear wavelength in the $z$
direction (axial) and $15$ grid points per linear wavelength in the $x$
direction (transverse).

\begin{figure}
	\begin{center}
	\parbox{0.41\textwidth}{\vspace*{-3mm}
	\includegraphics[clip,angle=270,width=0.41\textwidth]{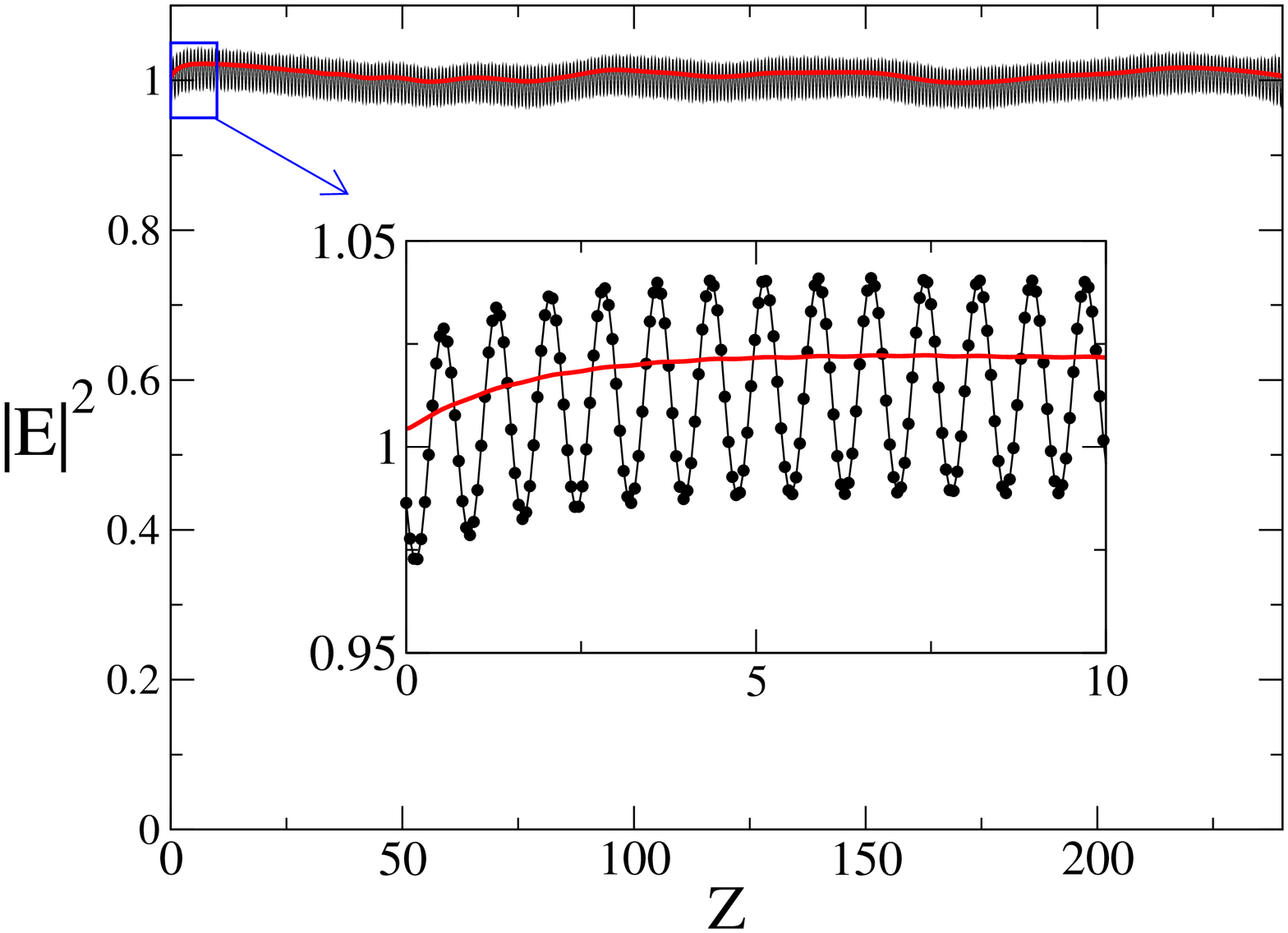}%
             \\ \centerline{A} 
	}\parbox{0.21\textwidth}{
		\includegraphics[clip,width=0.21\textwidth]{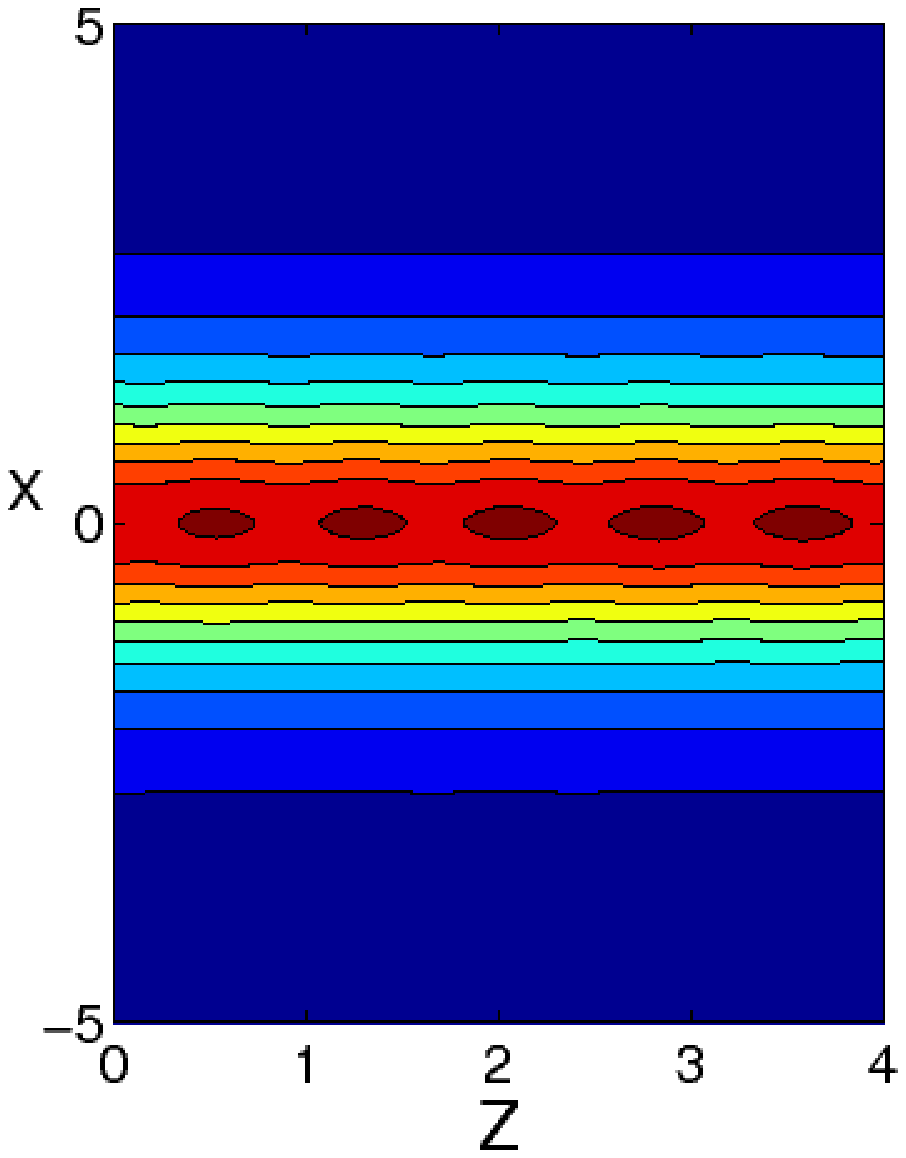}%
             \\ \centerline{B} 
	}\parbox{0.36\textwidth}{
		\includegraphics[clip,width=0.36\textwidth]{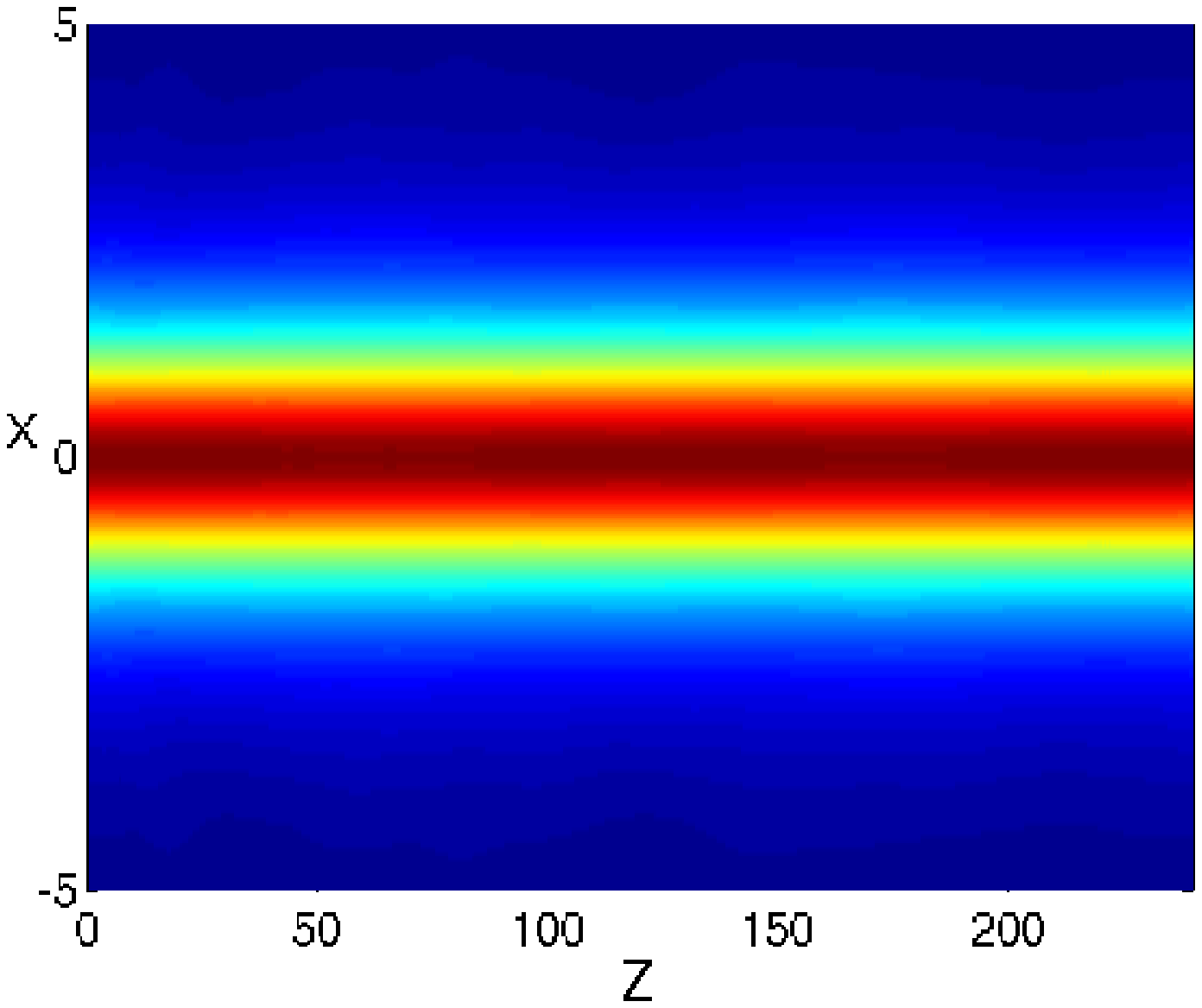}%
             \\ \centerline{C} 
	}	
	\end{center}
	\caption{(color online)\label{fig:single_sech-Sz-amp}
		A nonparaxial soliton for the 2D Cartesian NLH with cubic Kerr nonlinearity.
		A) Normalized on-axis $|E|^2$ (black, dotted) and $S_z$ (red).
		B) $|E|^2$ contour plot (zoom in on several oscillations).
		C) $S_z$ surface plot. 
	}
\end{figure}

In Figure~\ref{fig:single_sech-Sz-amp}A, we plot the on-axis amplitude of the
Cartesian NLH solution.
The square amplitude $|E|^2$ exhibits fast oscillations in the $z$ direction,
as can be seen in the insert of Figure~\ref{fig:single_sech-Sz-amp}A, and
in Figure~\ref{fig:single_sech-Sz-amp}B.
Although at a first glance 
these oscillations may appear a manifestation of numerical instability, in fact they
are physical and indicate the presence of a backward propagating component of the
field.
Indeed, for a field with both forward and backward propagating components: 
\begin{equation} \label{eq:NLH-ansatz}
	E \approx Ae^{ik_0 z} + B e^{ik_0z},
\end{equation} 
the square amplitude is given by the expression:
\begin{equation} \label{eq:amp-NLH-ansatz}
	|E|^2 \approx |A|^2 + 2 \RE \left( AB^*e^{2ik_0z}\right) + |B|^2,
\end{equation} 
which has a $\sim 2k_0$ oscillating term.
We note that the amplitude oscillations in Figures~\ref{fig:single_sech-Sz-amp}A
and \ref{fig:single_sech-Sz-amp}B are indeed $\sim 2k_0$, as predicted by
formula~(\ref{eq:amp-NLH-ansatz}).
We further note that these oscillations are also exhibited by the explicit 
solutions of the $1D$ NLH~\cite{chen-mills-PRB:87}.

We recall that for the NLS the square amplitude $|A|^2$ is proportional to
the energy flux density,\footnote{
	In the Gaussian system, the quantity 
$\frac{c}{4\pi}|E|^2$ has the units of energy flux:  
	$\frac{\text{erg}}{\text{cm}^2\cdot\text{sec}}$, i.e., 
of energy per unit area per unit of time.} 
and that the $L_2$ norm of the solution $
	\|A\|_2^2=\int|A|^2d\bvec{x}_\perp
$ is a conserved quantity  proportional to the total energy flux or,
equivalently, the beam power.
For the NLH, however, the proper measure of the energy flux density is the
Poynting vector: \[
	\bvec{S} = k_0^{-1} \IM\left(E^*\nabla E\right)
\]
rather than the square amplitude.
Accordingly,  the conserved beam power (i.e., the total energy flux) is the
integral  its $z$ component over the beam cross-section: 
\[
	N=\int S_z d\bvec{x}_\perp ,\qquad 
	S_z = k_0^{-1}\IM\left(E^*E_z\right).
\]
Then, for the field~(\ref{eq:NLH-ansatz}) with both forward and backward
propagating components, the energy flux density reduces to
\begin{equation} \label{eq:Sz-NLH-ansatz}
	S_z \approx \left(|A|^2 - |B|^2 \right),
\end{equation} 
i.e., to the difference of the forward and backward square amplitudes. 
Clearly, expression~(\ref{eq:Sz-NLH-ansatz}) 
contains no (rapidly) oscillating terms.
The Poynting flux $S_z$ for the $2D$ NLH solution is given in 
Figure~\ref{fig:single_sech-Sz-amp}C, and is indeed much smoother than the
amplitude, see Figure~\ref{fig:single_sech-Sz-amp}A.
We therefore suggest that the energy flux density provides a more adequate 
quantitative measure of the long-scale (collapse) dynamics in the NLH model.

The key physical question that the simulations in this section attempt to answer
was whether there exist any solitons beyond the paraxial limit, i.e., of  the
$\mathcal O (\lambda_0)$ radius.
Considering the energy flux of the $2D$ NLH solution with $\sigma=1$ shown in
Figure~\ref{fig:single_sech-Sz-amp}C, we see  that it indeed resembles a
soliton propagating essentially unchanged in the positive $z$ direction.
We can therefore conclude that such a nonparaxial soliton does exist.

Let us also note that nonparaxial solitons (solutions of the NLH, rather than NLS)
for a single collimated beam were obtained in \cite{Posada-McDonald-New:1}
for the case of a semi-infinite Kerr medium. Our formulation is different as it involves a
finite-width Kerr material slab with the interfaces that may partially reflect the waves.
Hence, a direct comparison of our results with those of 
 \cite{Posada-McDonald-New:1} is not appropriate. However, a comparison
from the standpoint of physics may be of interest
for the future.

\subsubsection{Grid Convergence Study}
\label{sssec:grid_conv-2D}

In order to demonstrate the fourth-order grid convergence in the nonlinear
regime, we simultaneously refine the grid in the transverse and longitudinal
direction, and  monitor the maximum difference between the computed fields for
each pair of consecutive grids, the coarser and the finer, that differ by a
factor of 2 in size.
For the grids with fewer than roughly $7$ points per linear wavelength, 
the iterations diverge, apparently due to insufficient resolution.
Hence, we choose our coarsest grid to have $\lambda_0/h_z=7.5$ nodes per
wavelength, and compare the results with those on the twice as fine grid,
$\lambda_0/h_z=15$.
Then, we keep  decreasing the size and hence increasing the dimension of the
grid, and the largest dimension that we can take is limited by the memory
requirements of the LU solver that we employ for inverting the Jacobians (see
Sections~\ref{sec:Newton} and \ref{sec:method-summary}).
Currently, it is close to $N\times M=4480\times112$, which corresponds to $30$
points per linear wavelength in the $z$ direction.
The results of the grid convergence study are summarized in
Table~\ref{tab:gridconv-2D}.
The convergence rate that we find is $\mathcal{O}\left(h^{3.8}\right)$, which is
close to the design rate of~$\mathcal{O}\left(h^{4}\right)$.

\newcommand{\respair}[2]{%
	{$\displaystyle \left(\frac{\lambda_0}{#1}, \frac{\lambda_0}{#2}\right)$}%
}
\begin{table}[ht!]
	\begin{center}
		\caption{\label{tab:gridconv-2D}
			Grid convergence study for the 2D Cartesian NLH with $\sigma=1$, 
			$k_0=4$, $\epsilon=k_0^{-2}$, $\Zmax =240$, and $\Xmax=40$. 
		}
		\begin{tabular}{|c||c|c|c|}
			\hline 
			$\left(
				h_z, h_\rho 
			\right)$ & 
				\respair{15}{7.5} &
				\respair{21}{10} &
				\respair{30}{15} \\
			\hline
			\hline 
			$\Vert E^{(2h)}-E^{(h)}\Vert_\infty$ & 
				$1.1$ & 
				$0.30$ & 
				$0.080$ \\
			\hline 
			$\log_{2} \Vert E^{(2h)}-E^{(h)}\Vert_\infty$ & 
				$0.20$ & 
				$-1.7$ & 
				$-3.6$ \\
			\hline
		\end{tabular}
	\end{center}
\end{table}

\subsubsection{\label{sssec:CP-angled-soliton} Collision of Nonparaxial Solitons}

The NLH is an elliptic equation with no preferred direction of propagation.
Therefore, it can be used to model the interaction of beams traveling at 
different angles, and specifically counter-propagating beams.
To demonstrate this capability, we solve the 
$2D$ NLH with $\sigma=1$ for two configurations. In the
first one, two perpendicular nonparaxial solitons collide, while in
the second one, two counter-propagating beams collide almost head-on,
at the angle of $150\degree$. 
Note that the paraxial approximation is invalid in
the region of interaction between the beams for either case. 

\begin{figure}[ht]
	\begin{center}
	\begin{tabular}{cc}
	\includegraphics[clip,width=0.525\textwidth]{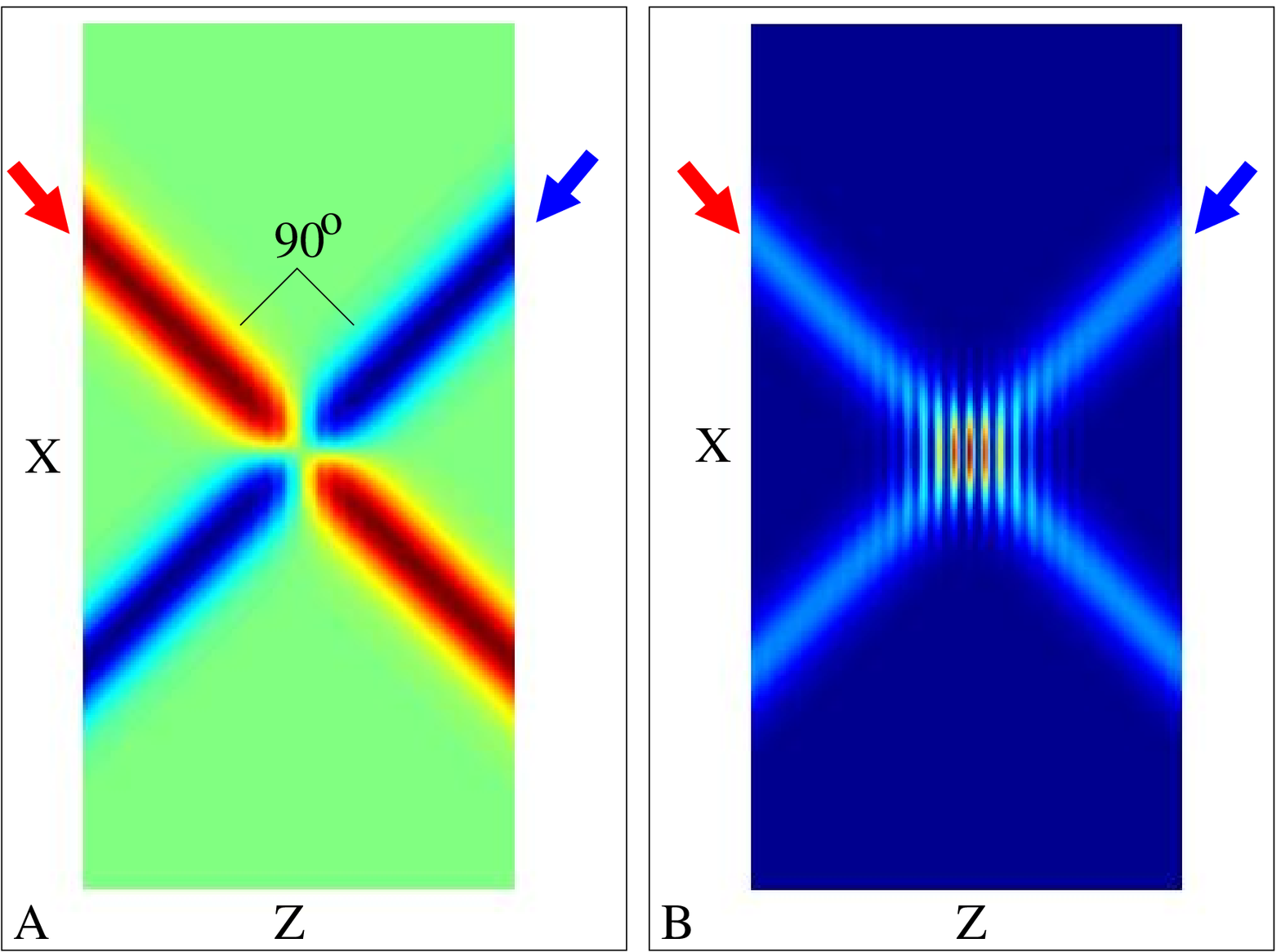} &
	\includegraphics[clip,width=0.45\textwidth]{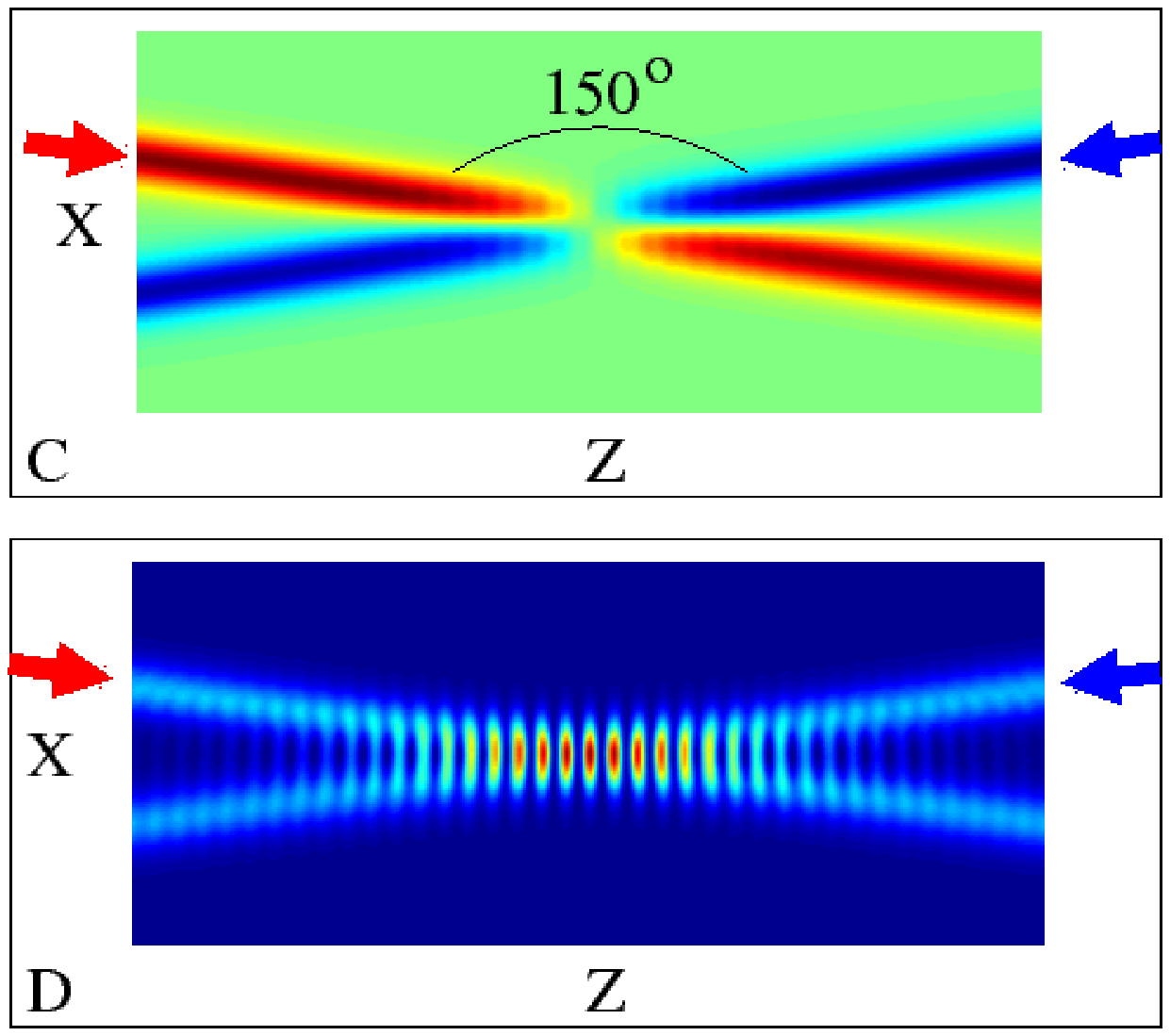}
     \end{tabular}
	\end{center}
	\caption{(color online)\label{fig:CP-soliton}
		Collisions of two nonparaxial solitons in the 2D Cartesian NLH with
		cubic Kerr nonlinearity.
		A) Collision angle $90\degree$, $S_z$ surface plot.
		B) Collision angle $90\degree$, $|E|^2$ surface plot.
		C) Collision angle $150\degree$, $S_z$ surface plot.
		D) Collision angle $150\degree$, $|E|^2$ surface plot.
	}
\end{figure}
For the perpendicular beam configuration, we solve the $2D$ NLH with $k_0=6$, 
$\Zmax=20$, $\Xmax=30$, $\lri^2=1$, and $\epsilon=k_0^{-2}$.
The forward-traveling incoming beam enters the material slab at $z=0$, $x=10$, and
propagates in the $-45\degree$ direction, while its counterpart enters at 
$z=\Zmax$, $x=10$, and propagates in the $-135\degree$ direction.
The resolutions were $\lambda_0/h_z=\lambda_0/h_x=10$ points per linear
wavelength.
A surface plot of the energy flux density $S_z$ is shown in
Figure~\ref{fig:CP-soliton}A.
Positive values of $S_z$ (forward propagation) are red, while negative values 
(backward propagation) are blue.
As in the paraxial NLS model, the two nonparaxial 
solitons are almost unchanged by the collision.
A surface plot of $|E|^2$ is shown in Figure~\ref{fig:CP-soliton}B;
the oscillations in the interaction region are due to the presence of
counter-propagation waves.

For the head-on collision configuration, we solve the $2D$ NLH with $k_0=4$, 
$\Zmax=30$, $\Xmax=12$, $\lri^2=1$ and $\epsilon=k_0^{-2}$.
The forward-traveling incoming beam enters the material slab at $z=0$, $x=4$, and
propagates in the $-15\degree$ direction, while its counterpart enters at 
$z=\Zmax$, $x=4$, and propagates in the $-165\degree$ direction, resulting
in a collision at the angle of $150\degree$.
The resolutions were $\lambda_0/h_z=\lambda_0/h_x=16$ points per linear
wavelength.
The results presented in Figures~\ref{fig:CP-soliton}C and
\ref{fig:CP-soliton}D show that as in the previous case,
the solitons re-emerge essentially intact after the collision.

\subsection{\label{ssec:arrest}Arrest of Collapse in the NLH}

\subsubsection{\label{sssec:arrest-3D} $3D$ Cylindrically Symmetric Case}

We solve the cylindrically symmetric NLH~(\ref{eqs:3DNLH-system}) for 
$\sigma=1$, $k_0=2\pi/\lambda_{0}=8$, $\lri=1$, $\Zmax=9$, and $\Rmax=3.5$. 
The problem is driven by the incoming beam~$
	\EincL(\rho)=\frac{ 1+\sqrt{ 1+\epsilon e^{-2\rho^2} } }{2}e^{-\rho^2},
$ for which the refracted beam is approximately a Gaussian: $
	E^0_\text{refracted} \approx e^{-\rho^2}
$, see formula~(\ref{eq:NLH-IC-approx}).
The grid dimension is $N\times M=1080\times360$, 
which translates into the resolution of $\lambda_0/h_z=94$ and
$\lambda_0/h_\rho=81$, i.e., $94$ grid points per linear wavelength in the $z$
direction (axial) and $81$ grid points per linear wavelength in the $\rho$
direction (radial).

While this estimate shows that the waves in the linear region are very well resolved,
we note that the NLH \[
		\Delta E + k_{\text{NL}}^2\left(|E|^2\right)E = 0
		,\qquad 
		k^2_{\text{NL}}=k_0^2\left(1+\epsilon|E|^2\right),
\] supports waves with {\em nonlinear} wavenumber $k_{\text{NL}}$.
In order to ensure that these nonlinear waves are also well resolved, a similar 
resolution estimate should be performed for the {\em nonlinear} wavelength 
$\lambda_{\text{NL}}=\lambda_0/\sqrt{1+\epsilon|E|^2}$.
Below, we will see experimentally that at the maximum self focusing point
(with the maximal amplitude), we have $\epsilon|E|^2\approx4.6$.
Hence, the nonlinear waves with the minimum wavelength of 
$\lambda_{\text{NL}}\approx\lambda_0/2.4$ are well resolved, with $\lambda_{\text{NL}}/h_z\approx40$
points per nonlinear wavelength in the $z$ direction, and $\lambda_{\text{NL}}/h_\rho\approx31$
points per nonlinear wavelength in the $\rho$ direction.\footnote{
	For the soliton simulations in Section~\ref{ssec:solitons}, the nonlinearity
	was smaller and $\lambda_{\text{NL}}\approx\lambda_0$, so that a separate
	resolution estimate for $\lambda_{\text{NL}}$ was not needed.
} 

The nonlinearity coefficient  was $\epsilon=0.15$.
The parameter that controls the beam collapse in the corresponding critical 
NLS~(\ref{eq:NLS}) is the ratio of the incoming beam power $
	P_0=\int_0^\infty\rho e^{-2\rho^2}d\rho=\frac{1}{4}
$ to the critical power $
	P_{\rm c}\approx1.8623 / (\epsilon k_0^2)
$, see~\cite{Fibich-Papanicolau:99}.
For the NLH~(\ref{eqs:3DNLH-system}) with the values of the parameters we have
chosen, this power ratio is related to the nonlinearity coefficient~$\epsilon$
as
\[	
	p  =  \frac{P_0}{P_{\rm c}}\approx\frac{\epsilon}{4\cdot1.8623} k_0=1.29.
\]

\begin{figure}
	\begin{center}
	\includegraphics[clip=,width=0.37\textwidth,angle=-90]{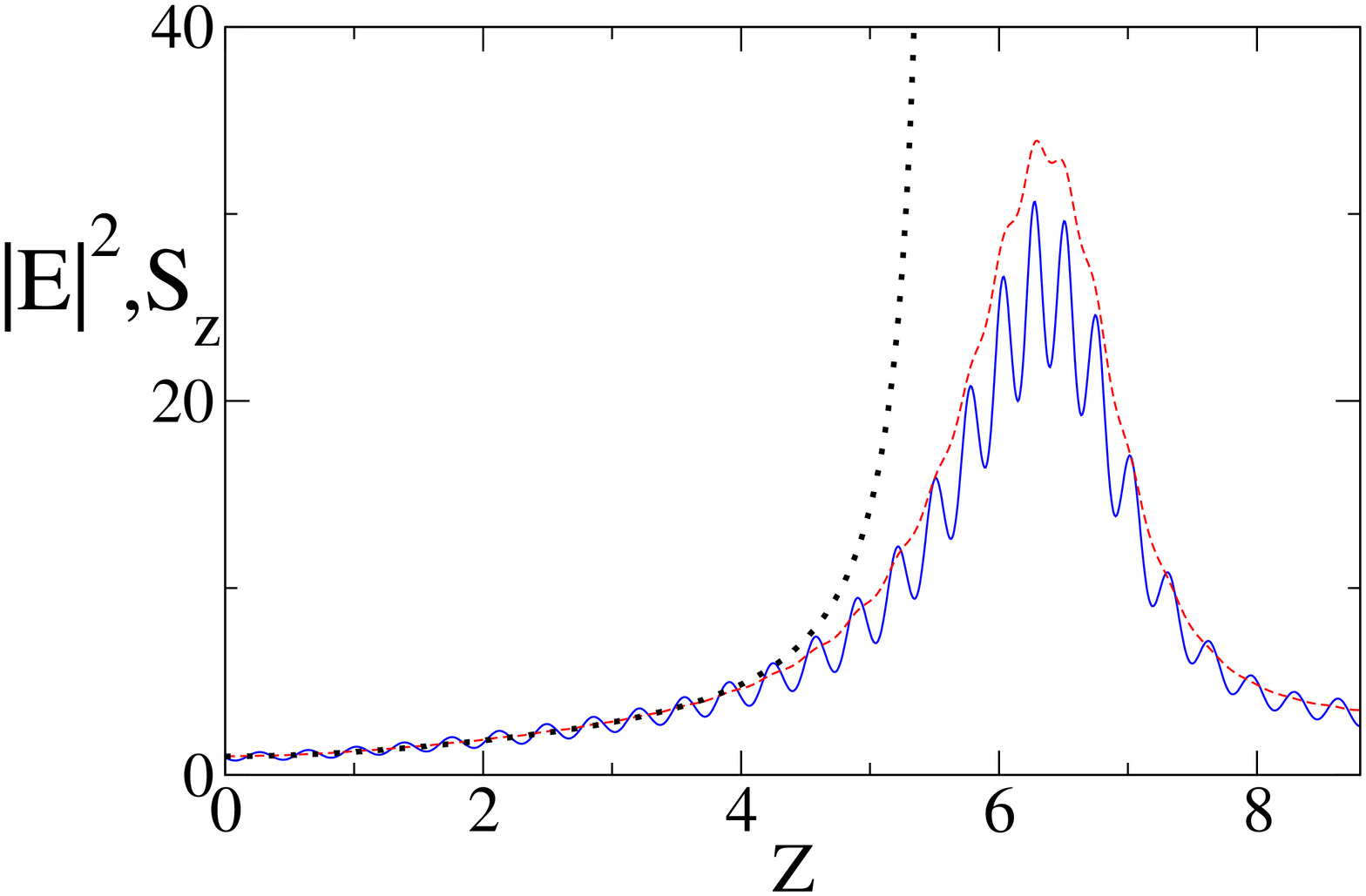}~\includegraphics[clip=,width=0.37\textwidth,angle=-90]{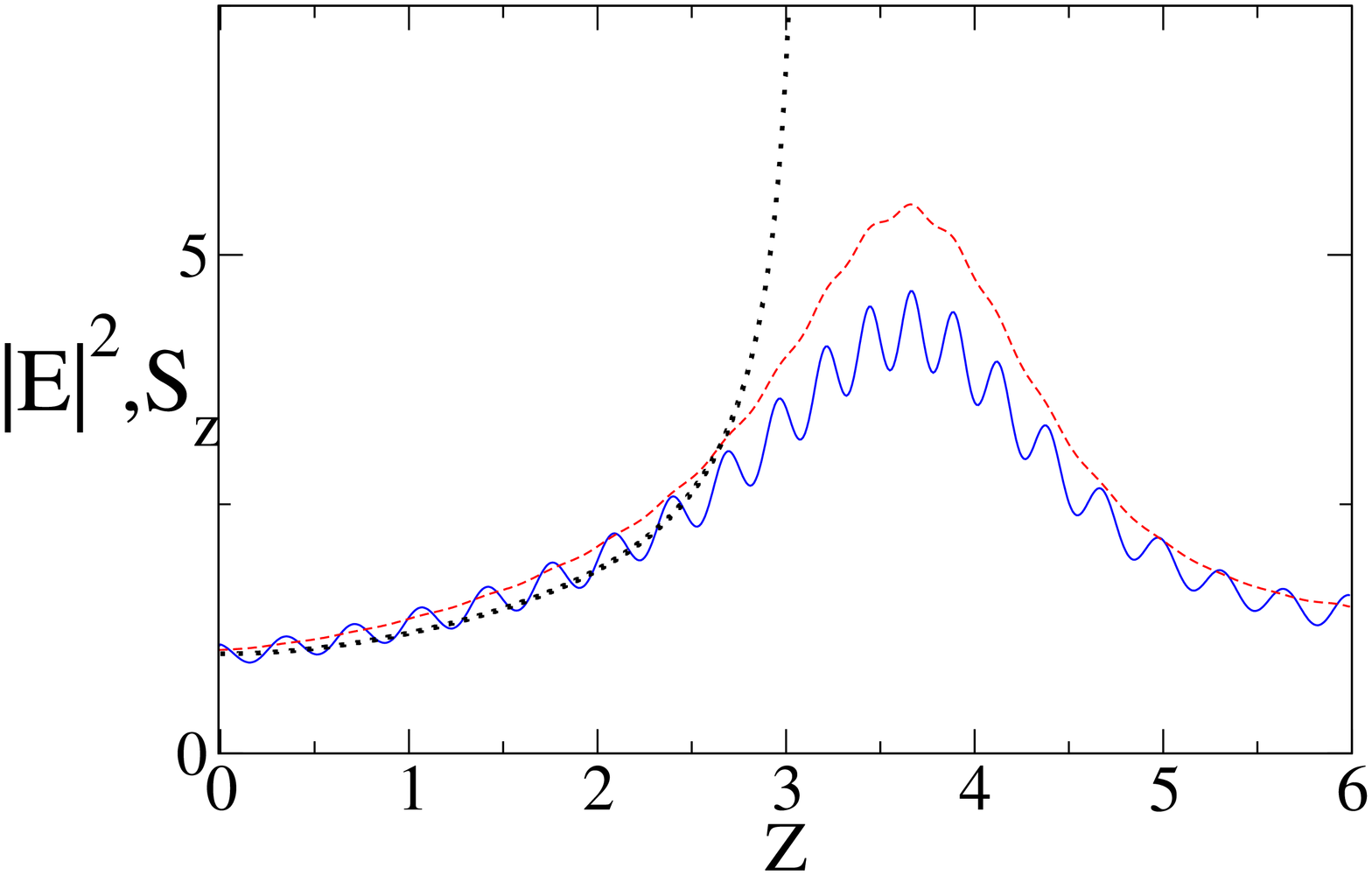} \\
    \hspace*{10mm}A\hspace*{85mm}B
	\caption{(color online)\label{fig:collapse-3D-oa}
		Arrest of collapse in the NLH:
		Normalized on-axis square-amplitude $|E|^2$ (blue solid), the Poynting
		vector $S_z$~(red dashed), and the NLS solution (black dotted) on the
		axis. 
		A) $D=3$ and $\sigma=1$.
		B) $D=2$ and $\sigma=2$.
	}
	\end{center}
\end{figure}

In Figure~\ref{fig:collapse-3D-oa}A, we compare the cylindrically symmetric
NLH solution with the 
corresponding NLS solution at the axis of symmetry $\rho=0$.
Since the beam power is $29\%$ above $P_{\rm c}$, the solution to the NLS blows
up and its on-axis amplitude tends to infinity at $z\approx 5.5$.
The corresponding NLH solution, however, remains bounded and its amplitude
attains its maximum 
	$\displaystyle \max_{n,m}|E_{n,m}|\approx5.5
$ at $z\approx 6.25$. This yields the maximum Kerr nonlinearity of $
	\displaystyle \max_{n,m} \{\epsilon |E_{n,m}|^2\}\approx4.6
$.

\begin{figure}[t]
	\begin{center} 
		\begin{tabular}[c]{cc}
			\includegraphics[clip,width=0.5\textwidth]{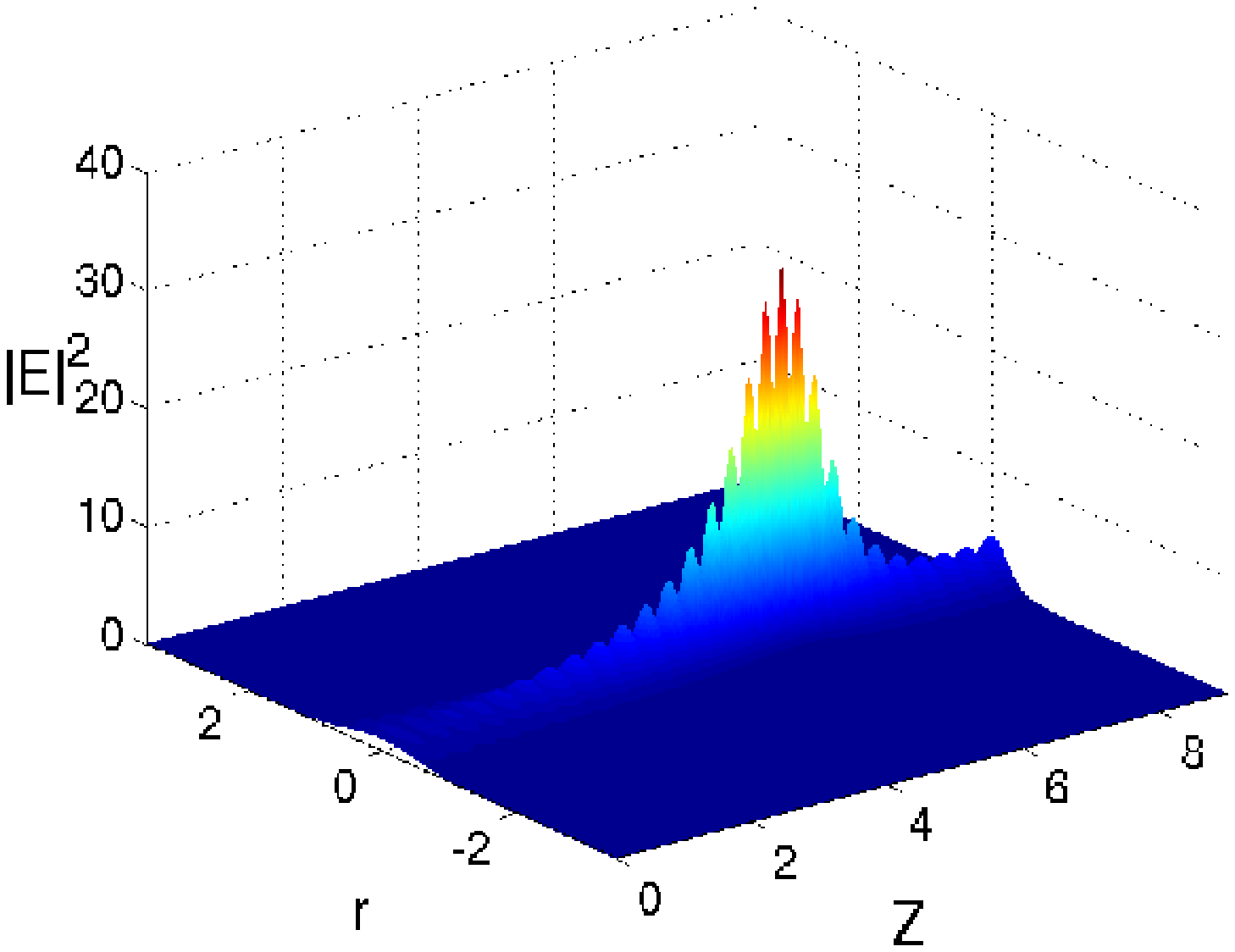}
			& 
			\includegraphics[clip,width=0.5\textwidth]{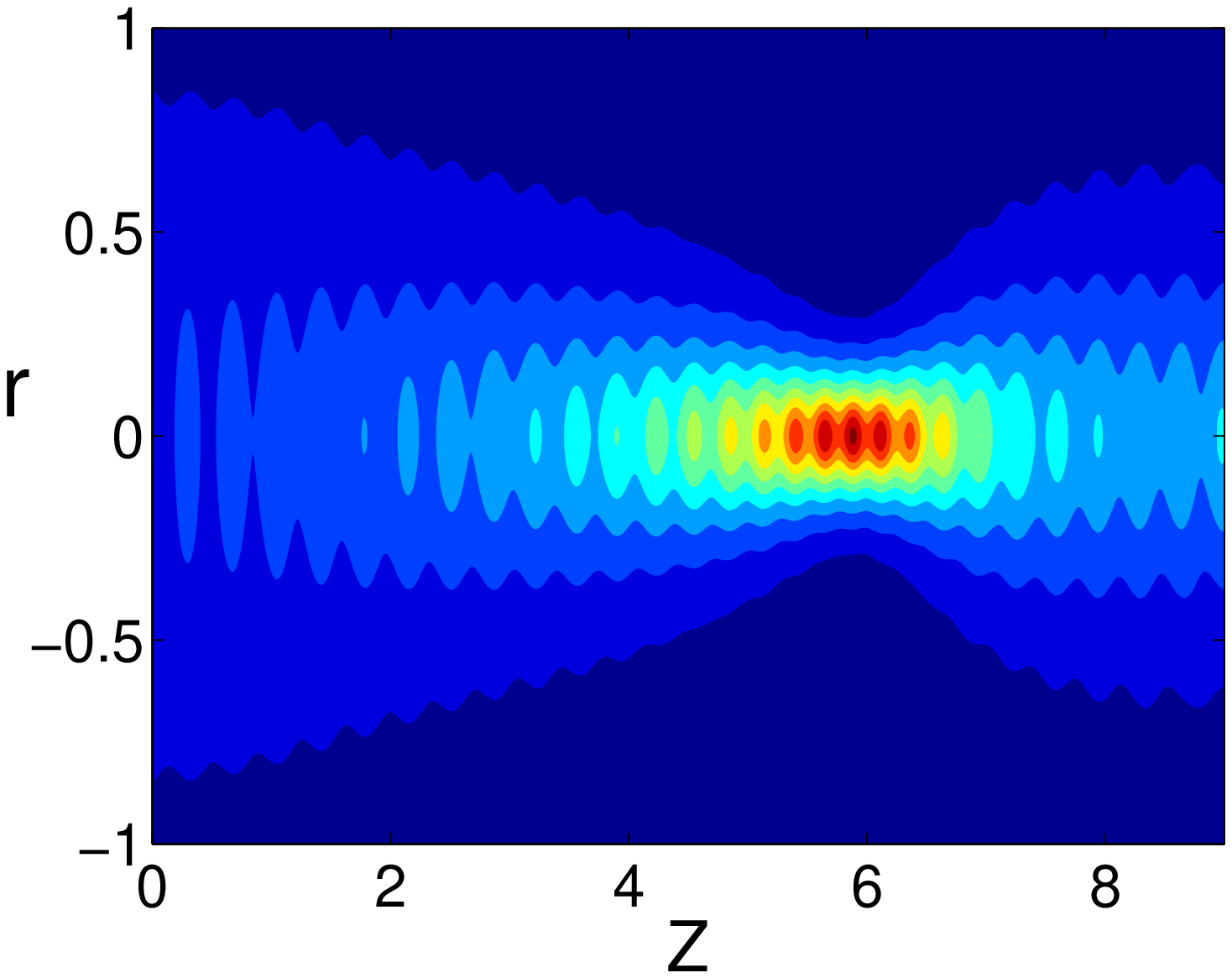} \\
			\includegraphics[clip,width=0.5\textwidth]{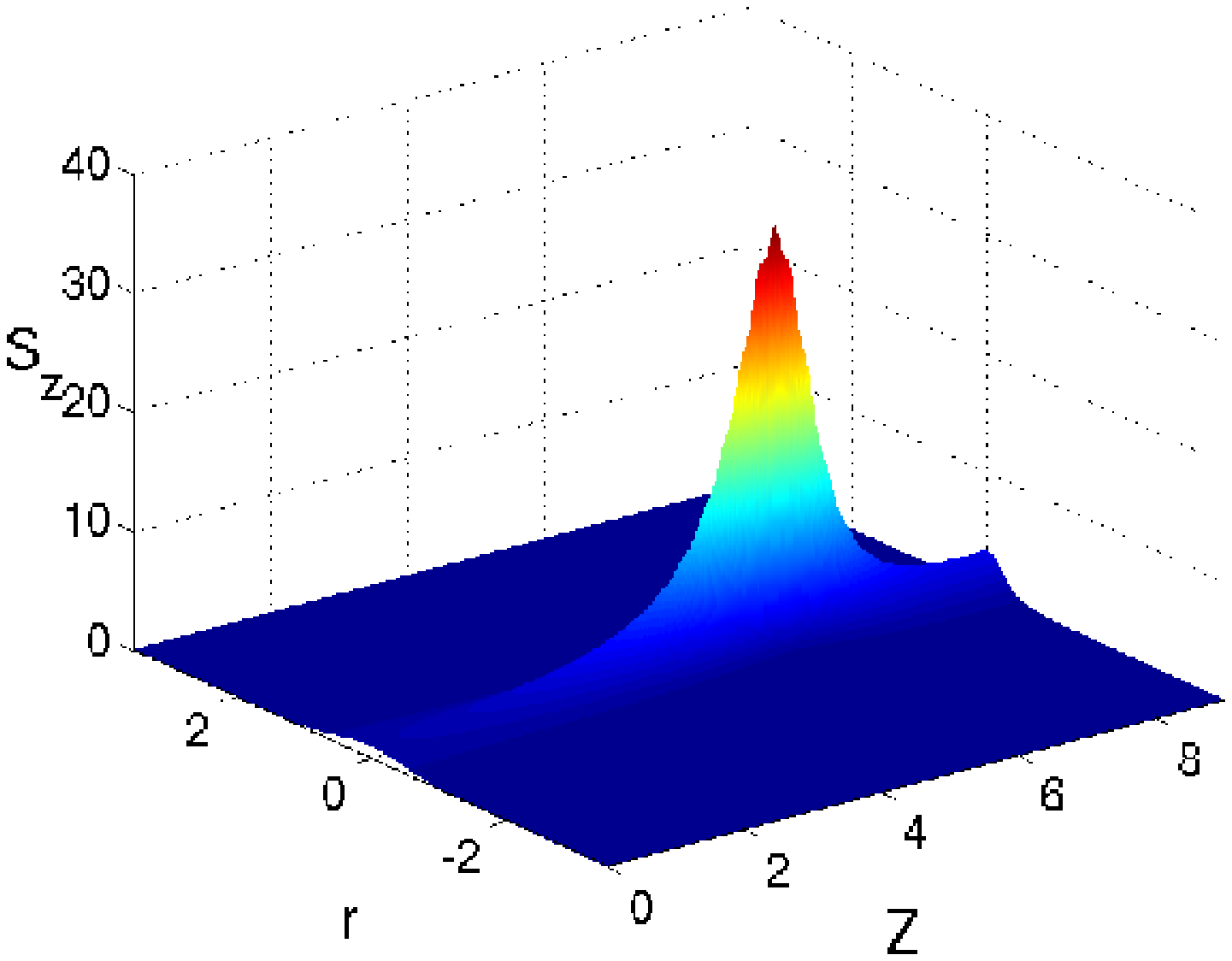}
			& 
			\includegraphics[clip,width=0.5\textwidth]{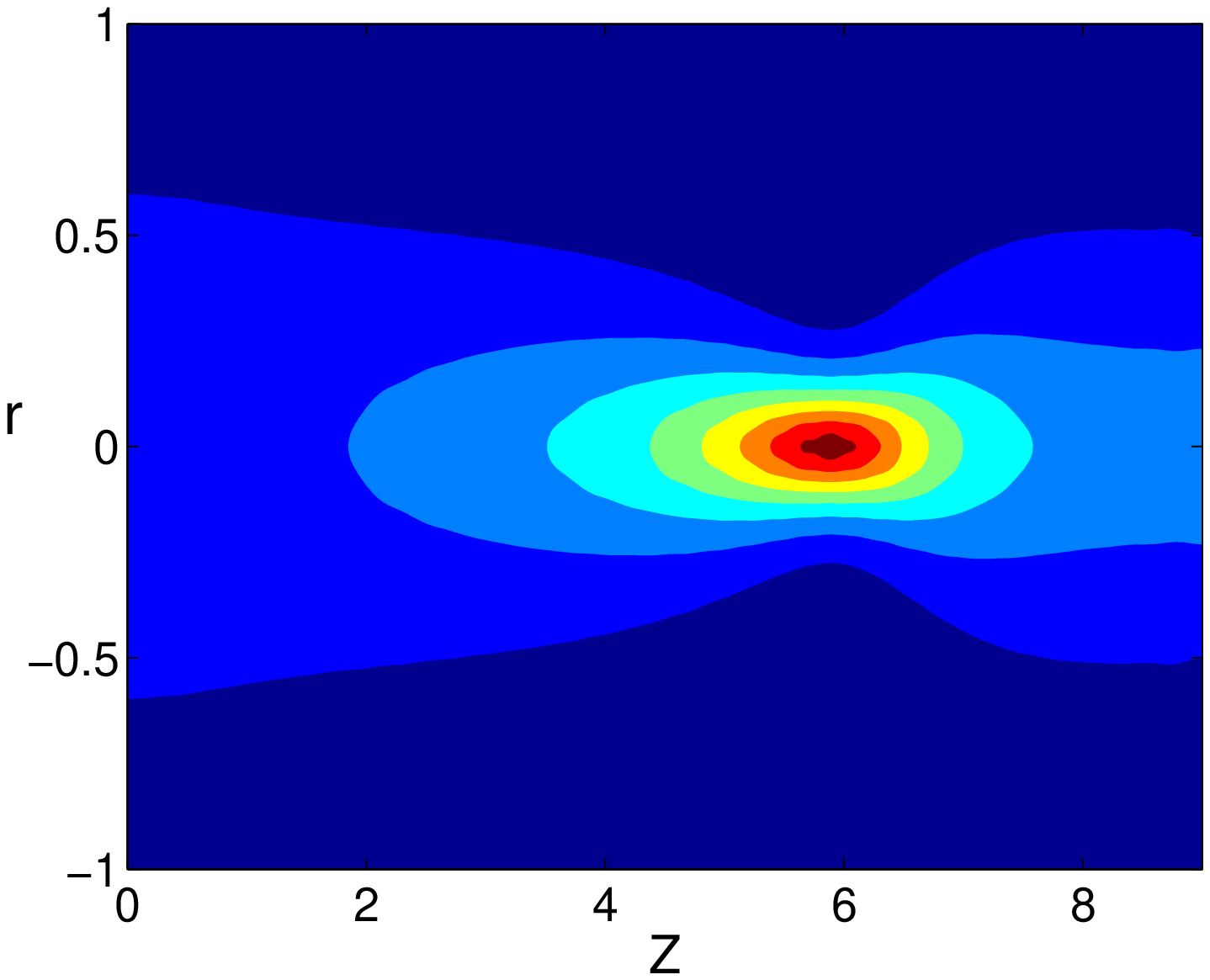}
		\end{tabular}

	\caption{(color online)\label{fig:collapse-3D-amp-Vs-Sz}
		Arrest of collapse in the cylindrically symmetric NLH.
		Plots of the square amplitude (top) and the energy flux density (bottom).
	}
	\end{center}
\end{figure}

The square amplitude and energy flux density of the cylindrically symmetric
NLH solution are displayed
in Figure~\ref{fig:collapse-3D-amp-Vs-Sz}.
As in the ``soliton" case, fast oscillations in the $z$ direction are clearly 
observed for the square amplitude, but not for the energy flux, which appears
smooth.

\paragraph{Grid Convergence Study}
\label{sssec:grid_conv-3D}

In order to demonstrate the fourth-order grid convergence for the cylindrical 
geometry case, we conduct a grid convergence study similar to that of
Section~\ref{sssec:grid_conv-2D}.
For the grids with fewer than roughly $18$ points per linear wavelength, 
the iterations diverge. This is apparently due to insufficient resolution in the
region of strong focusing, where it will only be about $18/2.4=7.5$ nodes per
wavelength. As such,
the coarsest grid we have taken had $17.5$ points per linear wavelength in the $z$
direction, and the finest grid was $N\times M=1140\times380$, which corresponds
to $100$ points per linear wavelength in the $z$ direction.
The results of the grid convergence study are summarized in
Table~\ref{tab:gridconv-3D}.
The convergence rate that we find is $\mathcal{O}\left(h^{4.88}\right)$,
which is even somewhat better than the $\mathcal{O}\left(h^{4}\right)$
theoretical rate.

\begin{table}[H]
	\begin{center}
		\caption{\label{tab:gridconv-3D}
			Grid convergence study for the cylindrically symmetric NLH 
			with $\sigma=1$, $p=1.29$, $\Zmax =9$, and $\Rmax=3.5$. 
		}
		\begin{tabular}{|c||c|c|c|c|}
			\hline 
			$\left(
				h_z, h_\rho 
			\right)$ & 
				\respair{35}{30} &
				\respair{50}{43} &
				\respair{70}{60} &
				\respair{100}{86} \\
			\hline
			\hline 
			$\Vert E^{(2h)}-E^{(h)}\Vert_\infty$ & 
				$3.63$ & 
				$0.965$ & 
				$0.176$ & 
				$0.0225$ \\
			\hline 
			$\log_{2} \Vert E^{(2h)}-E^{(h)}\Vert_\infty$ & 
				$1.86$ & 
				$-0.051$ & 
				$-2.51$ & 
				$-5.47$ \\
			\hline
		\end{tabular}
	\end{center}
\end{table}

\paragraph{\label{sssec:domain-size}Effect of the Domain Size}

Our simulations show that the convergence of Newton's iterations depends on the
domain size, and specifically on the length $\Zmax$ of the Kerr material slab.
To investigate this dependence, we attempt to solve the $(2+1)D$ NLH for several
domain sizes $\Zmax=1,2,3,\dots,15$.
In order to limit possible effects of the transverse boundaries on the
convergence, it was positioned relatively far from the axis, at $\Rmax=5$.
In order to limit possible effects of under-resolution, we have chosen
moderate resolutions of $\lambda_0/h_z=53$ points per linear wavelength in the
longitudinal $z$ direction and $\lambda_0/h_\rho=31$ points per linear wavelength
in the transverse $\rho$ direction.
The results are displayed in Table~\ref{tab:convergence-3D}.
It can be seen that for some domain lengths the algorithm converges, while for
others it diverges.
It may be possible that the divergence observed for the domain lengths between
$4$ and $7$ is related to the boundary $z=\Zmax$ being positioned too close to
the region of maximum self-focusing, see Figure~\ref{fig:collapse-3D-oa}A.

\begin{table}[H]
	\begin{center}
		\caption{\label{tab:convergence-3D}
			Convergence of Newton's method for the cylindrically symmetric 
NLH with $\sigma=1$, $k_0=8$, and $\epsilon=0.15$ on the series of domains with
			$\Zmax = 1,2,\dots,15$ and 
			$\Rmax=5$. The criterion of convergence is 
			$\left|\delta \E^{(j)}\right|<10^{-12}$ and $\omega=0.5$.
		}
		\begin{tabular}{|c||c|c|c|c|} \hline 
			$\Zmax$ 	& 1-3  & 4-7		& 8, 9  & 10-15 	\\ \hline
			Convergence	& YES & NO & YES & NO \\ \hline
		\end{tabular}
	\end{center}
\end{table}

\subsubsection{\label{sssec:arrest-2D} The $2D$ Quintic Nonlinearity Case}

We solve the Cartesian NLH~(\ref{eqs:2DNLH-system}) for
$\sigma=2$, $k_0=2\pi/\lambda_{0}=8$, $\lri=1$, $\Zmax=6$, and $\Xmax=3$. 
The problem is driven by the collimated incoming beam~$
	\EincL(x)=\frac{ 1+\sqrt{ 1+\epsilon e^{-4x^2} } }{2}e^{-x^2},
$ for which the refracted beam is approximately a Gaussian: $
	E^0_\text{refracted} \approx e^{-x^2}
$, see formula~(\ref{eq:NLH-IC-approx}).
The grid dimension is $N\times M=900\times300$, which translates into the
resolution of $\lambda_0/h_z=120$ grid points per linear wavelength in the $z$
direction and $\lambda_0/h_x=80$, grid points per linear wavelength in the $x$
direction.
The shortest nonlinear wavelength was $
	\lambda_{\text{NL}}=\lambda_0/\sqrt{1+\epsilon\max|E|^4}
		\sim \lambda_0/1.85.
$ The nonlinear waves are therefore still well resolved, with
$\lambda_{\text{NL}}/h_z=65$ and $\lambda_{\text{NL}}/h_x=43$.
The nonlinearity coefficient was chosen $\epsilon=0.125$, and the ratio of
the incoming beam power was $P_0/P_{\rm c} \approx 1.30$.
The results are displayed in Figure~\ref{fig:collapse-3D-oa}B, and are similar
to the $3D$ cylindrically symmetric critical case.

\subsubsection{\label{sssec:arrest-2D-angle} An Inclined Beam: Focusing-Defocusing
Oscillations}

\begin{figure}[t]
	\begin{center} 
		\begin{tabular}[c]{cc}
			\includegraphics[clip,width=0.5\textwidth]{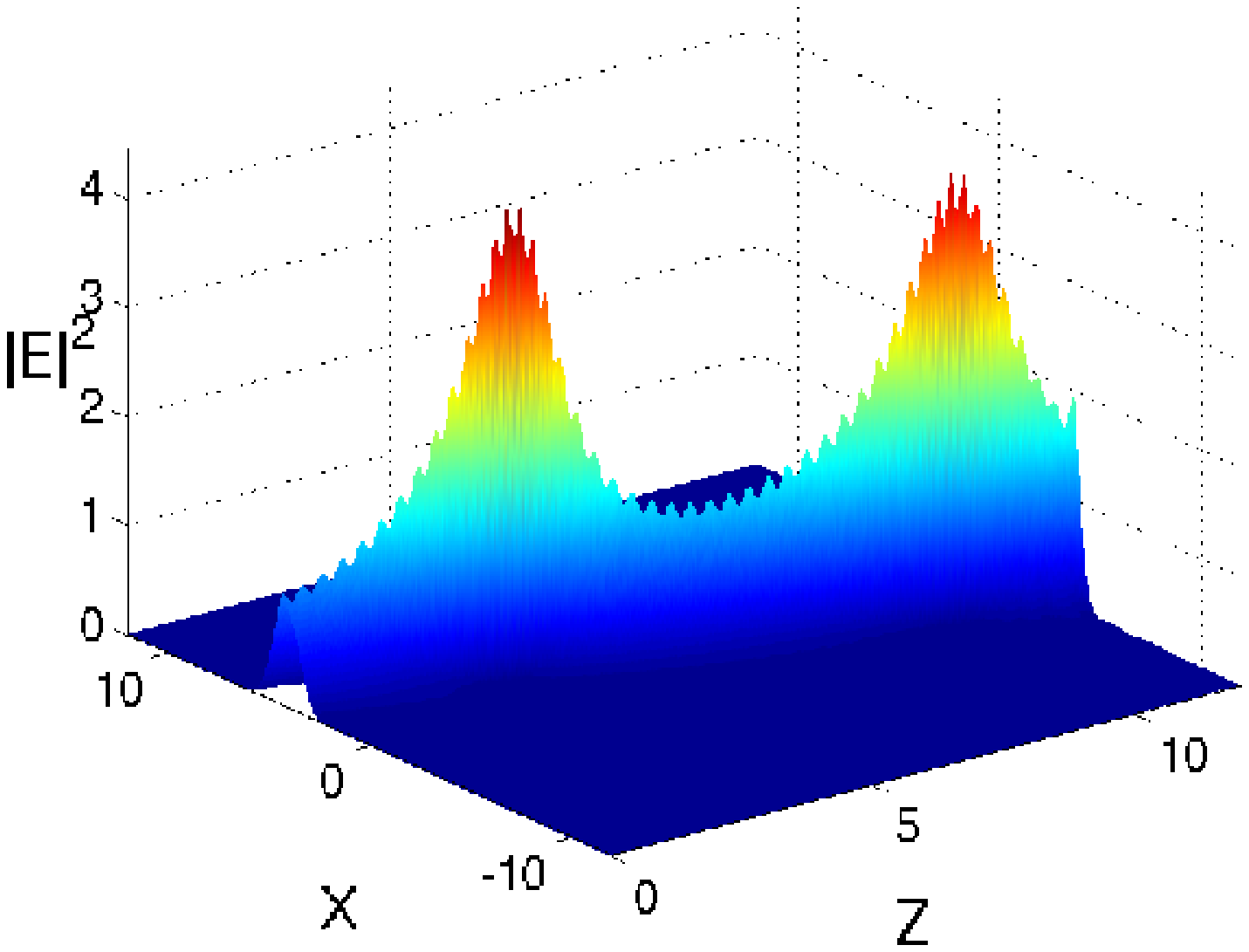}
			&
			\includegraphics[clip,width=0.5\textwidth]{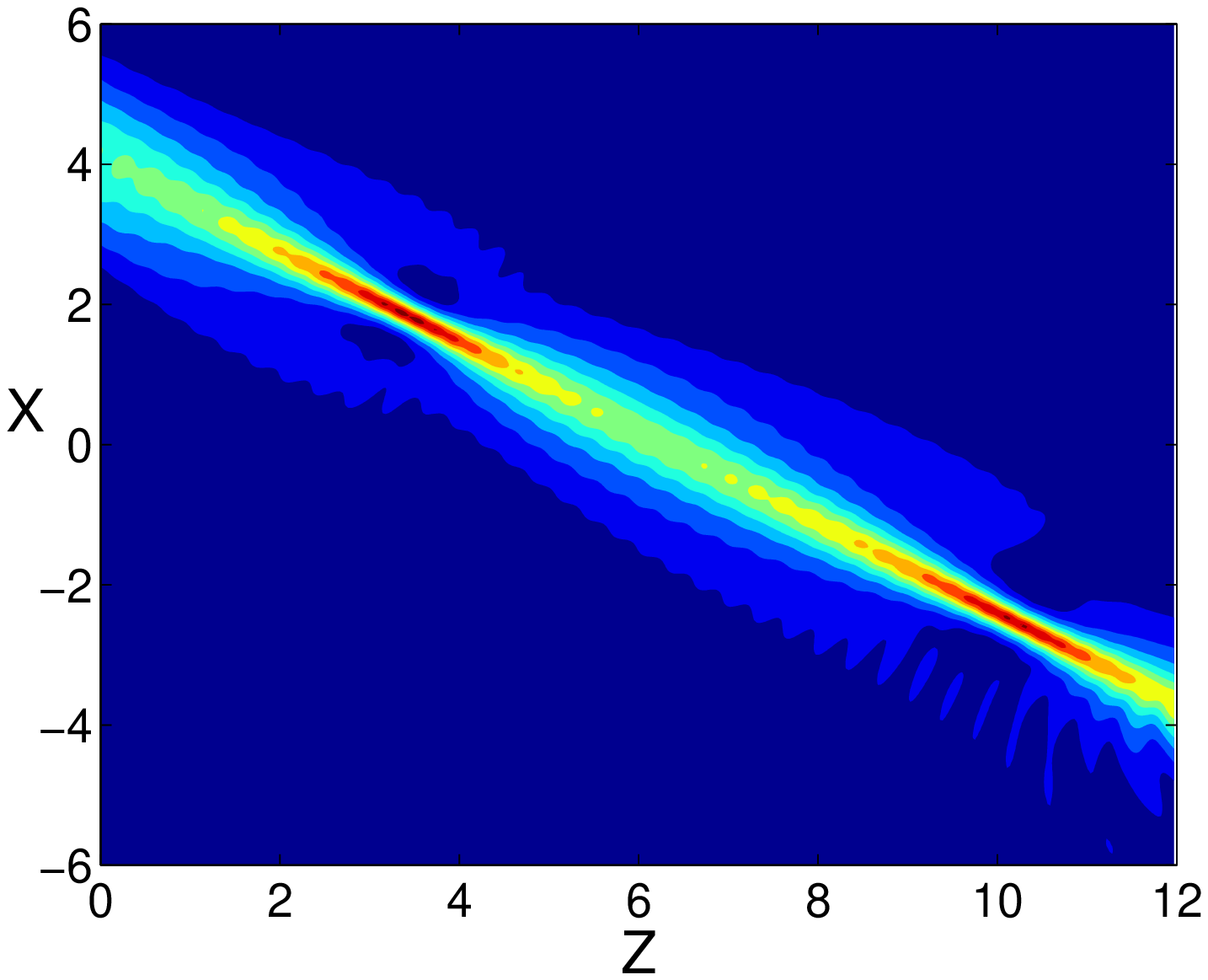} \\
			\includegraphics[clip,width=0.5\textwidth]{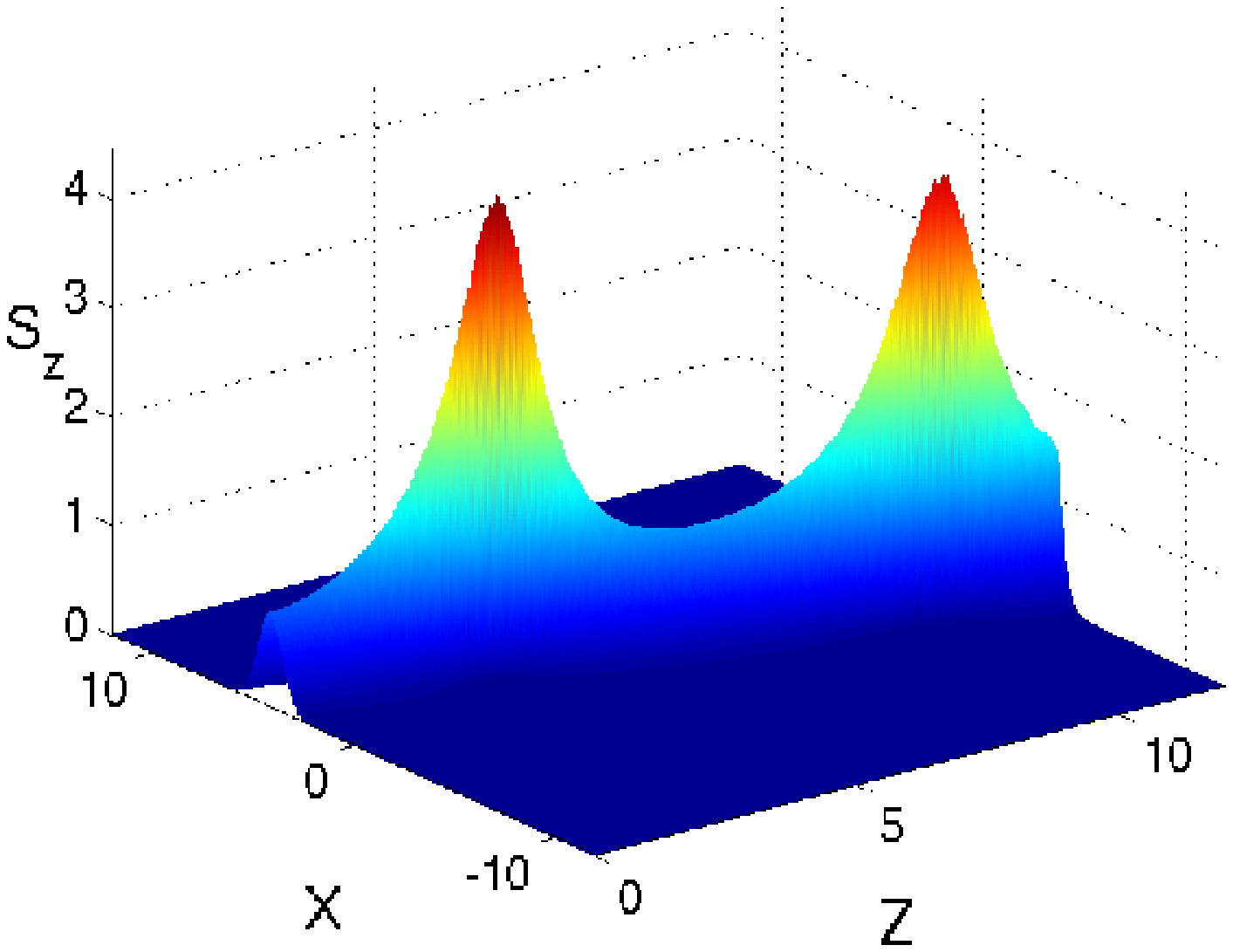}
			&
			\includegraphics[clip,width=0.5\textwidth]{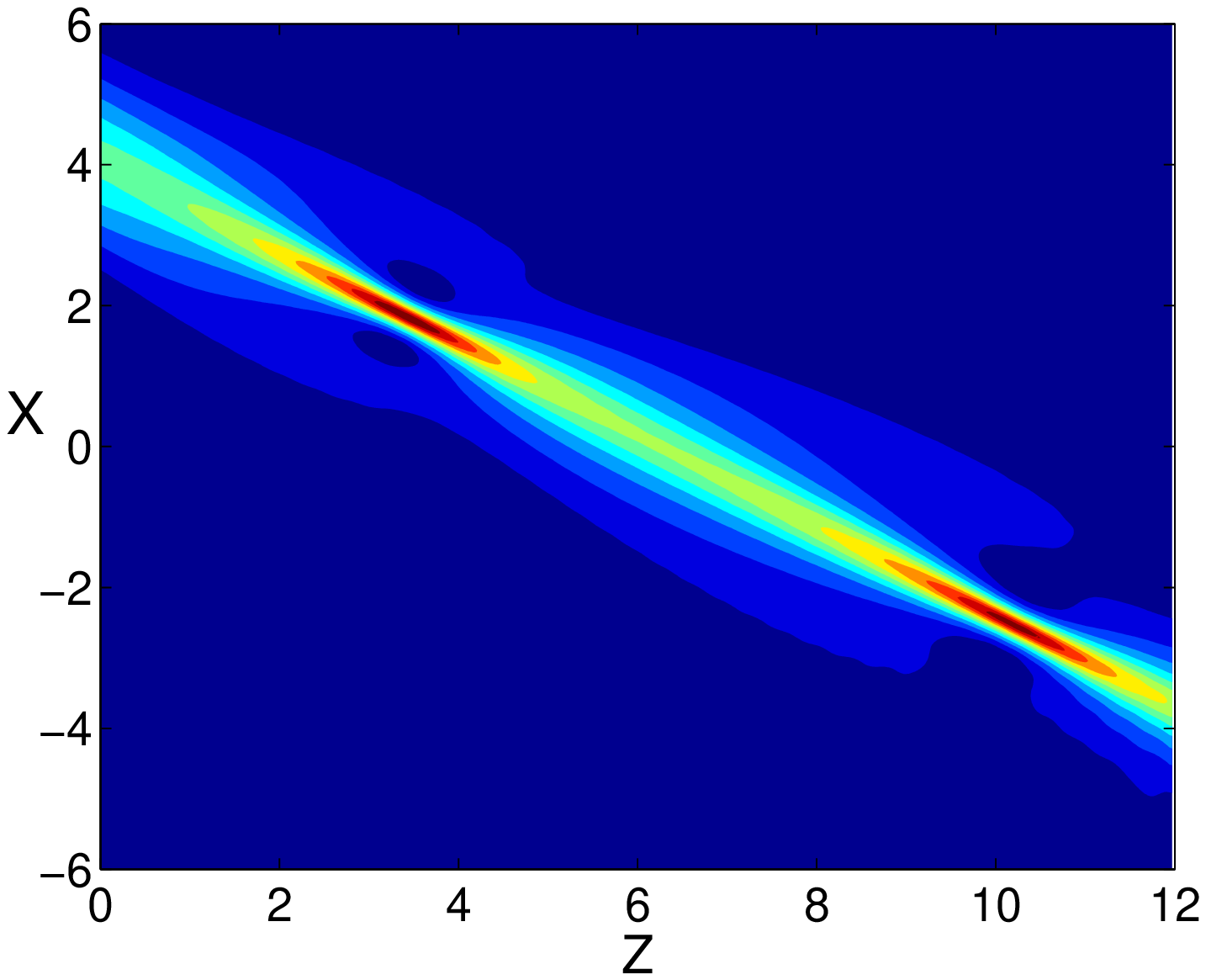}
		\end{tabular}

	\caption{(color online)\label{fig:collapse-2D-angle-amp-Vs-Sz}
		Arrest of collapse for an inclined beam in the 2D Cartesian NLH.
		Plots of the square amplitude (top) and the energy flux density (bottom).
	}
	\end{center}
\end{figure}
We solve the $2D$ NLH with $\sigma=2$, $k_0=8$ and $\lri^2=1$ on the domain with
$\Zmax=12$ and  $\Xmax=12$ for a Gaussian incoming beam entering the Kerr material
at $z=0$, $x=4$ and propagating at the angle of $\pi/5.3$.
The nonlinearity coefficient was $\epsilon=0.12$, which yields the input power of
$28\%$ above critical.
The grid was $N\times M=400\times 800$, which corresponds to resolutions of 
$\lambda_0/h_z=\lambda_0/h_x=26$ points per linear wavelength, and $14$
points per nonlinear wavelength $\lambda_{\text{NL}}$.

As shown in Figure~\ref{fig:collapse-2D-angle-amp-Vs-Sz}, 
the beam undergoes {\em two} focusing-defocusing oscillations, which qualitatively
agrees with the predictions of the modulation theory for the NLS~\cite{Fibich-PRL:96}.
This is the first time that two focusing-defocusing oscillations are observed in
a critical NLH model.

\subsection{The Effect of Adjusting the Incoming Beam}
\label{sec:inc_adj}

\begin{figure}[ht!]
        \begin{center}
\includegraphics[clip,width=0.4\textwidth,angle=270]{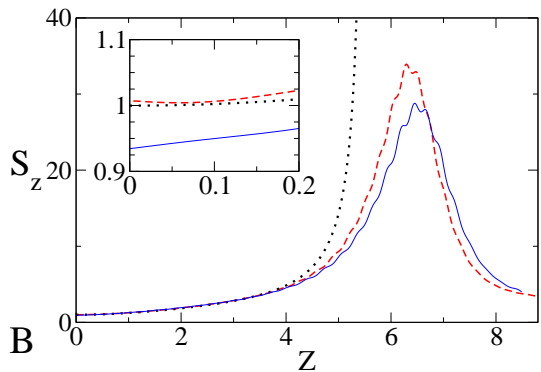} 
		\caption{(color online)\label{fig:adj}
 			Comparison of the NLH solutions with and without the adjustment of incoming beam
                         described in
			Section~\ref{sec:interface} (dashed red lines and solid blue lines,
			respectively), and the NLS solution (dotted black line).
        }
        \end{center}
\end{figure}

As indicated in Section~\ref{sec:interface}, the incoming beam for the NLH needs
to be adjusted so that to enable a more accurate comparison of the results with
those obtained for the corresponding NLS.
In this section, we investigate the difference between the NLH solutions
obtained with or without adjusting the incoming beam.
Namely, we analyze the critical case $D=3,\ \sigma=1$, and rerun the simulation of
Section~\ref{sssec:arrest-3D}  with $\Zmax=8.5$ and for the incoming beam
$\EincL=e^{-\rho^2}$, i.e., without adjusting the incoming beam.
The resolutions are $\lambda_0/h_z=83$ and $\lambda_0/h_x=67$ points per linear
wavelength.
The results presented in Figure~\ref{fig:adj} show that in this case the
collapse occurs later and achieves a smaller maximum self-focusing than for the
adjusted incoming beam.
The insert of Figure~\ref{fig:adj} also shows that near the boundary (after
the refraction by the interface) the solution with the adjusted incoming beam is indeed
much closer to the corresponding NLS profile.

\subsection{\label{ssec:method-comparison}Comparison with the Previous Method}

In the nested iteration scheme of \cite{FT:01,FT:04,BFT:05},
at each outer iteration the Kerr nonlinearity is considered fixed, or frozen,
which yields the linear homogeneous variable coefficient equation
\begin{equation}\label{eq:freezing-iterations}
	\left( 
		\Delta +k_0^2+ \epsilon k_0^2 |E^{(j)}|^{2\sigma}
	\right)	E^{(j+1)} =0.
\end{equation}
Equation~(\ref{eq:freezing-iterations}) is also solved iteratively, by building
a sequence of Born approximations.
In doing so, at each inner iteration an inhomogeneous linear constant
coefficient equation
\begin{equation}\label{eq:inner-iterations}
	\left( \Delta +k_0^2 \right) E^{(j+1,k+1)}
		= - \epsilon k_0^2 |E^{(j,K)}|^{2\sigma} E^{(j+1,k)}
\end{equation}
is solved using the separation of variables.
We will call this approach the ``nested iterations method.''
The efficacy of this method can be improved by getting rid of the inner
iterations~(\ref{eq:inner-iterations}) and solving
equation~(\ref{eq:freezing-iterations}) by the Gaussian elimination.
We will call this the ``freezing iterations method.''

In the one-dimensional case of \cite{BFT:07}, the freezing iterations
diverged above a certain nonlinearity threshold, while Newton's iterations
converged for the entire range of nonlinearities of interest.
In the current multi-D cases that correspond to the critical NLS, i.e., 
$D=3,\ \sigma=1$ and $D=2,\ \sigma=2$, both the nested iterations method and the 
freezing iterations method diverge when the NLS solution collapses, i.e., when 
the input power is above $P_{\text{c}}$, while Newton's algorithm converges, 
at least for some configurations, thereby showing a much better efficacy.

Another case of interest from the standpoint of applications is the subcritical 
NLS, $D=2,\ \sigma=1$, which admits solutions in the form of spatial solitons.
To compare the three methods in this case, we use each of them to repeat the
simulation of Section~\ref{sssec:single-soliton} while varying the domain size
$\Zmax$.
The quantity of interest is the threshold value
$\Zmax=Z_{\max}^{\text{threshold}}$, below which a given solver converges and
above which it diverges.
The results are given in Table~\ref{tab:efficacy-2D}.
We can see that the nested iterations method of \cite{FT:01,FT:04,BFT:05}
converges only for relatively short domains $
	Z_{\max}<Z_{\max}^{\text{threshold}}=42
$.
Replacing the inner iteration by a direct solver
brings along a certain improvement: $
	Z_{\max}<Z_{\max}^{\text{threshold}}=135
$.
However, similarly to the one-dimensional case, Newton's iterations converge for 
the widest selection of cases, at least until 
	$Z_{\max} = 500$.
Moreover, this limit is due to the memory constraints rather than
divergence, and the actual $Z_{\max}^{\text{threshold}}$ may be even larger.

\begin{table}[H]
	\begin{center}
		\caption{\label{tab:efficacy-2D}
			A comparison of the efficacy of the three methods for the soliton
			case $D=2,\ \sigma=1$.
			Each method converges for $Z_{\max}<Z_{\max}^{\text{threshold}}$
			and diverges for $Z_{\max}\geq Z_{\max}^{\text{threshold}}$.
		}
		\begin{tabular}{|c||c|c|c|}
			\hline 
			Method & 
nested freezing~(\ref{eq:freezing-iterations}), (\ref{eq:inner-iterations}) &
freezing~(\ref{eq:freezing-iterations}), LU solver &
Newton's \\
			\hline
			\hline 
			$Z_{\max}^{\text{threshold}}$ & 
				$42$ & 
				$135$ &  
				$>500$ \\
			\hline
		\end{tabular}
	\end{center}
\end{table}

\section{\label{sec:discussion}Discussion and Future Plans}

In this study, we  propose a novel numerical method for solving the scalar 
nonlinear Helmholtz equation, which governs the propagation of linearly polarized 
monochromatic light in Kerr dielectrics.
The NLH is the simplest model in nonlinear optics that allows for the
propagation of electromagnetic waves in all directions and, in particular, 
for backscattering, and accounts for nonparaxial effects.
Our key result is that the NLH eliminates the singularity that characterizes
solutions of the nonlinear Schr\"odinger equation, which is a reduced
model based on the paraxial approximation.
Another important finding is the discovery of narrow nonparaxial solitons and 
the development of numerical capability for simulating their collisions.

Mathematically, the NLH is an elliptic equation, and must be solved as a
nonlinear boundary-value problem. 
This presents additional difficulties for both analysis and computations
compared to the traditional treatment based on the NLS.
The latter has a predominant direction of propagation and requires a Cauchy
problem.
Physically, we consider the propagation of laser light in a layered medium with
interfaces across which both the linear and nonlinear components of
the refraction index may undergo jumps.
The presence of material discontinuities necessitates setting the condition
that the field and its first normal derivative be continuous at the interface.

To solve the NLH numerically, we develop a fourth-order finite difference scheme 
for one, two, and three space dimensions (in the latter case we assume
cylindrical symmetry). Finite differences are chosen over other possible
approximation strategies because of their simplicity and ease of implementation.
Indeed, the geometry of the problem enables a straightforward discretization on
a uniform rectangular grid.
On the other hand, having a high order scheme is important because it alleviates
the point-per-wavelength constraint for large domains and also helps resolve
the small-scale phenomenon of backscattering. In particular, high order accuracy
must be maintained across the material discontinuities. This is achieved by
using special one-sided differences. In doing so, to simplify the overall 
discretization we move the outer boundaries away from the interfaces so that the
artificial boundary conditions do not ``interfere'' with the interface
treatment.
The scheme used in the interior is of a semi-compact type, it is written on 
three nodes in the longitudinal direction and five nodes in the transverse
direction. 
Having a compact three-node stencil in the longitudinal direction greatly
simplifies both the treatment of the interfaces (no special ``near interface''
nodes) and the treatment of the
outer boundaries (no non-physical evanescent modes). At the same time, 
a compact stencil in the transverse direction is not required because
there are no material discontinuities. This circumstance greatly simplifies the
design of the overall scheme.

The second key component of the proposed algorithm is the nonlinear solver, 
which is based on Newton's method.
The simulations of~\cite{BFT:07} have demonstrated a clear superiority of 
Newton's method in the one-dimensional case.
In this paper, we generalize our Newton's solver to the multi-dimensional case,
with the expectation that it will let us solve the NLH for those settings when
the NLS breaks down, namely, when the NLS solution becomes singular
($\sigma(D-1)=2$ with input powers above critical), or when the beam width
becomes very narrow in the subcritical case ($D=2,\ \sigma=1$), or when
counter-propagating nonparaxial solitons interact.

The Newton's solver that we developed has indeed lived up to the promise.
In the critical cases, it enables the central result of this work, which is the
discovery of bounded NLH solutions for those cases when the corresponding
NLS solution blows up.
Physically, it shows that nonparaxiality can suppress the singularity formation
and hence arrest the collapse of focusing nonlinear waves.
While there may be other physical mechanisms that also help arrest the collapse
(neglected along the way when the NLH was derived from the Maxwell's equations),
it was not known until now whether the solution becomes regular already in the
framework of the scalar NLH model, which is the simplest nonparaxial model that 
incorporates the backward traveling waves.

Predictions of the NLH in the subcritical case include the existence
of narrow nonparaxial solitons, and analysis of the interactions (collisions) 
of such beams, specifically in counter-propagation.
These results may be of of relevance to potential applications, e.g., the design
of the next generation of all-optical circuits.
Note that in our previous work \cite{FT:04} we have already been able to compute
narrow spatial solitons. 
However, the new method proposed in this paper allows us to do that over much
longer propagation distances, see Section~\ref{ssec:method-comparison}.

Let us also note that a different configuration with counter-propagating
solitons has been studied by Cohen et.~al. in \cite{Cohen-counterprop:02} using
a system of coupled NLS equations which approximates the NLH.
As, however, mentioned in~\cite{Cohen-counterprop:02}, the coupled NLS model is
not problem free as it is neither an initial-value problem nor a boundary-value
problem.
In contrast, since the NLH is solved as a boundary-value problem, it is a
natural mathematical setting for such counter-propagating configurations.

The computational cost of the proposed algorithm still remains relatively 
high; it is dominated (both in memory and CPU time) by the cost of inverting
the Jacobian matrix using a direct method.
This cost can be reduced if the LU decomposition is replaced with
an iterative method. As, however, the Helmholtz operator subject to
the radiation boundary conditions is not self-adjoint, the only viable
choice of an iteration scheme will be a method of the Krylov subspace type.
For this method to work, the system must be preconditioned, and it is
the design of a good preconditioner that will be in the focus of our future
work on the linear solver. Several candidate techniques will be investigated,
including the constant coefficient Helmholtz operator to be inverted by the
separation of variables and a paraxial preconditioner based on the Schr\"odinger
operator. 

As far as the dependence of Newton's convergence on the domain size, 
see Section~\ref{sssec:domain-size},
we attribute it to the generally known ``fragility'' and,
in particular, sensitivity of Newton's convergence to the choice of the initial guess.
On one hand, it is intuitively reasonable to expect that if the outer boundary
is located in the region of maximum self-focusing, then the iterations may
experience difficulties to converge, see Table~\ref{tab:convergence-3D}.
On the other hand, at the moment we do not have a clear and unambiguous
mathematical explanation as to why exactly that happens.
We have tried a few simple remedies, such as using a continuation approach in the
nonlinearity coefficient $\epsilon$ and using a damped NLS solution as the
initial guess, but none of those has made a substantial difference.
We note that in the one-dimensional case the exact solution was available
in the closed form \cite{BFT:07} and hence we could at least test Newton's
convergence by substituting this exact solution as the initial guess. In multi-D,
however, we are not aware of any closed form solutions for the slab of finite
thickness and therefore, a similar validation procedure becomes problematic.

The piecewise constant formulation that we have considered in the paper in
fact presents no loss of generality, at least from the standpoint of numerical
solution.
It can be very easily extended to the NLH with piecewise smooth material
coefficients $\lri^2(\bvec x)$ and $\epsilon(\bvec x)$.
All one needs to do is replace the constants $\lri$ and $\epsilon$ in the
definition of the scheme with the values at the corresponding grid nodes:
$\lri_{n,m}\equiv\lri(z_n,x_{\perp,m})$ and  
$\epsilon_{n,m}\equiv\epsilon(z_n,x_{\perp,m})$. 
However, while the resulting scheme will approximate the variable coefficient
scalar NLH~(\ref{eq:NLH}) with fourth-order accuracy, the validity of
equation~(\ref{eq:NLH}) itself from the standpoint of physics may be in question.
Indeed, the derivation of the scalar NLH from Maxwell's equations in the case of
variable coefficients introduces additional terms (spatial derivatives of $\lri$
and $\epsilon|E|^2$) which are not included in equation~(\ref{eq:NLH}).

The layered structure and simple geometry that we have adopted present no
substantial loss of generality, because this formulation corresponds to many
actual physical (e.g., laboratory) settings.
The  plain-parallel setup studied in the paper certainly simplifies the
discretization. 
At the same time, we are reasonably confident that the proposed scheme can be
generalized to more elaborate geometries without compromising its high order
accuracy, which is of key importance.
One natural approach to doing that is to use Calderon's
projections and the method of difference potentials~\cite{ryab-02}.

From the standpoint of physics, the scalar NLH is certainly not the most
comprehensive model. 
It is rather a reduced model based on a number of simplifications. 
Most notably, the vector nature of electromagnetic field is not taken into
account by the scalar NLH because of the assumption of linear polarization.
Vectorial effects, on the other hand, are known to become important close to
 when the
nonparaxiality does, i.e., once the beam width becomes comparable to the carrier
wavelength.
Moreover, the scalar NLH governs monochromatic fields (continuous-wave laser), 
whereas the actual fields are always time-dependent (typically, pulses of 
certain duration).
Nonetheless, if the duration of the pulse is sufficiently long (many oscillation
periods), then the time-periodic model will provide 
a good approximation.

To take into account the entire range of relevant physical phenomena one needs,
of course, to go back and solve the full nonlinear Maxwell's equations.
This, however, is a very challenging computational task and besides, the
solutions of full Maxwell's equations may be hard to analyze or verify precisely
because of all too many additional physical effects.
That's why the analysis of the simplest nonparaxial model (i.e., the NLH) may
provide a very useful insight into the relevant physics as, in particular, it
allows to study the important phenomenon of nonlinear backscattering.

Given the previous considerations, we believe that in the context of physics, 
the next most natural and most beneficial extension of the work presented in
this paper will be taking into account the vectorial effects.
The current work provides a solid foundation for this extension as many key
elements of the algorithm, e.g., the nonlocal artificial boundary conditions,
will only require technical rather than conceptual changes.
On the pure numerical side, in addition to the previously mentioned major
modifications to the linear solver, we can consider a number of strategies aimed
at further improving the numerical resolution in the regions of foremost
interest (e.g., around the maximum self-focusing) while not increasing the
overall computational cost.
Examples include local grid requirement and/or combined approaches when most of
the domain is to be done using the NLS whereas the local area of collision
between the solitons is computed using the NLH.

\appendix

\section{Continuity Conditions at Material Interfaces}
\label{app:interface}

For optical frequencies, we can disregard all magnetization effects in the
medium (see \cite[Chapter IX]{land8}) and write down the time-harmonic Maxwell's
equations as follows:
\begin{equation}
\label{eq:max}
\frac{i\omega}{c}\B=\curl\E,\qquad\qquad -\frac{i\omega}{c}\D=\curl\B,
\end{equation}
where the specific form of how the electric induction $\D$ depends on the field $\E$
is not important for the derivation of the interface conditions. 
Note, however, that as our medium is a dielectric, both fields $\E$ and $\B$, as well as 
the induction $\D$, remain finite everywhere including the interfaces.

Let an interface plane be normal to the coordinate $z$ of the Cartesian system 
$(x,y,z)$. Then, the first equation of (\ref{eq:max}) 
implies that the quantity
$(\curl\E)_x=\frac{\partial E_z}{\partial y}-\frac{\partial E_y}{\partial z}$ is
bounded at the interface. As the derivative $\frac{\partial E_z}{\partial y}$, which is
taken along the interface, is bounded in its own right, we conclude that 
$\frac{\partial E_y}{\partial z}$ is bounded. This immediately yields the continuity
of $E_y$ across the interface. The continuity of $E_x$ can be established the same way,
by taking into account the boundedness of 
$(\curl\E)_y=\frac{\partial E_x}{\partial z}-\frac{\partial E_z}{\partial x}$.
Altogether, this means that the tangential component of the electric field $\E$ must
remain continuous.
Likewise, the continuity of the 
tangential component of $\B$ across
the interface can be derived by employing the  second equation of (\ref{eq:max})
and the boundedness of $\D$.

Next, consider the case of linear polarization:
\begin{equation*}
\E=[E_x,0,0]\qquad\text{and}\qquad\B=[0,B_y,0].
\end{equation*}
Then, the continuity of $B_y$ immediately implies the continuity of 
$\frac{\partial E_x}{\partial z}$, because from the Faraday law 
(the first equation of (\ref{eq:max})) we now have: $\frac{i\omega}{c}B_y=
\frac{\partial E_x}{\partial z}$.
Altogether, we conclude that for the linearly polarized light propagating
through a (transparent) dielectric with material discontinuities, both the
electric field $\E$ and its first normal derivative must be continuous at
all the interfaces.

\section{\label{app:CDOs} Notation for Central Difference Operators}

We denote the central difference operators by the letter $D$ with the order of
differentiation in the subscript and the order of accuracy in the
superscript.
The full list for the finite differences in the $x$ (or $\rho$) direction is as
follows:
{\allowdisplaybreaks
\begin{eqnarray*} 
	\CDO{x}{2} E & \myeqdef &
		{\displaystyle \frac{
			E_{n,m+1} - E_{n,m-1} 
		}{ 
			2 h_x
		}}
		= \partial_x E_{n,m} + \oh{2},  \\
	\CDO{xx}{2} E & \myeqdef &
		{\displaystyle \frac{
			E_{n,m+1} -2 E_{n,m} +E_{n,m-1} 
		}{
			h_x^2
		}}
		= \partial_{xx} E_{n,m} + \oh{2},  \\
	\CDO{xxx}{2} E & \myeqdef &
		{\displaystyle \frac{
			E_{n,m+2} -2 E_{n,m+1} +2 E_{n,m-1} - E_{n,m-2}
		}{ 
			2 h_x^3
		}}
		= \partial_{xxx} E_{n,m} + \oh{2}, \\
	\CDO{xxxx}{2} E & \myeqdef &
		{\displaystyle \frac{
			E_{n,m+2}-4E_{n,m+1}+6E_{n,m}-4E_{n,m-1}+E_{n,m-2}
		}{ 
			h_x^4
		}}
		= \partial_{xxxx} E_{n,m} + \oh{2}, \\
	\CDO{x}{4} E & \myeqdef &
		{\displaystyle \frac{
			- E_{n,m+2} +8 E_{n,m+1}  - 8 E_{n,m-1} +E_{n,m-2}
		}{
			12h_x 
		}}
		= \partial_x E_{n,m} + \oh{4}, \\
	\CDO{xx}{4} E & \myeqdef &
		{\displaystyle \frac{
			-E_{n,m+2}+16 E_{n,m+1} - 30 E_{n,m} +16 E_{n,m-1} -E_{n,m-2}
		}{ 
			12 h_x^2
		}}
		= \partial_{xx} E_{n,m} + \oh{4}. 
\end{eqnarray*}
}

Because of the semi-compact approximation we use, only the second-order
operator is required in the $z$ direction
\[
	\CDO{zz}{2} E  \myeqdef 
		{\displaystyle \frac{
			E_{n+1,m}-2E_{n,m}+E_{n-1,m}
		}{
			h_z^2
		}}
		= \partial_{zz} E_{n,m} + \oh{2}.  
\]

\bibliographystyle{elsarticle-num}
\bibliography{NLH,NLS}

\end{document}